\documentclass[12pt]{article}

\usepackage{newtxtext,newtxmath}

\usepackage{graphicx}

\usepackage[letterpaper,margin=1in]{geometry}

\renewenvironment{abstract}
	{\quotation}
	{\endquotation}

\date{}

\makeatletter
\renewcommand{\fnum@figure}{\textbf{Figure \thefigure}}
\renewcommand{\fnum@table}{\textbf{Table \thetable}}
\makeatother

\usepackage{scicite}

\usepackage{url}

\usepackage[T1]{fontenc} 

\def\scititle{Helium escaping from the atmosphere of a nearby rocky exoplanet orbiting in a habitable zone}
\title{\bfseries \boldmath \scititle}

\author{
Collin Cherubim$^{1,2\ast}$,
Shreyas Vissapragada$^{3}$,
Tim Cunningham$^{2}$,
Annabella G. Meech$^{2}$, \\
David Charbonneau$^{2}$,
Robin Wordsworth$^{4,1}$,
Aaron Householder$^{5, 6}$,
Johanna Teske$^{3,7}$, \\
Leonardo A. Dos Santos$^{8,9}$,
Nicole L. Wallack$^{7}$,
William Misener$^{7}$,
Zifan Lin$^{10,11}$, \\
Andrew McWilliam$^{3}$,
Michael Zhang$^{12}$,
Jason A. Dittmann$^{13}$,
Mercedes L\'opez-Morales$^{8}$ 
\and
\small$^{1}$Department of Earth and Planetary Sciences, Harvard University, Cambridge, USA. 
\and
\small$^{2}$Center for Astrophysics, Harvard \& Smithsonian, Cambridge, USA.
\and
\small$^{3}$Carnegie Science Observatories, Pasadena, USA.
\and
\small$^{4}$School of Engineering and Applied Sciences, Harvard University, Cambridge, USA.
\and
\small$^{5}$Kavli Institute for Astrophysics and Space Research, Massachusetts Institute of Technology, Cambridge, USA.
\and
\small$^{6}$Department of Earth, Atmospheric and Planetary Sciences, Massachusetts Institute of Technology, Cambridge, USA.
\and
\small$^{7}$Earth and Planets Laboratory, Carnegie Institution for Science, Washington, USA.
\and
\small$^{8}$Space Telescope Science Institute, Baltimore, USA.
\and
\small$^{9}$Department of Physics \& Astronomy, Johns Hopkins University, Baltimore, USA.
\and
\small$^{10}$Department of Physics, Washington University, St. Louis, USA.
\and
\small$^{11}$McDonnell Center for the Space Sciences, Washington University, St. Louis, USA.
\and
\small$^{12}$Department of Astronomy and Astrophysics, University of Chicago, Chicago, USA.
\and
\small$^{13}$Department of Astronomy, University of Florida, Gainesville, USA
\and
\small$^\ast$Corresponding author. Email: ccherubim@g.harvard.edu
}

\begin{document} 
\maketitle

\begin{abstract} \bfseries \boldmath
Observations of highly irradiated gas giant exoplanets have shown helium escaping from their atmospheres. 
There is limited evidence for atmospheres on rocky exoplanets, perhaps because they have already escaped. We report spectroscopic observations of LHS~1140b, a rocky exoplanet that orbits in the habitable zone of a nearby low-mass star. The near-infrared transit spectra show absorption by helium escaping from the planet's atmosphere. Helium absorption is detected in 2024 but not in 2025, indicating time-variable atmospheric escape. We interpret these results as indicating an upper atmosphere dominated by helium and depleted in hydrogen, with other volatile species trapped at lower altitudes, consistent with atmospheric fractionation models. No helium absorption is detected for LHS~1140c, a smaller and more heavily irradiated exoplanet in the same system.
\end{abstract}

\noindent

Theoretical models predict that atmospheres foster conditions for life on rocky exoplanets by regulating the climate, shielding the surface from ionizing radiation, and enabling the presence of liquid water~\cite{Kasting_1993, Wordsworth_2022}.
Atmospheres have been observed on large, gas-rich, highly irradiated exoplanets~\cite{Charbonneau_2002, Madhusudhan_2019}, but
observing atmospheres on smaller, cooler, rocky exoplanets is technically challenging because they are dwarfed (in size and brightness) by the stars they orbit. 
Those challenges can be reduced by studying planets that orbit red dwarf stars (M dwarfs), whose small sizes and low brightness reduce the contrast between the star and any orbiting planet.
However, M dwarfs are more active than Sun-like stars, emitting high-energy radiation that can drive atmospheric escape from closely orbiting planets. It is therefore unclear whether such planets can retain their atmospheres for billions of years~\cite{Zahnle_2017, Pass_2025}. Observations of small, rocky exoplanets have mostly revealed airless worlds or atmospheres too tenuous to detect, with some debated evidence for atmospheres \cite{Diamond-Lowe_2023, Hu_2024, Glidden_2025, Fortune_2025, Teske_2025}.

\subsection*{The LHS~1140 system}

The transiting rocky exoplanet LHS~1140b 
has a mass of 5.60 $\pm\ 0.19$ Earth masses ($\mathrm{M}_\oplus$) and a radius of 1.730 $\pm\ 0.025$ Earth radii ($\mathrm{R}_\oplus$), consistent with a rocky, Earth-like bulk composition plus a low-density component such as an atmosphere or a high abundance of water.
It has an orbital period of 24.7 days and receives 42\% of the stellar irradiation received by Earth, giving it an equilibrium temperature of $T_\mathrm{eq}$ = 226 ± 4 K (assuming zero albedo), placing it in the liquid-water habitable zone~\cite{Dittmann_2017, Cadieux_2024a}. There is another transiting rocky planet in the same system, LHS~1140c (1.91 $\pm\ 0.06\ M_\oplus$ and 1.272 $\pm\ 0.026\ R_\oplus$) with an orbital period of 3.78 days, which receives about five times the irradiation received by Earth~\cite{Ment_2019}. The host star LHS~1140 (also cataloged as GJ 3053) is an old [$> 3$ Gyr~\cite{Dittmann_2017, Pass_2024, Cadieux_2024a}], inactive~\cite{Dittmann_2017, Medina_2022} M dwarf located 14.96 ± 0.01 parsecs away from the Sun~\cite{Cadieux_2024a}.

\subsection*{Spectroscopic observations of LHS~1140}

We observed the LHS 1140 system using the Warm Infrared Echelle Spectrograph to Realize Extreme Dispersion (WINERED) mounted on the Magellan Clay telescope at Las Campanas Observatory as part of the WINERED Helium Consortium. On 2024 September 23, we observed the system for 6.5 hours, covering one transit of each planet, separated by 39 minutes. A total of 70 spectra were collected in total: 35 out-of-transit, 12 during the transit of LHS~1140c, and 23 during the transit of LHS~1140b. We used the WINERED Automatic Reduction Pipeline [WARP~\cite{Hamano_2024}] for the initial data reduction~\cite{methods}. 
To construct time series spectra showing wavelength-dependent absorption of stellar radiation by the planet's atmosphere (excess absorption), we divided each spectrum by a stellar template, where $F_\mathrm{out}$ is the mean stellar flux of all out-of-transit spectra (Fig. \ref{fig1}A). The resulting time series (\ref{fig1}B) shows an absorption feature near 10,833 $\textnormal{\AA}$ during the transit, pre-ingress, and post-egress of LHS~1140b, consistent with the presence of metastable helium.

We produced a planetary transmission spectrum of LHS~1140b (Fig. \ref{fig2})
by taking the mean of all in-transit excess absorption spectra in the planetary rest frame~\cite{methods}.
This transmission spectrum contains correlated noise, which we modeled using a Gaussian process [GP;~\cite{methods}]. 
Metastable helium is expected to produce a triplet of closely spaced absorption lines, which we modeled with three Gaussian profiles at 
10,832.057 \AA\, 10,833.217, and 10,833.306 \AA\ 
[rest wavelengths in vacuum~\cite{Drake_1996}]; the latter two lines are blended at the resolution of the WINERED spectra. We used a Markov Chain Monte Carlo (MCMC) analysis to fit the model to the data and determine the uncertainties in our measurements. The MCMC process followed a Bayesian retrieval framework with five free parameters: the three peak amplitudes, a shared peak width, and a shared Doppler shift~\cite{methods}.

We report results as the median values and 16 to 84\% confidence intervals of the MCMC posterior probability distributions. The excess absorption depth is $1.24^{+0.22}_{-0.23}$\% at the position of the two blended long-wavelength peaks and $0.25^{+0.14}_{-0.12}\%$ for the single short-wavelength peak, with a Doppler shift of $0.072^{+0.080}_{-0.073}$ \AA\ relative to the planet velocity, equivalent to $2.0^{+2.0}_{-2.2}$ km s$^{-1}$. This corresponds to an equivalent opaque radius (the planetary radius including an opaque atmospheric layer that would produce the observed absorption feature) of 1.52 times the radius of LHS~1140b. The full-width at half-maximum (FWHM) of the blended helium absorption lines is $0.86^{+0.15}_{-0.27}$ \AA, corresponding to $23.9^{+4.2}_{-7.5}$ km s$^{-1}$. 
The measured ratio of the blended red peak amplitude to the blue peak amplitude is $6.7^{+12.7}_{-3.1}$. This is consistent with the ratio of 8 we expect due to the fine structure of the helium triplet~\cite{de_Jager_1966, Drake_1971}; this assumes negligible optical depth ($\tau \ll 1$) in the thin upper atmosphere where metastable helium can persist. 

We also fitted the nine pre-ingress spectra with apparent helium absorption in the time series (Fig. \ref{fig1}) using the same model and MCMC process. The resulting absorption depth is $1.01^{+0.31}_{-0.34} \%$ and the Doppler shift is $0.33^{+3.52}_{-7.53}$ km s$^{-1}$. This is consistent with escaping helium ahead of the planet in its orbit (a leading tail). Leading tails have been previously observed for some gas giant exoplanets with escaping atmospheres~\cite{Czesla_2022, Gully-Santiago_2024, Krishnamurthy_2026} and has been interpreted as resulting from stellar winds or interactions between the magnetic fields of the star and planet~\cite{Matsakos_2015, McCann_2019}. We also performed the same analysis for the eight post-egress spectra, finding an absorption depth of $\leq 0.76 \%$ (1$\sigma$ upper limit) and a Doppler shift of $0.53^{+0.33}_{-0.42}$ \AA\ equivalent to $14.7^{+11.8}_{-9.0}$ km s$^{-1}$, 
indicating tentative evidence of a trailing tail. We calculated the mean helium absorption between 10,833 and 10,834 \AA\ to construct a transit light curve of the blended red absorption lines (Fig. \ref{fig3}A). The transmission spectra of the LHS~1140b transit, leading tail, and possible trailing tail are shown in Fig. \ref{fig3}B-D. 

We repeated the same analysis for the transit of LHS~1140c in 2024 (see Supplementary Text), but found no evidence of helium absorption (Fig. \ref{fig:LHS1140c_2024}). The transit of LHS~1140b in 2025 also shows no evidence of helium absorption (Figs. \ref{fig4}, \ref{fig:timeseries_2025}). To investigate whether these differences between transits could be explained by choices in the data reduction process, the 2024 and 2025 datasets for LHS~1140b were independently re-reduced using a different reduction code; we found no difference in the results [Fig. \ref{fig:MZ_reduction}; \cite{methods}].



\subsection*{Interpretation as an atmospheric outflow} 
\label{sec:composition}

We attribute the helium absorption observed for LHS~1140b in 2024 to a hydrodynamic outflow from an atmosphere, driven by heating due to stellar X-ray and extreme-ultraviolet (XUV, collectively) radiation~\cite{Murray-Clay_2009}. We considered several alternative explanations, including stellar activity or contamination of the spectra by Earth's atmosphere, but exclude each of them (see Supplementary Text).

To determine the physical properties of the atmospheric outflow, we modeled the observed spectra of LHS~1140b using the \texttt{p-winds} code~\cite{Dos_Santos_2022}. This code models an escaping atmosphere as a one-dimensional, isothermal outflow~\cite{Parker_1958} then forward models a predicted transmission spectrum using a radiative transfer model. We used an MCMC process to estimate the atmospheric mass-loss rate, the temperature of the outflow $T_\mathrm{wind}$, the hydrogen to helium atomic number ratio (H:He) of the outflow, and the line-of-sight bulk velocity shift $v_\mathrm{wind}$, corresponding to the Doppler shift of the helium absorption lines. This calculation requires an input stellar spectral energy distribution (SED) to determine the ionization rates and number densities of hydrogen and helium as functions of altitude. We analyzed archival X-ray observations of LHS~1140 using the X-ray Multi-Mirror Mission (XMM-Newton), to determine the X-ray flux of the host star. This was then used to scale the previously published semi-empirical SEDs 
of GJ~1132 and GJ~699 (Barnard's star), two M dwarfs with similar spectral types, masses, and rotation rates to LHS~1140 [Fig. \ref{fig:SEDs}; \cite{methods}]. 

For the GJ~1132 (GJ~699) SED, the MCMC modeling indicates an atmospheric mass-loss rate of $2.03^{+0.67}_{-0.58} \times 10^8$ ($4.22^{+1.14}_{-1.00} \times 10^8$) g s$^{-1}$, H:He ratio of $1.01^{+0.85}_{-0.50} \times 10^{-3}$ ($1.32^{+0.89}_{-0.56} \times 10^{-3}$), $T_\mathrm{wind}$ of $5160^{+46}_{-50}$ ($5850^{+78.1}_{-81.9}$), and velocity shift of $2260^{+330}_{-300}$ ($2270^{+330}_{-290}$) m s$^{-1}$ (Figs. \ref{fig5}, \ref{fig:pwinds_corner}, and \ref{fig:pwinds_corner_699}). 
For comparison, we used previous methods~\cite{Caldiroli_2022} to calculate the mass-loss rate expected for an outflow limited by XUV energy from stellar irradiation, finding 6.2 to $29 \times 10^7$ g s$^{-1}$. To estimate the incident XUV flux for this calculation, we integrated the scaled SEDs for GJ~1132 and GJ~699 between 10 and 1300 \AA. 
An alternative mechanism for atmospheric escape, core-powered mass loss, predicts a much smaller mass-loss rate \cite{Ginzburg_2016, Ginzburg_2018}, so we conclude that the atmospheric escape is predominantly driven by the XUV stellar radiation. These calculations are not sensitive to the composition of the outflow 
assumed in the core-powered mass loss scenario.

The age of the LHS~1140 system is not well constrained. The star's long rotation period of 131 days~\cite{Dittmann_2017} implies an age $\gtrsim$ 3.1 Gyr, assuming the typical spin-down rates of M dwarfs~\cite{Pass_2024}. Assuming the atmosphere was initially 1.5\% of the planetary mass~\cite{Ginzburg_2016}, the atmosphere of LHS~1140 b would have been completely removed if the mean escape rate was $\gtrsim 5 \times 10^9$ g s$^{-1}$. We interpret this as an upper limit on the present-day escape rate because the XUV flux was probably orders of magnitude higher in the past~\cite{Peacock_2020}.

LHS~1140 has an X-ray luminosity of $6.5 \times 10^{-6}$ times its bolometric (all wavelengths) luminosity~\cite{methods}, which is low for an M dwarf. The bolometric luminosity is 0.0038 $\pm$ 0.0003 times the Sun's luminosity, and LHS~1140b receives 42\% of the stellar energy received by Earth. The X-ray flux received by LHS~1140b is therefore 
2.7 to 16 times that received by Earth, where the range corresponds to periods of high and low solar activity, over the Sun's 11-year cycle~\cite{methods}. Our \texttt{p-winds} models indicate that for H:He ratios $\gtrsim$0.01 in the escaping wind, atmospheric hydrogen would attenuate the relatively low XUV flux $\leq 911$ \AA, which would prevent the ionization of neutral helium and the subsequent production of metastable helium. We therefore conclude that the upper atmosphere of LHS~1140b has a low abundance of hydrogen relative to helium.

\subsection*{Atmospheric composition}

LHS~1140b's transmission spectrum has been constrained by previous observations~\cite{Diamond-Lowe_2020, Edwards_2021, Cadieux_2024b, Damiano_2024, Biagini_2024}. Those studies did not detect an atmosphere but ruled out cloud-free, hydrogen-dominated atmospheres with metallicities (abundance of elements heavier than helium) up to 1,000 times that of the Sun. This constraint and our retrieved H:He ratio of $\sim 1 \times 10^{-3}$ are consistent with predictions from a previous model of planetary atmospheric escape and magma ocean evolution~\cite{Cherubim_2025}. That study predicted that LHS~1140b might have a helium-dominated atmosphere due to mass fractionation by hydrodynamic atmospheric escape. Other models have also predicted escape-driven formation of helium-dominated atmospheres for exoplanets similar in size and temperature~\cite{Hu_2015, Lammer_2025}. 

The measured mass of LHS~1140b~\cite{Cadieux_2024a} implies a bulk density slightly less than that of an airless rocky planet. A helium-dominated atmosphere would have a height $\sim 1.7 \times$ smaller than that of a solar-composition, hydrogen-dominated atmosphere, and likely would have been detected in previous studies~\cite{Damiano_2024, Cadieux_2024b} if it were free of high-altitude clouds or hazes, which can obscure features in a transmission spectrum. Clouds and hazes are unlikely to form at the equilibrium temperature of LHS~1140b's atmosphere~\cite{Brande_2024}. Previous observations of LHS~1140b have ruled out water clouds and, tentatively, hazes composed of methane or hydrogen sulfide~\cite{Damiano_2024, Cadieux_2024b}. While our observations indicate that the upper atmosphere is probably helium-dominated, it is possible that the bulk atmosphere (at lower altitudes) is more metal-rich.

For a helium-dominated wind escaping at a rate of 3 $\times\ 10^8$ g s$^{-1}$, the mass below which atmospheric species can be dragged into the escaping outflow (crossover mass), is $< 9$ amu at any temperature $< 10,000$ K \cite{Hunten_1987}. In other words, our inferred escape rate of helium is insufficient to carry off any species with mass $\geq 9$ amu. We calculate that the minimum atmospheric escape rate required to drag atomic oxygen is $2 \times 10^9$ g s$^{-1}$ (Fig. \ref{fig5}). We therefore expect the observed outflow to enrich the atmosphere in heavier elements like O, C and N.

An alternative explanation for the bulk density of LHS~1140b is an Earth-like ratio of iron and rock plus 9 to 19\% water by mass~\cite{Cadieux_2024a}. This explanation assumed an Earth-like atmosphere with 1 bar surface pressure, and did not account for escaping helium. However, if the planet does contain a substantial amount of water, it would probably be 
shielded from escape via condensation (cold trapping) in the upper part of the atmosphere’s convective layer (tropopause), given the cool equilibrium temperature of 226 K.
A common approximation of the tropopause temperature is the skin temperature, which is $T_\mathrm{skin} \equiv 2^{-1/4} T_\mathrm{eq} = 194 $ K for LHS~1140b~\cite{Pierrehumbert_2010}. Atmospheric species with boiling points above this value are expected to be cold trapped and are less likely to escape. For example, we estimate the molar concentration of water at the tropopause as $f_\mathrm{H_2O} = P_\mathrm{sat, trop}/P_\mathrm{trop}$, where $P_\mathrm{sat, trop}$ is the water saturation vapor pressure at the skin temperature and $P_\mathrm{trop}$ is the tropopause pressure, which we assume is 0.1 bar, based on its value for the Solar System planets and previous radiative transfer models of LHS~1140b~\cite{Robinson_2012, Cadieux_2024a, Damiano_2024}. We find $f_\mathrm{H_2O} \approx 7$ ppm at the tropopause of LHS~1140b (compared to $\sim$3 ppm on Earth). This is consistent with an outflow that lacks hydrogen, which would be produced by photodissociation of water if it reached the upper atmosphere. The H:He ratio we estimated for the upper atmosphere implies that highly volatile, reduced species, such as CH$_\mathrm{4}$, could be depleted in the lower atmosphere because their vertical transport would not be inhibited by the cold trap.


\subsection*{Variable helium escape}

The non-detection of helium in 2025 could be due to variable atmospheric escape from LHS~1140b.
To investigate this possibility, we generated helium line models with the \texttt{p-winds} software, assuming 
XUV fluxes and escape rates from 1\% to 33\% of the fiducial values (0.033 W m$^2$ and $2.03^{+0.67}_{-0.58} \times 10^8$ g s$^{-1}$) for the SED derived using GJ~1132 spectra, and $T_\mathrm{wind}$ from 5160 K (the best fitting temperature from the 2024 data) to 6400 K. Several of these models predict helium absorption depths $\lesssim$0.6\% (Fig. \ref{fig4}), the detection limit of our 2025 observations (see Supplementary Text). 
We therefore conclude that there could have been undetectable helium escape in 2025 for XUV flux or upper atmosphere temperature variations observed for other stars [\cite{Qian_2012, Levine_2024, Pillitteri_2026} see Supplementary Text].

\subsection*{Constraints on the cosmic shoreline}

The cosmic shoreline is a proposed boundary that separates airless rocky planets from those that retain atmospheres for billions of years~\cite{Zahnle_2017, Pass_2025}. The two planets in the LHS~1140 system are on either side of the proposed cosmic shoreline.
Our non-detection of helium absorption by LHS~1140c is consistent with the previously measured dayside emission, which indicates that the planet has little to no atmosphere~\cite{Fortune_2025}. 
Therefore, this system is consistent with the proposed position of the cosmic shoreline.

\begin{figure}
\centering
\includegraphics[width=0.9\textwidth]{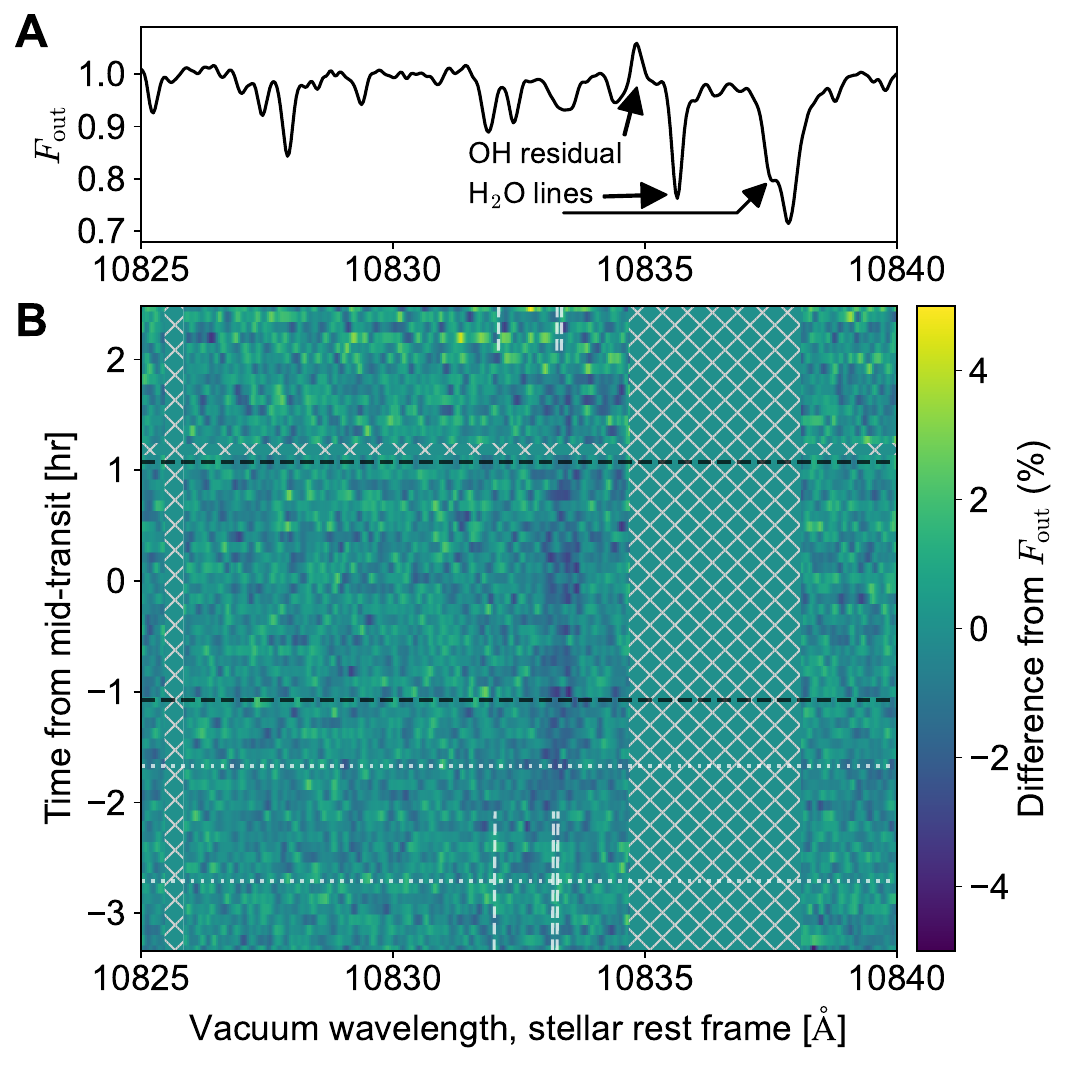}
\caption{\textbf{Time series spectra for LHS~1140b observed in 2024.} (A) The average out-of-transit stellar template spectrum constructed from exposures with no helium absorption apparent in the time series (see Fig. \ref{fig3}). Arrows indicate absorption features from Earth's atmosphere that are masked in panel B. (B) Time series spectra of LHS~1140b in the stellar rest frame. Colors indicate the percentage difference from the stellar template in panel A. The black dashed lines show the expected transit of LHS~1140b and the white dotted lines show that of LHS~1140c. The vertical dashed white lines indicate the expected positions of helium absorption lines moving with the the same velocity as LHS~1140b. The vertical spaces with grid lines show the mask for Earth's atmospheric features and the horizontal space with grid lines shows a noisy exposure excluded from our analysis.}\label{fig1}
\end{figure}

\begin{figure}
\centering
\includegraphics[width=0.8\textwidth]{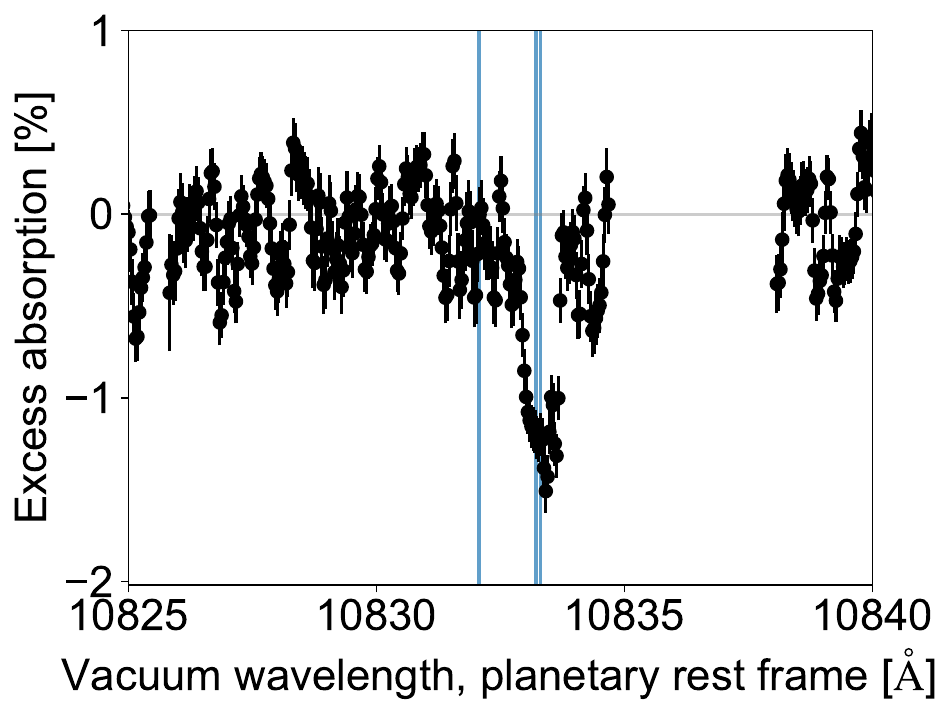}
\caption{\textbf{Raw transmission spectrum for LHS~1140b in 2024.} Excess absorption (black dots) is plotted as the mean of the in-transit spectra shown in Fig. \ref{fig1}. Pre-ingress and post-egress spectra have been excluded, even when they contain evidence for helium absorption. The vertical blue lines indicate the rest wavelengths of the helium absorption lines. The horizontal gray line indicates zero absorption. The gap in the data is due to the same data exclusion as in Fig. \ref{fig1}B. Error bars show $1\sigma$ uncertainties.}\label{fig2}
\end{figure}

\begin{figure}
\centering
\includegraphics[width=1.0\textwidth]{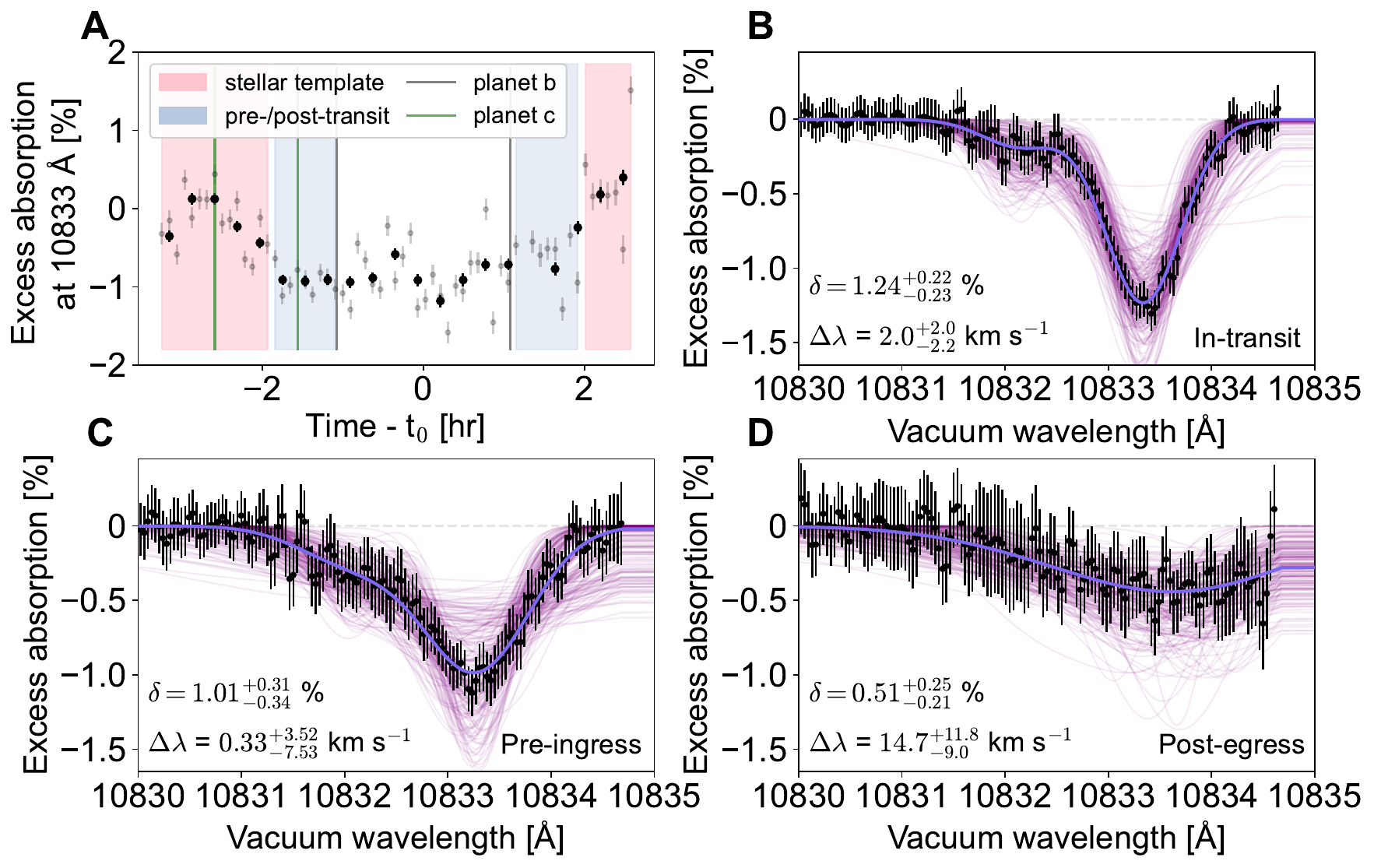}
\caption{\textbf{Helium absorption as a function of time and line profiles observed in 2024.} (A) Mean helium absorption calculated between 10,833 to 10,834 \AA\ as a function of time relative to the mid-transit ($t_0$) of LHS~1140b. Gray points are all observed spectra, and black points have been binned by a factor of three (17 minutes). The pink shaded regions indicate the data used to construct the stellar template (Fig. \ref{fig1}). The blue shaded regions show the pre-ingress and post-egress data used to construct the transmission spectra shown in panels C and D, which were chosen by eye from Fig. \ref{fig1}B. Vertical lines show the expected transit times of LHS~1140b (gray) and LHS~1140c (green). (B-D) Black dots show the transmission spectrum constructed from the in-transit data (B), pre-ingress data (C), and post-egress data (D), after subtraction of the Gaussian process model.
The thick blue line shows the best-fitting absorption line model, which has the labeled values of the absorption depth $\delta$ and Doppler shift $\Delta \lambda$. The thin purple lines show 100 random samples drawn from the MCMC fitting process. All error bars show $1\sigma$ uncertainties.}\label{fig3}
\end{figure}

\begin{figure}
\centering
\includegraphics[width=1.0\textwidth]{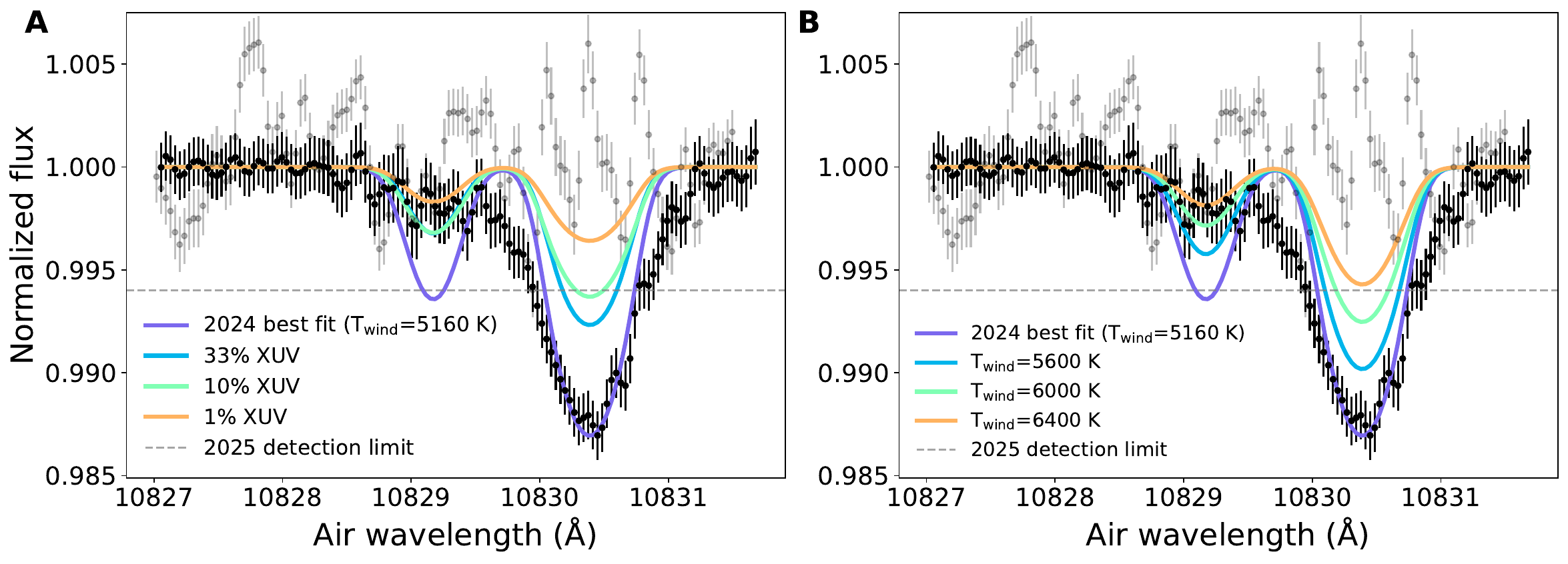}
\caption{\textbf{Comparison between the 2024 and 2025 observations of LHS~1140b.} Data points are the raw transmission spectrum in 2025 (gray points) and the GP-subtracted transmission spectrum in 2024 (black points, as in Fig. \ref{fig2}). Error bars show 1~$\sigma$ uncertainties. (A) Colored lines are \texttt{p-winds} models with different XUV flux (see legend) and escape rates (assumed to be proportional to the XUV flux), relative to best-fitting model to the 2024 data (blue line). The gray dashed line shows the detection limit of 0.6\%, determined through injection-recovery tests (see Supplementary Text). (B) Same as panel A, but varying the outflow temperature at fixed XUV flux. Several models with low XUV flux or high temperature predict helium absorption depths at or below the detection limit.}\label{fig4}
\end{figure}

\begin{figure}
\centering
\includegraphics[width=1.0\textwidth]{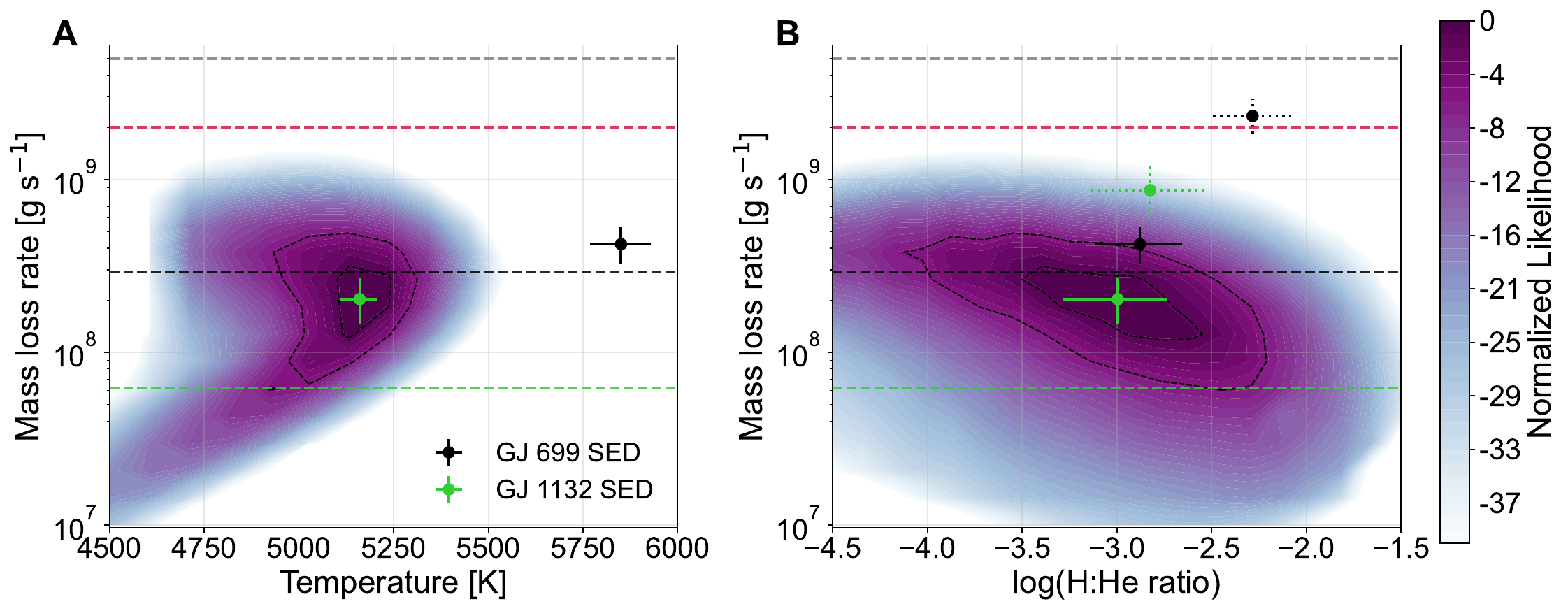}
\caption{\textbf{Models of mass-loss rate, outflow temperature, and H:He ratio for LHS~1140b in 2024.} (A) Models of atmospheric mass-loss rate as a function of outflow temperature, $T_\mathrm{wind}$. Color indicates the normalized likelihood calculation, where zero corresponds to the maximum likelihood. Black contours indicate $1\sigma$ and $3\sigma$ derivations from the maximum likelihood. 
Data points with 1$\sigma$ error bars are the best-fitting values from the MCMC analysis of the 2024 spectrum of LHS~1140b, assuming an SED normalized to GJ~699 (black) or GJ~1132 (green). All models assumed the GJ~1132-normalized SED and fixed $v_\mathrm{wind}$ the best-fitting value from the MCMC analysis. The dashed gray line shows a theoretical upper limit on the escape rate (see text). The dashed red line shows the theoretical minimum escape rate required to drag atomic oxygen in the outflow. The dashed black and green lines show the energy-limited escape rates calculated with the SEDs normalized by GJ~699 and GJ~1132, respectively. (B) Same as panel A, but as a function of the H:He ratio. Data points with dotted error bars assumed 10 times higher stellar XUV flux.}\label{fig5}
\end{figure}




	


\clearpage 

%
\bibliography{LHS1140_manuscript} 

@ARTICLE{Kasting_1993,
       author = {{Kasting}, James F. and {Whitmire}, Daniel P. and {Reynolds}, Ray T.},
        title = "{Habitable Zones around Main Sequence Stars}",
      journal = {\icarus},
         year = 1993,
        month = jan,
       volume = {101},
       number = {1},
        pages = {108-128},
          doi = {10.1006/icar.1993.1010},
       adsurl = {https://ui.adsabs.harvard.edu/abs/1993Icar..101..108K}
}

@ARTICLE{Zahnle_2017,
       author = {{Zahnle}, Kevin J. and {Catling}, David C.},
        title = "{The Cosmic Shoreline: The Evidence that Escape Determines which Planets Have Atmospheres, and what this May Mean for Proxima Centauri B}",
      journal = {\apj},
         year = 2017,
        month = jul,
       volume = {843},
       number = {2},
          eid = {122},
        pages = {122},
          doi = {10.3847/1538-4357/aa7846},
archivePrefix = {arXiv},
       eprint = {1702.03386},
 primaryClass = {astro-ph.EP},
       adsurl = {https://ui.adsabs.harvard.edu/abs/2017ApJ...843..122Z}
}

@ARTICLE{Pass_2025,
       author = {{Pass}, Emily K. and {Charbonneau}, David and {Vanderburg}, Andrew},
        title = "{The Receding Cosmic Shoreline of Mid-to-late M Dwarfs: Measurements of Active Lifetimes Worsen Challenges for Atmosphere Retention by Rocky Exoplanets}",
      journal = {\apjl},
         year = 2025,
        month = jun,
       volume = {986},
       number = {1},
          eid = {L3},
        pages = {L3},
          doi = {10.3847/2041-8213/adda39},
archivePrefix = {arXiv},
       eprint = {2504.01182},
 primaryClass = {astro-ph.EP},
       adsurl = {https://ui.adsabs.harvard.edu/abs/2025ApJ...986L...3P}
}

@ARTICLE{Edwards_2021,
       author = {{Edwards}, Billy and {Changeat}, Quentin and {Mori}, Mayuko and {Anisman}, Lara O. and {Morvan}, Mario and {Yip}, Kai Hou and {Tsiaras}, Angelos and {Al-Refaie}, Ahmed and {Waldmann}, Ingo and {Tinetti}, Giovanna},
        title = "{Hubble WFC3 Spectroscopy of the Habitable-zone Super-Earth LHS 1140 b}",
      journal = {\aj},
         year = 2021,
        month = jan,
       volume = {161},
       number = {1},
          eid = {44},
        pages = {44},
          doi = {10.3847/1538-3881/abc6a5},
archivePrefix = {arXiv},
       eprint = {2011.08815},
 primaryClass = {astro-ph.EP},
       adsurl = {https://ui.adsabs.harvard.edu/abs/2021AJ....161...44E}
}

@ARTICLE{Biagini_2024,
       author = {{Biagini}, Alfredo and {Cracchiolo}, Gianluca and {Petralia}, Antonino and {Maldonado}, Jes{\'u}s and {Di Maio}, Claudia and {Micela}, Giuseppina},
        title = "{A reanalysis of the LHS 1140 b atmosphere observed with the Hubble Space Telescope}",
      journal = {\mnras},
         year = 2024,
        month = may,
       volume = {530},
       number = {1},
        pages = {1054-1065},
          doi = {10.1093/mnras/stae823},
archivePrefix = {arXiv},
       eprint = {2403.20285},
 primaryClass = {astro-ph.EP},
       adsurl = {https://ui.adsabs.harvard.edu/abs/2024MNRAS.530.1054B}
}

@ARTICLE{Cadieux_2024a,
       author = {{Cadieux}, Charles and {Plotnykov}, Mykhaylo and {Doyon}, Ren{\'e} and {Valencia}, Diana and {Jahandar}, Farbod and {Dang}, Lisa and {Turbet}, Martin and {Fauchez}, Thomas J. and {Cloutier}, Ryan and {Cherubim}, Collin and {Artigau}, {\'E}tienne and {Cook}, Neil J. and {Edwards}, Billy and {Hallatt}, Tim and {Charnay}, Benjamin and {Bouchy}, Fran{\c{c}}ois and {Allart}, Romain and {Mignon}, Lucile and {Baron}, Fr{\'e}d{\'e}rique and {Barros}, Susana C.~C. and {Benneke}, Bj{\"o}rn and {Canto Martins}, B.~L. and {Cowan}, Nicolas B. and {De Medeiros}, J.~R. and {Delfosse}, Xavier and {Delgado-Mena}, Elisa and {Dumusque}, Xavier and {Ehrenreich}, David and {Frensch}, Yolanda G.~C. and {Gonz{\'a}lez Hern{\'a}ndez}, J.~I. and {Hara}, Nathan C. and {Lafreni{\`e}re}, David and {Lo Curto}, Gaspare and {Malo}, Lison and {Melo}, Claudio and {Mounzer}, Dany and {Passeger}, Vera Maria and {Pepe}, Francesco and {Poulin-Girard}, Anne-Sophie and {Santos}, Nuno C. and {Sosnowska}, Danuta and {Su{\'a}rez Mascare{\~n}o}, Alejandro and {Thibault}, Simon and {Vaulato}, Valentina and {Wade}, Gregg A. and {Wildi}, Fran{\c{c}}ois},
        title = "{New Mass and Radius Constraints on the LHS 1140 Planets: LHS 1140 b Is either a Temperate Mini-Neptune or a Water World}",
      journal = {\apjl},
         year = 2024,
        month = jan,
       volume = {960},
       number = {1},
          eid = {L3},
        pages = {L3},
          doi = {10.3847/2041-8213/ad1691},
archivePrefix = {arXiv},
       eprint = {2310.15490},
 primaryClass = {astro-ph.EP},
       adsurl = {https://ui.adsabs.harvard.edu/abs/2024ApJ...960L...3C}
}

@ARTICLE{Cadieux_2024b,
       author = {{Cadieux}, Charles and {Doyon}, Ren{\'e} and {MacDonald}, Ryan J. and {Turbet}, Martin and {Artigau}, {\'E}tienne and {Lim}, Olivia and {Radica}, Michael and {Fauchez}, Thomas J. and {Salhi}, Salma and {Dang}, Lisa and {Albert}, Lo{\"\i}c and {Coulombe}, Louis-Philippe and {Cowan}, Nicolas B. and {Lafreni{\`e}re}, David and {L'Heureux}, Alexandrine and {Piaulet-Ghorayeb}, Caroline and {Benneke}, Bj{\"o}rn and {Cloutier}, Ryan and {Charnay}, Benjamin and {Cook}, Neil J. and {Fournier-Tondreau}, Marylou and {Plotnykov}, Mykhaylo and {Valencia}, Diana},
        title = "{Transmission Spectroscopy of the Habitable Zone Exoplanet LHS 1140 b with JWST/NIRISS}",
      journal = {\apjl},
         year = 2024,
        month = jul,
       volume = {970},
       number = {1},
          eid = {L2},
        pages = {L2},
          doi = {10.3847/2041-8213/ad5afa},
archivePrefix = {arXiv},
       eprint = {2406.15136},
 primaryClass = {astro-ph.EP},
       adsurl = {https://ui.adsabs.harvard.edu/abs/2024ApJ...970L...2C}
}

@ARTICLE{Damiano_2024,
       author = {{Damiano}, Mario and {Bello-Arufe}, Aaron and {Yang}, Jeehyun and {Hu}, Renyu},
        title = "{LHS 1140 b Is a Potentially Habitable Water World}",
      journal = {\apjl},
         year = 2024,
        month = jun,
       volume = {968},
       number = {2},
          eid = {L22},
        pages = {L22},
          doi = {10.3847/2041-8213/ad5204},
archivePrefix = {arXiv},
       eprint = {2403.13265},
 primaryClass = {astro-ph.EP},
       adsurl = {https://ui.adsabs.harvard.edu/abs/2024ApJ...968L..22D}
}

@ARTICLE{Cherubim_2025,
       author = {{Cherubim}, Collin and {Wordsworth}, Robin and {Bower}, Dan J. and {Sossi}, Paolo A. and {Adams}, Danica and {Hu}, Renyu},
        title = "{An Oxidation Gradient Straddling the Small Planet Radius Valley}",
      journal = {\apj},
         year = 2025,
        month = apr,
       volume = {983},
       number = {2},
          eid = {97},
        pages = {97},
          doi = {10.3847/1538-4357/adbca9},
archivePrefix = {arXiv},
       eprint = {2503.05055},
 primaryClass = {astro-ph.EP},
       adsurl = {https://ui.adsabs.harvard.edu/abs/2025ApJ...983...97C}
}

@ARTICLE{Pass_2024,
       author = {{Pass}, Emily K. and {Charbonneau}, David and {Latham}, David W. and {Berlind}, Perry and {Calkins}, Michael L. and {Esquerdo}, Gilbert A. and {Mink}, Jessica},
        title = "{The Mass Dependence of H{\ensuremath{\alpha}} Emission and Stellar Spindown for Fully Convective M Dwarfs}",
      journal = {\apj},
         year = 2024,
        month = may,
       volume = {966},
       number = {2},
          eid = {231},
        pages = {231},
          doi = {10.3847/1538-4357/ad3631},
archivePrefix = {arXiv},
       eprint = {2401.10167},
 primaryClass = {astro-ph.SR},
       adsurl = {https://ui.adsabs.harvard.edu/abs/2024ApJ...966..231P}
}

@Article{Dittmann_2017,
       author = {{Dittmann}, Jason A. and {Irwin}, Jonathan M. and {Charbonneau}, David and {Bonfils}, Xavier and {Astudillo-Defru}, Nicola and {Haywood}, Rapha{\"e}lle D. and {Berta-Thompson}, Zachory K. and {Newton}, Elisabeth R. and {Rodriguez}, Joseph E. and {Winters}, Jennifer G. and {Tan}, Thiam-Guan and {Almenara}, Jose-Manuel and {Bouchy}, Fran{\c{c}}ois and {Delfosse}, Xavier and {Forveille}, Thierry and {Lovis}, Christophe and {Murgas}, Felipe and {Pepe}, Francesco and {Santos}, Nuno C. and {Udry}, Stephane and {W{\"u}nsche}, Ana{\"e}l and {Esquerdo}, Gilbert A. and {Latham}, David W. and {Dressing}, Courtney D.},
        title = "{A temperate rocky super-Earth transiting a nearby cool star}",
      journal = {\nat},
         year = 2017,
        month = apr,
       volume = {544},
       number = {7650},
        pages = {333-336},
          doi = {10.1038/nature22055},
archivePrefix = {arXiv},
       eprint = {1704.05556},
 primaryClass = {astro-ph.EP},
       adsurl = {https://ui.adsabs.harvard.edu/abs/2017Natur.544..333D}
}

@ARTICLE{Medina_2022,
       author = {{Medina}, Amber A. and {Winters}, Jennifer G. and {Irwin}, Jonathan M. and {Charbonneau}, David},
        title = "{Galactic Kinematics and Observed Flare Rates of a Volume-complete Sample of Mid-to-late M Dwarfs: Constraints on the History of the Stellar Radiation Environment of Planets Orbiting Low-mass Stars}",
      journal = {\apj},
         year = 2022,
        month = aug,
       volume = {935},
       number = {2},
          eid = {104},
        pages = {104},
          doi = {10.3847/1538-4357/ac77f9},
archivePrefix = {arXiv},
       eprint = {2205.02331},
 primaryClass = {astro-ph.SR},
       adsurl = {https://ui.adsabs.harvard.edu/abs/2022ApJ...935..104M}
}

@ARTICLE{Czesla_2022,
       author = {{Czesla}, S. and {Lamp{\'o}n}, M. and {Sanz-Forcada}, J. and {Garc{\'\i}a Mu{\~n}oz}, A. and {L{\'o}pez-Puertas}, M. and {Nortmann}, L. and {Yan}, D. and {Nagel}, E. and {Yan}, F. and {Schmitt}, J.~H.~M.~M. and {Aceituno}, J. and {Amado}, P.~J. and {Caballero}, J.~A. and {Casasayas-Barris}, N. and {Henning}, Th. and {Khalafinejad}, S. and {Molaverdikhani}, K. and {Montes}, D. and {Pall{\'e}}, E. and {Reiners}, A. and {Schneider}, P.~C. and {Ribas}, I. and {Quirrenbach}, A. and {Zapatero Osorio}, M.~R. and {Zechmeister}, M.},
        title = "{H{\ensuremath{\alpha}} and He I absorption in HAT-P-32 b observed with CARMENES. Detection of Roche lobe overflow and mass loss}",
      journal = {\aap},
         year = 2022,
        month = jan,
       volume = {657},
          eid = {A6},
        pages = {A6},
          doi = {10.1051/0004-6361/202039919},
archivePrefix = {arXiv},
       eprint = {2110.13582},
 primaryClass = {astro-ph.EP},
       adsurl = {https://ui.adsabs.harvard.edu/abs/2022A&A...657A...6C}
}

@ARTICLE{Gully-Santiago_2024,
       author = {{Gully-Santiago}, Michael and {Morley}, Caroline V. and {Luna}, Jessica and {MacLeod}, Morgan and {Oklop{\v{c}}i{\'c}}, Antonija and {Ganesh}, Aishwarya and {Tran}, Quang H. and {Zhang}, Zhoujian and {Bowler}, Brendan P. and {Cochran}, William D. and {Krolikowski}, Daniel M. and {Mahadevan}, Suvrath and {Ninan}, Joe P. and {Stef{\'a}nsson}, Gu{\dj}mundur and {Vanderburg}, Andrew and {Zalesky}, Joseph A. and {Zeimann}, Gregory R.},
        title = "{A Large and Variable Leading Tail of Helium in a Hot Saturn Undergoing Runaway Inflation}",
      journal = {\aj},
         year = 2024,
        month = apr,
       volume = {167},
       number = {4},
          eid = {142},
        pages = {142},
          doi = {10.3847/1538-3881/ad1ee8},
archivePrefix = {arXiv},
       eprint = {2307.08959},
 primaryClass = {astro-ph.EP},
       adsurl = {https://ui.adsabs.harvard.edu/abs/2024AJ....167..142G}
}

@ARTICLE{Krishnamurthy_2026,
       author = {{Krishnamurthy}, Vigneshwaran and {Carteret}, Yann and {Piaulet-Ghorayeb}, Caroline and {Splinter}, Jared and {Doshi}, Dhvani and {Radica}, Michael and {Coulombe}, Louis-Philippe and {Allart}, Romain and {Bourrier}, Vincent and {Cowan}, Nicolas B. and {Doyon}, Ren{\'e} and {Lafreni{\`e}re}, David and {Albert}, Lo{\"\i}c and {Benneke}, Bj{\"o}rn and {Dang}, Lisa and {Jayawardhana}, Ray and {Johnstone}, Doug and {Kaltenegger}, Lisa and {Langeveld}, Adam B. and {Pelletier}, Stefan and {Rowe}, Jason F. and {Roy}, Pierre-Alexis and {Taylor}, Jake and {Turner}, Jake D.},
        title = "{Continuous helium absorption from both the leading and trailing tails of WASP-107 b}",
      journal = {\natast},
         year = 2026,
        month = feb,
       volume = {10},
        pages = {258-270},
          doi = {10.1038/s41550-025-02710-8},
archivePrefix = {arXiv},
       eprint = {2505.20588},
 primaryClass = {astro-ph.EP},
       adsurl = {https://ui.adsabs.harvard.edu/abs/2026NatAs..10..258K}
}

@ARTICLE{Matsakos_2015,
       author = {{Matsakos}, Titos and {Uribe}, Ana and {K{\"o}nigl}, Arieh},
        title = "{Classification of magnetized star-planet interactions: bow shocks, tails, and inspiraling flows}",
      journal = {\aap},
         year = 2015,
        month = jun,
       volume = {578},
          eid = {A6},
        pages = {A6},
          doi = {10.1051/0004-6361/201425593},
archivePrefix = {arXiv},
       eprint = {1503.03551},
 primaryClass = {astro-ph.EP},
       adsurl = {https://ui.adsabs.harvard.edu/abs/2015A&A...578A...6M}
}

@ARTICLE{Rackham_2018,
       author = {{Rackham}, Benjamin V. and {Apai}, D{\'a}niel and {Giampapa}, Mark S.},
        title = "{The Transit Light Source Effect: False Spectral Features and Incorrect Densities for M-dwarf Transiting Planets}",
      journal = {\apj},
         year = 2018,
        month = feb,
       volume = {853},
       number = {2},
          eid = {122},
        pages = {122},
          doi = {10.3847/1538-4357/aaa08c},
archivePrefix = {arXiv},
       eprint = {1711.05691},
 primaryClass = {astro-ph.EP},
       adsurl = {https://ui.adsabs.harvard.edu/abs/2018ApJ...853..122R}
}

@ARTICLE{Drake_1971,
       author = {{Drake}, G.~W.},
        title = "{Theory of Relativistic Magnetic Dipole Transitions: Lifetime of the Metastable {}2$^{3}$S State of the Heliumlike Ions}",
      journal = {\pra},
         year = 1971,
        month = mar,
       volume = {3},
       number = {3},
        pages = {908-915},
          doi = {10.1103/PhysRevA.3.908},
       adsurl = {https://ui.adsabs.harvard.edu/abs/1971PhRvA...3..908D}
}

@ARTICLE{Andretta_2008,
       author = {{Andretta}, Vincenzo and {Mauas}, Pablo J.~D. and {Falchi}, Ambretta and {Teriaca}, Luca},
        title = "{Helium Line Formation and Abundance during a C-Class Flare}",
      journal = {\apj},
         year = 2008,
        month = jul,
       volume = {681},
       number = {1},
        pages = {650-663},
          doi = {10.1086/587933},
archivePrefix = {arXiv},
       eprint = {0803.0418},
 primaryClass = {astro-ph},
       adsurl = {https://ui.adsabs.harvard.edu/abs/2008ApJ...681..650A}
}

@ARTICLE{Judge_2015,
       author = {{Judge}, Philip G. and {Kleint}, Lucia and {Sainz Dalda}, Alberto},
        title = "{On Helium 1083 nm Line Polarization during the Impulsive Phase of an X1 Flare}",
      journal = {\apj},
         year = 2015,
        month = dec,
       volume = {814},
       number = {2},
          eid = {100},
        pages = {100},
          doi = {10.1088/0004-637X/814/2/100},
archivePrefix = {arXiv},
       eprint = {1510.09218},
 primaryClass = {astro-ph.SR},
       adsurl = {https://ui.adsabs.harvard.edu/abs/2015ApJ...814..100J}
}

@ARTICLE{Parker_1958,
       author = {{Parker}, E.~N.},
        title = "{Dynamics of the Interplanetary Gas and Magnetic Fields.}",
      journal = {\apj},
         year = 1958,
        month = nov,
       volume = {128},
        pages = {664},
          doi = {10.1086/146579},
       adsurl = {https://ui.adsabs.harvard.edu/abs/1958ApJ...128..664P}
}

@ARTICLE{Vissapragada_2021,
       author = {{Vissapragada}, Shreyas and {Stef{\'a}nsson}, Gudmundur and {Greklek-McKeon}, Michael and {Oklop{\v{c}}i{\'c}}, Antonija and {Knutson}, Heather A. and {Ninan}, Joe P. and {Mahadevan}, Suvrath and {Ca{\~n}as}, Caleb I. and {Chachan}, Yayaati and {Cochran}, William D. and {Collins}, Karen A. and {Dai}, Fei and {David}, Trevor J. and {Halverson}, Samuel and {Hawley}, Suzanne L. and {Hebb}, Leslie and {Kanodia}, Shubham and {Kowalski}, Adam F. and {Livingston}, John H. and {Maney}, Marissa and {Metcalf}, Andrew J. and {Morley}, Caroline and {Ramsey}, Lawrence W. and {Robertson}, Paul and {Roy}, Arpita and {Spake}, Jessica and {Schwab}, Christian and {Terrien}, Ryan C. and {Tinyanont}, Samaporn and {Vasisht}, Gautam and {Wisniewski}, John},
        title = "{A Search for Planetary Metastable Helium Absorption in the V1298 Tau System}",
      journal = {\aj},
         year = 2021,
        month = nov,
       volume = {162},
       number = {5},
          eid = {222},
        pages = {222},
          doi = {10.3847/1538-3881/ac1bb0},
archivePrefix = {arXiv},
       eprint = {2108.05358},
 primaryClass = {astro-ph.EP},
       adsurl = {https://ui.adsabs.harvard.edu/abs/2021AJ....162..222V}
}

@ARTICLE{Wang_2021,
       author = {{Wang}, Lile and {Dai}, Fei},
        title = "{Metastable Helium Absorptions with 3D Hydrodynamics and Self-consistent Photochemistry. I. WASP-69b, Dimensionality, X-Ray and UV Flux Level, Spectral Types, and Flares}",
      journal = {\apj},
         year = 2021,
        month = jun,
       volume = {914},
       number = {2},
          eid = {98},
        pages = {98},
          doi = {10.3847/1538-4357/abf1ee},
archivePrefix = {arXiv},
       eprint = {2101.00042},
 primaryClass = {astro-ph.EP},
       adsurl = {https://ui.adsabs.harvard.edu/abs/2021ApJ...914...98W}
}

@ARTICLE{Dos_Santos_2022,
       author = {{Dos Santos}, Leonardo A. and {Vidotto}, Aline A. and {Vissapragada}, Shreyas and {Alam}, Munazza K. and {Allart}, Romain and {Bourrier}, Vincent and {Kirk}, James and {Seidel}, Julia V. and {Ehrenreich}, David},
        title = "{p-winds: An open-source Python code to model planetary outflows and upper atmospheres}",
      journal = {\aap},
         year = 2022,
        month = mar,
       volume = {659},
          eid = {A62},
        pages = {A62},
          doi = {10.1051/0004-6361/202142038},
archivePrefix = {arXiv},
       eprint = {2111.11370},
 primaryClass = {astro-ph.EP},
       adsurl = {https://ui.adsabs.harvard.edu/abs/2022A&A...659A..62D}
}

@ARTICLE{Brande_2024,
       author = {{Brande}, Jonathan and {Crossfield}, Ian J.~M. and {Kreidberg}, Laura and {Morley}, Caroline V. and {Barman}, Travis and {Benneke}, Bj{\"o}rn and {Christiansen}, Jessie L. and {Dragomir}, Diana and {Fortney}, Jonathan J. and {Greene}, Thomas P. and {Hardegree-Ullman}, Kevin K. and {Howard}, Andrew W. and {Knutson}, Heather A. and {Lothringer}, Joshua D. and {Mikal-Evans}, Thomas},
        title = "{Clouds and Clarity: Revisiting Atmospheric Feature Trends in Neptune-size Exoplanets}",
      journal = {\apjl},
         year = 2024,
        month = jan,
       volume = {961},
       number = {1},
          eid = {L23},
        pages = {L23},
          doi = {10.3847/2041-8213/ad1b5c},
archivePrefix = {arXiv},
       eprint = {2310.07714},
 primaryClass = {astro-ph.EP},
       adsurl = {https://ui.adsabs.harvard.edu/abs/2024ApJ...961L..23B}
}

@ARTICLE{Fortune_2025,
       author = {{Fortune}, Mark and {Gibson}, Neale P. and {Diamond-Lowe}, Hannah and {Mendon{\c{c}}a}, Jo{\~a}o M. and {Gressier}, Am{\'e}lie and {Kitzmann}, Daniel and {Allen}, Natalie H. and {August}, Prune C. and {Ih}, Jegug and {Meier Vald{\'e}s}, Erik and {Zgraggen}, Merlin and {Buchhave}, Lars A. and {Demory}, Brice-Olivier and {Espinoza}, N{\'e}stor and {Heng}, Kevin and {Jones}, Kathryn and {Rathcke}, Alexander D.},
        title = "{Hot Rocks Survey: III. A deep eclipse for LHS 1140c and a new Gaussian process method to account for correlated noise in individual pixels}",
      journal = {\aap},
         year = 2025,
        month = sep,
       volume = {701},
          eid = {A25},
        pages = {A25},
          doi = {10.1051/0004-6361/202554198},
archivePrefix = {arXiv},
       eprint = {2505.22186},
 primaryClass = {astro-ph.EP},
       adsurl = {https://ui.adsabs.harvard.edu/abs/2025A&A...701A..25F}
}

@ARTICLE{Hamano_2024,
       author = {{Hamano}, Satoshi and {Ikeda}, Yuji and {Otsubo}, Shogo and {Katoh}, Haruki and {Fukue}, Kei and {Matsunaga}, Noriyuki and {Taniguchi}, Daisuke and {Kawakita}, Hideyo and {Takenaka}, Keiichi and {Kondo}, Sohei and {Sameshima}, Hiroaki},
        title = "{WARP: The Data Reduction Pipeline for the WINERED Spectrograph}",
      journal = {\pasp},
         year = 2024,
        month = jan,
       volume = {136},
       number = {1},
          eid = {014504},
        pages = {014504},
          doi = {10.1088/1538-3873/ad1b38},
archivePrefix = {arXiv},
       eprint = {2401.04876},
 primaryClass = {astro-ph.IM},
       adsurl = {https://ui.adsabs.harvard.edu/abs/2024PASP..136a4504H}
}

@INPROCEEDINGS{Ikeda_2016,
       author = {{Ikeda}, Yuji and {Kobayashi}, Naoto and {Kondo}, Sohei and {Otsubo}, Shogo and {Hamano}, Satoshi and {Sameshima}, Hiroaki and {Yoshikawa}, Tomoshiro and {Fukue}, Kei and {Nakanishi}, Kenshi and {Kawanishi}, Takafumi and {Nakaoka}, Tetsuya and {Kinoshita}, Masaomi and {Kitano}, Ayaka and {Asano}, Akira and {Takenaka}, Keiichi and {Watase}, Ayaka and {Mito}, Hiroyuki and {Yasui}, Chikako and {Minami}, Atsushi and {Izumu}, Natsuko and {Yamamoto}, Ryo and {Mizumoto}, Misaki and {Arasaki}, Takayuki and {Arai}, Akira and {Matsunaga}, Noriyuki and {Kawakita}, Hideyo},
        title = "{High sensitivity, wide coverage, and high-resolution NIR non-cryogenic spectrograph, WINERED}",
    booktitle = {Ground-based and Airborne Instrumentation for Astronomy VI},
         year = 2016,
       editor = {{Evans}, Christopher J. and {Simard}, Luc and {Takami}, Hideki},
       series = {Society of Photo-Optical Instrumentation Engineers (SPIE) Conference Series},
       volume = {9908},
        month = aug,
          eid = {99085Z},
        pages = {99085Z},
          doi = {10.1117/12.2230886},
       adsurl = {https://ui.adsabs.harvard.edu/abs/2016SPIE.9908E..5ZI}
}

@ARTICLE{Ikeda_2022,
       author = {{Ikeda}, Yuji and {Kondo}, Sohei and {Otsubo}, Shogo and {Hamano}, Satoshi and {Yasui}, Chikako and {Matsunaga}, Noriyuki and {Sameshima}, Hiroaki and {Yoshikawa}, Tomohiro and {Fukue}, Kei and {Nakanishi}, Kenshi and {Kawanishi}, Takafumi and {Watase}, Ayaka and {Nakaoka}, Tetsuya and {Arai}, Akira and {Kinoshita}, Masaomi and {Kitano}, Ayaka and {Nakamura}, Kazuki and {Asano}, Akira and {Takenaka}, Keiichi and {Murai}, Taichi and {Kawakita}, Hideyo and {Minami}, Atsushi and {Izumi}, Natsuko and {Yamamoto}, Ryo and {Mizumoto}, Misaki and {Taniguchi}, Daisuke and {Tsujimoto}, Takuji},
        title = "{Highly Sensitive, Non-cryogenic NIR High-resolution Spectrograph, WINERED}",
      journal = {\pasp},
         year = 2022,
        month = jan,
       volume = {134},
       number = {1031},
          eid = {015004},
        pages = {015004},
          doi = {10.1088/1538-3873/ac1c5f},
       adsurl = {https://ui.adsabs.harvard.edu/abs/2022PASP..134a5004I}
}

@INPROCEEDINGS{Otsubo_2016,
       author = {{Otsubo}, Shogo and {Ikeda}, Yuji and {Kobayashi}, Naoto and {Sukegawa}, Takashi and {Kondo}, Sohei and {Hamano}, Satoshi and {Sameshima}, Hiroaki and {Fukue}, Kei and {Yoshikawa}, Tomohiro and {Nakanishi}, Kenshi and {Watase}, Ayaka and {Takenaka}, Keiichi and {Asano}, Akira and {Yasui}, Chikako and {Matsunaga}, Noriyuki and {Kawakita}, Hideyo},
        title = "{First high-efficiency and high-resolution (R=80,000) NIR spectroscopy with high-blazed Echelle grating: WINERED HIRES modes}",
    booktitle = {Ground-based and Airborne Instrumentation for Astronomy VI},
         year = 2016,
       editor = {{Evans}, Christopher J. and {Simard}, Luc and {Takami}, Hideki},
       series = {Society of Photo-Optical Instrumentation Engineers (SPIE) Conference Series},
       volume = {9908},
        month = aug,
          eid = {990879},
        pages = {990879},
          doi = {10.1117/12.2233845},
       adsurl = {https://ui.adsabs.harvard.edu/abs/2016SPIE.9908E..79O}
}

@ARTICLE{Kochanov_2016,
       author = {{Kochanov}, R.~V. and {Gordon}, I.~E. and {Rothman}, L.~S. and {Wcis{\l}o}, P. and {Hill}, C. and {Wilzewski}, J.~S.},
        title = "{HITRAN Application Programming Interface (HAPI): A comprehensive approach to working with spectroscopic data}",
      journal = {\jqsrt},
         year = 2016,
        month = jul,
       volume = {177},
        pages = {15-30},
          doi = {10.1016/j.jqsrt.2016.03.005},
       adsurl = {https://ui.adsabs.harvard.edu/abs/2016JQSRT.177...15K}
}

@ARTICLE{Gordon_2022,
       author = {{Gordon}, I.~E. and {Rothman}, L.~S. and {Hargreaves}, R.~J. and {Hashemi}, R. and {Karlovets}, E.~V. and {Skinner}, F.~M. and {Conway}, E.~K. and {Hill}, C. and {Kochanov}, R.~V. and {Tan}, Y. and {Wcis{\l}o}, P. and {Finenko}, A.~A. and {Nelson}, K. and {Bernath}, P.~F. and {Birk}, M. and {Boudon}, V. and {Campargue}, A. and {Chance}, K.~V. and {Coustenis}, A. and {Drouin}, B.~J. and {Flaud}, J. -M. and {Gamache}, R.~R. and {Hodges}, J.~T. and {Jacquemart}, D. and {Mlawer}, E.~J. and {Nikitin}, A.~V. and {Perevalov}, V.~I. and {Rotger}, M. and {Tennyson}, J. and {Toon}, G.~C. and {Tran}, H. and {Tyuterev}, V.~G. and {Adkins}, E.~M. and {Baker}, A. and {Barbe}, A. and {Can{\`e}}, E. and {Cs{\'a}sz{\'a}r}, A.~G. and {Dudaryonok}, A. and {Egorov}, O. and {Fleisher}, A.~J. and {Fleurbaey}, H. and {Foltynowicz}, A. and {Furtenbacher}, T. and {Harrison}, J.~J. and {Hartmann}, J. -M. and {Horneman}, V. -M. and {Huang}, X. and {Karman}, T. and {Karns}, J. and {Kassi}, S. and {Kleiner}, I. and {Kofman}, V. and {Kwabia-Tchana}, F. and {Lavrentieva}, N.~N. and {Lee}, T.~J. and {Long}, D.~A. and {Lukashevskaya}, A.~A. and {Lyulin}, O.~M. and {Makhnev}, V. Yu. and {Matt}, W. and {Massie}, S.~T. and {Melosso}, M. and {Mikhailenko}, S.~N. and {Mondelain}, D. and {M{\"u}ller}, H.~S.~P. and {Naumenko}, O.~V. and {Perrin}, A. and {Polyansky}, O.~L. and {Raddaoui}, E. and {Raston}, P.~L. and {Reed}, Z.~D. and {Rey}, M. and {Richard}, C. and {T{\'o}bi{\'a}s}, R. and {Sadiek}, I. and {Schwenke}, D.~W. and {Starikova}, E. and {Sung}, K. and {Tamassia}, F. and {Tashkun}, S.~A. and {Vander Auwera}, J. and {Vasilenko}, I.~A. and {Vigasin}, A.~A. and {Villanueva}, G.~L. and {Vispoel}, B. and {Wagner}, G. and {Yachmenev}, A. and {Yurchenko}, S.~N.},
        title = "{The HITRAN2020 molecular spectroscopic database}",
      journal = {\jqsrt},
         year = 2022,
        month = jan,
       volume = {277},
          eid = {107949},
        pages = {107949},
          doi = {10.1016/j.jqsrt.2021.107949},
       adsurl = {https://ui.adsabs.harvard.edu/abs/2022JQSRT.27707949G}
}

@ARTICLE{astropy_2013,
       author = {{Astropy Collaboration} and {Robitaille}, Thomas P. and {Tollerud}, Erik J. and {Greenfield}, Perry and {Droettboom}, Michael and {Bray}, Erik and {Aldcroft}, Tom and {Davis}, Matt and {Ginsburg}, Adam and {Price-Whelan}, Adrian M. and {Kerzendorf}, Wolfgang E. and {Conley}, Alexander and {Crighton}, Neil and {Barbary}, Kyle and {Muna}, Demitri and {Ferguson}, Henry and {Grollier}, Fr{\'e}d{\'e}ric and {Parikh}, Madhura M. and {Nair}, Prasanth H. and {Unther}, Hans M. and {Deil}, Christoph and {Woillez}, Julien and {Conseil}, Simon and {Kramer}, Roban and {Turner}, James E.~H. and {Singer}, Leo and {Fox}, Ryan and {Weaver}, Benjamin A. and {Zabalza}, Victor and {Edwards}, Zachary I. and {Azalee Bostroem}, K. and {Burke}, D.~J. and {Casey}, Andrew R. and {Crawford}, Steven M. and {Dencheva}, Nadia and {Ely}, Justin and {Jenness}, Tim and {Labrie}, Kathleen and {Lim}, Pey Lian and {Pierfederici}, Francesco and {Pontzen}, Andrew and {Ptak}, Andy and {Refsdal}, Brian and {Servillat}, Mathieu and {Streicher}, Ole},
        title = "{Astropy: A community Python package for astronomy}",
      journal = {\aap},
         year = 2013,
        month = oct,
       volume = {558},
          eid = {A33},
        pages = {A33},
          doi = {10.1051/0004-6361/201322068},
archivePrefix = {arXiv},
       eprint = {1307.6212},
 primaryClass = {astro-ph.IM},
       adsurl = {https://ui.adsabs.harvard.edu/abs/2013A&A...558A..33A}
}

@ARTICLE{astropy_2018,
       author = {{Astropy Collaboration} and {Price-Whelan}, A.~M. and {Sip{\H{o}}cz}, B.~M. and {G{\"u}nther}, H.~M. and {Lim}, P.~L. and {Crawford}, S.~M. and {Conseil}, S. and {Shupe}, D.~L. and {Craig}, M.~W. and {Dencheva}, N. and {Ginsburg}, A. and {VanderPlas}, J.~T. and {Bradley}, L.~D. and {P{\'e}rez-Su{\'a}rez}, D. and {de Val-Borro}, M. and {Aldcroft}, T.~L. and {Cruz}, K.~L. and {Robitaille}, T.~P. and {Tollerud}, E.~J. and {Ardelean}, C. and {Babej}, T. and {Bach}, Y.~P. and {Bachetti}, M. and {Bakanov}, A.~V. and {Bamford}, S.~P. and {Barentsen}, G. and {Barmby}, P. and {Baumbach}, A. and {Berry}, K.~L. and {Biscani}, F. and {Boquien}, M. and {Bostroem}, K.~A. and {Bouma}, L.~G. and {Brammer}, G.~B. and {Bray}, E.~M. and {Breytenbach}, H. and {Buddelmeijer}, H. and {Burke}, D.~J. and {Calderone}, G. and {Cano Rodr{\'\i}guez}, J.~L. and {Cara}, M. and {Cardoso}, J.~V.~M. and {Cheedella}, S. and {Copin}, Y. and {Corrales}, L. and {Crichton}, D. and {D'Avella}, D. and {Deil}, C. and {Depagne}, {\'E}. and {Dietrich}, J.~P. and {Donath}, A. and {Droettboom}, M. and {Earl}, N. and {Erben}, T. and {Fabbro}, S. and {Ferreira}, L.~A. and {Finethy}, T. and {Fox}, R.~T. and {Garrison}, L.~H. and {Gibbons}, S.~L.~J. and {Goldstein}, D.~A. and {Gommers}, R. and {Greco}, J.~P. and {Greenfield}, P. and {Groener}, A.~M. and {Grollier}, F. and {Hagen}, A. and {Hirst}, P. and {Homeier}, D. and {Horton}, A.~J. and {Hosseinzadeh}, G. and {Hu}, L. and {Hunkeler}, J.~S. and {Ivezi{\'c}}, {\v{Z}}. and {Jain}, A. and {Jenness}, T. and {Kanarek}, G. and {Kendrew}, S. and {Kern}, N.~S. and {Kerzendorf}, W.~E. and {Khvalko}, A. and {King}, J. and {Kirkby}, D. and {Kulkarni}, A.~M. and {Kumar}, A. and {Lee}, A. and {Lenz}, D. and {Littlefair}, S.~P. and {Ma}, Z. and {Macleod}, D.~M. and {Mastropietro}, M. and {McCully}, C. and {Montagnac}, S. and {Morris}, B.~M. and {Mueller}, M. and {Mumford}, S.~J. and {Muna}, D. and {Murphy}, N.~A. and {Nelson}, S. and {Nguyen}, G.~H. and {Ninan}, J.~P. and {N{\"o}the}, M. and {Ogaz}, S. and {Oh}, S. and {Parejko}, J.~K. and {Parley}, N. and {Pascual}, S. and {Patil}, R. and {Patil}, A.~A. and {Plunkett}, A.~L. and {Prochaska}, J.~X. and {Rastogi}, T. and {Reddy Janga}, V. and {Sabater}, J. and {Sakurikar}, P. and {Seifert}, M. and {Sherbert}, L.~E. and {Sherwood-Taylor}, H. and {Shih}, A.~Y. and {Sick}, J. and {Silbiger}, M.~T. and {Singanamalla}, S. and {Singer}, L.~P. and {Sladen}, P.~H. and {Sooley}, K.~A. and {Sornarajah}, S. and {Streicher}, O. and {Teuben}, P. and {Thomas}, S.~W. and {Tremblay}, G.~R. and {Turner}, J.~E.~H. and {Terr{\'o}n}, V. and {van Kerkwijk}, M.~H. and {de la Vega}, A. and {Watkins}, L.~L. and {Weaver}, B.~A. and {Whitmore}, J.~B. and {Woillez}, J. and {Zabalza}, V. and {Astropy Contributors}},
        title = "{The Astropy Project: Building an Open-science Project and Status of the v2.0 Core Package}",
      journal = {\aj},
         year = 2018,
        month = sep,
       volume = {156},
       number = {3},
          eid = {123},
        pages = {123},
          doi = {10.3847/1538-3881/aabc4f},
archivePrefix = {arXiv},
       eprint = {1801.02634},
 primaryClass = {astro-ph.IM},
       adsurl = {https://ui.adsabs.harvard.edu/abs/2018AJ....156..123A}
}

@ARTICLE{astropy_2022,
       author = {{Astropy Collaboration} and {Price-Whelan}, Adrian M. and {Lim}, Pey Lian and {Earl}, Nicholas and {Starkman}, Nathaniel and {Bradley}, Larry and {Shupe}, David L. and {Patil}, Aarya A. and {Corrales}, Lia and {Brasseur}, C.~E. and {N{\"o}the}, Maximilian and {Donath}, Axel and {Tollerud}, Erik and {Morris}, Brett M. and {Ginsburg}, Adam and {Vaher}, Eero and {Weaver}, Benjamin A. and {Tocknell}, James and {Jamieson}, William and {van Kerkwijk}, Marten H. and {Robitaille}, Thomas P. and {Merry}, Bruce and {Bachetti}, Matteo and {G{\"u}nther}, H. Moritz and {Aldcroft}, Thomas L. and {Alvarado-Montes}, Jaime A. and {Archibald}, Anne M. and {B{\'o}di}, Attila and {Bapat}, Shreyas and {Barentsen}, Geert and {Baz{\'a}n}, Juanjo and {Biswas}, Manish and {Boquien}, M{\'e}d{\'e}ric and {Burke}, D.~J. and {Cara}, Daria and {Cara}, Mihai and {Conroy}, Kyle E. and {Conseil}, Simon and {Craig}, Matthew W. and {Cross}, Robert M. and {Cruz}, Kelle L. and {D'Eugenio}, Francesco and {Dencheva}, Nadia and {Devillepoix}, Hadrien A.~R. and {Dietrich}, J{\"o}rg P. and {Eigenbrot}, Arthur Davis and {Erben}, Thomas and {Ferreira}, Leonardo and {Foreman-Mackey}, Daniel and {Fox}, Ryan and {Freij}, Nabil and {Garg}, Suyog and {Geda}, Robel and {Glattly}, Lauren and {Gondhalekar}, Yash and {Gordon}, Karl D. and {Grant}, David and {Greenfield}, Perry and {Groener}, Austen M. and {Guest}, Steve and {Gurovich}, Sebastian and {Handberg}, Rasmus and {Hart}, Akeem and {Hatfield-Dodds}, Zac and {Homeier}, Derek and {Hosseinzadeh}, Griffin and {Jenness}, Tim and {Jones}, Craig K. and {Joseph}, Prajwel and {Kalmbach}, J. Bryce and {Karamehmetoglu}, Emir and {Ka{\l}uszy{\'n}ski}, Miko{\l}aj and {Kelley}, Michael S.~P. and {Kern}, Nicholas and {Kerzendorf}, Wolfgang E. and {Koch}, Eric W. and {Kulumani}, Shankar and {Lee}, Antony and {Ly}, Chun and {Ma}, Zhiyuan and {MacBride}, Conor and {Maljaars}, Jakob M. and {Muna}, Demitri and {Murphy}, N.~A. and {Norman}, Henrik and {O'Steen}, Richard and {Oman}, Kyle A. and {Pacifici}, Camilla and {Pascual}, Sergio and {Pascual-Granado}, J. and {Patil}, Rohit R. and {Perren}, Gabriel I. and {Pickering}, Timothy E. and {Rastogi}, Tanuj and {Roulston}, Benjamin R. and {Ryan}, Daniel F. and {Rykoff}, Eli S. and {Sabater}, Jose and {Sakurikar}, Parikshit and {Salgado}, Jes{\'u}s and {Sanghi}, Aniket and {Saunders}, Nicholas and {Savchenko}, Volodymyr and {Schwardt}, Ludwig and {Seifert-Eckert}, Michael and {Shih}, Albert Y. and {Jain}, Anany Shrey and {Shukla}, Gyanendra and {Sick}, Jonathan and {Simpson}, Chris and {Singanamalla}, Sudheesh and {Singer}, Leo P. and {Singhal}, Jaladh and {Sinha}, Manodeep and {Sip{\H{o}}cz}, Brigitta M. and {Spitler}, Lee R. and {Stansby}, David and {Streicher}, Ole and {{\v{S}}umak}, Jani and {Swinbank}, John D. and {Taranu}, Dan S. and {Tewary}, Nikita and {Tremblay}, Grant R. and {de Val-Borro}, Miguel and {Van Kooten}, Samuel J. and {Vasovi{\'c}}, Zlatan and {Verma}, Shresth and {de Miranda Cardoso}, Jos{\'e} Vin{\'\i}cius and {Williams}, Peter K.~G. and {Wilson}, Tom J. and {Winkel}, Benjamin and {Wood-Vasey}, W.~M. and {Xue}, Rui and {Yoachim}, Peter and {Zhang}, Chen and {Zonca}, Andrea and {Astropy Project Contributors}},
        title = "{The Astropy Project: Sustaining and Growing a Community-oriented Open-source Project and the Latest Major Release (v5.0) of the Core Package}",
      journal = {\apj},
         year = 2022,
        month = aug,
       volume = {935},
       number = {2},
          eid = {167},
        pages = {167},
          doi = {10.3847/1538-4357/ac7c74},
archivePrefix = {arXiv},
       eprint = {2206.14220},
 primaryClass = {astro-ph.IM},
       adsurl = {https://ui.adsabs.harvard.edu/abs/2022ApJ...935..167A}
}

@ARTICLE{Oklopcic_2018,
       author = {{Oklop{\v{c}}i{\'c}}, Antonija and {Hirata}, Christopher M.},
        title = "{A New Window into Escaping Exoplanet Atmospheres: 10830 {\r{A}} Line of Helium}",
      journal = {\apjl},
         year = 2018,
        month = mar,
       volume = {855},
       number = {1},
          eid = {L11},
        pages = {L11},
          doi = {10.3847/2041-8213/aaada9},
archivePrefix = {arXiv},
       eprint = {1711.05269},
 primaryClass = {astro-ph.EP},
       adsurl = {https://ui.adsabs.harvard.edu/abs/2018ApJ...855L..11O}
}

@ARTICLE{Ginzburg_2016,
       author = {{Ginzburg}, Sivan and {Schlichting}, Hilke E. and {Sari}, Re'em},
        title = "{Super-Earth Atmospheres: Self-consistent Gas Accretion and Retention}",
      journal = {\apj},
         year = 2016,
        month = jul,
       volume = {825},
       number = {1},
          eid = {29},
        pages = {29},
          doi = {10.3847/0004-637X/825/1/29},
archivePrefix = {arXiv},
       eprint = {1512.07925},
 primaryClass = {astro-ph.EP},
       adsurl = {https://ui.adsabs.harvard.edu/abs/2016ApJ...825...29G}
}

@ARTICLE{McCreery_2025,
       author = {{McCreery}, Patrick and {Dos Santos}, Leonardo A. and {Espinoza}, N{\'e}stor and {Allart}, Romain and {Kirk}, James},
        title = "{Tracing the Winds: A Uniform Interpretation of Helium Escape in Exoplanets from Archival Spectroscopic Observations}",
      journal = {\apj},
         year = 2025,
        month = feb,
       volume = {980},
       number = {1},
          eid = {125},
        pages = {125},
          doi = {10.3847/1538-4357/ada6b9},
archivePrefix = {arXiv},
       eprint = {2501.03998},
 primaryClass = {astro-ph.EP},
       adsurl = {https://ui.adsabs.harvard.edu/abs/2025ApJ...980..125M}
}

@ARTICLE{jansen2001,
       author = {{Jansen}, F. and {Lumb}, D. and {Altieri}, B. and {Clavel}, J. and {Ehle}, M. and {Erd}, C. and {Gabriel}, C. and {Guainazzi}, M. and {Gondoin}, P. and {Much}, R. and {Munoz}, R. and {Santos}, M. and {Schartel}, N. and {Texier}, D. and {Vacanti}, G.},
        title = "{XMM-Newton observatory. I. The spacecraft and operations}",
      journal = {\aap},
         year = 2001,
        month = jan,
       volume = {365},
        pages = {L1-L6},
          doi = {10.1051/0004-6361:20000036},
       adsurl = {https://ui.adsabs.harvard.edu/abs/2001A&A...365L...1J}
}

@ARTICLE{struder2001,
       author = {{Str{\"u}der}, L. and {Briel}, U. and {Dennerl}, K. and {Hartmann}, R. and {Kendziorra}, E. and {Meidinger}, N. and {Pfeffermann}, E. and {Reppin}, C. and {Aschenbach}, B. and {Bornemann}, W. and {Br{\"a}uninger}, H. and {Burkert}, W. and {Elender}, M. and {Freyberg}, M. and {Haberl}, F. and {Hartner}, G. and {Heuschmann}, F. and {Hippmann}, H. and {Kastelic}, E. and {Kemmer}, S. and {Kettenring}, G. and {Kink}, W. and {Krause}, N. and {M{\"u}ller}, S. and {Oppitz}, A. and {Pietsch}, W. and {Popp}, M. and {Predehl}, P. and {Read}, A. and {Stephan}, K.~H. and {St{\"o}tter}, D. and {Tr{\"u}mper}, J. and {Holl}, P. and {Kemmer}, J. and {Soltau}, H. and {St{\"o}tter}, R. and {Weber}, U. and {Weichert}, U. and {von Zanthier}, C. and {Carathanassis}, D. and {Lutz}, G. and {Richter}, R.~H. and {Solc}, P. and {B{\"o}ttcher}, H. and {Kuster}, M. and {Staubert}, R. and {Abbey}, A. and {Holland}, A. and {Turner}, M. and {Balasini}, M. and {Bignami}, G.~F. and {La Palombara}, N. and {Villa}, G. and {Buttler}, W. and {Gianini}, F. and {Lain{\'e}}, R. and {Lumb}, D. and {Dhez}, P.},
        title = "{The European Photon Imaging Camera on XMM-Newton: The pn-CCD camera}",
      journal = {\aap},
         year = 2001,
        month = jan,
       volume = {365},
        pages = {L18-L26},
          doi = {10.1051/0004-6361:20000066},
       adsurl = {https://ui.adsabs.harvard.edu/abs/2001A&A...365L..18S}
}

@ARTICLE{turner2001,
       author = {{Turner}, M.~J.~L. and {Abbey}, A. and {Arnaud}, M. and {Balasini}, M. and {Barbera}, M. and {Belsole}, E. and {Bennie}, P.~J. and {Bernard}, J.~P. and {Bignami}, G.~F. and {Boer}, M. and {Briel}, U. and {Butler}, I. and {Cara}, C. and {Chabaud}, C. and {Cole}, R. and {Collura}, A. and {Conte}, M. and {Cros}, A. and {Denby}, M. and {Dhez}, P. and {Di Coco}, G. and {Dowson}, J. and {Ferrando}, P. and {Ghizzardi}, S. and {Gianotti}, F. and {Goodall}, C.~V. and {Gretton}, L. and {Griffiths}, R.~G. and {Hainaut}, O. and {Hochedez}, J.~F. and {Holland}, A.~D. and {Jourdain}, E. and {Kendziorra}, E. and {Lagostina}, A. and {Laine}, R. and {La Palombara}, N. and {Lortholary}, M. and {Lumb}, D. and {Marty}, P. and {Molendi}, S. and {Pigot}, C. and {Poindron}, E. and {Pounds}, K.~A. and {Reeves}, J.~N. and {Reppin}, C. and {Rothenflug}, R. and {Salvetat}, P. and {Sauvageot}, J.~L. and {Schmitt}, D. and {Sembay}, S. and {Short}, A.~D.~T. and {Spragg}, J. and {Stephen}, J. and {Str{\"u}der}, L. and {Tiengo}, A. and {Trifoglio}, M. and {Tr{\"u}mper}, J. and {Vercellone}, S. and {Vigroux}, L. and {Villa}, G. and {Ward}, M.~J. and {Whitehead}, S. and {Zonca}, E.},
        title = "{The European Photon Imaging Camera on XMM-Newton: The MOS cameras}",
      journal = {\aap},
         year = 2001,
        month = jan,
       volume = {365},
        pages = {L27-L35},
          doi = {10.1051/0004-6361:20000087},
archivePrefix = {arXiv},
       eprint = {astro-ph/0011498},
 primaryClass = {astro-ph},
       adsurl = {https://ui.adsabs.harvard.edu/abs/2001A&A...365L..27T}
}

@INPROCEEDINGS{gabriel2004-sas,
       author = {{Gabriel}, C. and {Denby}, M. and {Fyfe}, D.~J. and {Hoar}, J. and {Ibarra}, A. and {Ojero}, E. and {Osborne}, J. and {Saxton}, R.~D. and {Lammers}, U. and {Vacanti}, G.},
        title = "{The XMM-Newton SAS - Distributed Development and Maintenance of a Large Science Analysis System: A Critical Analysis}",
    booktitle = {Astronomical Data Analysis Software and Systems (ADASS) XIII},
         year = 2004,
       editor = {{Ochsenbein}, Francois and {Allen}, Mark G. and {Egret}, Daniel},
       series = {Astronomical Society of the Pacific Conference Series},
       volume = {314},
        month = jul,
        pages = {759},
       adsurl = {https://ui.adsabs.harvard.edu/abs/2004ASPC..314..759G}
}

@ARTICLE{buchner2016-bxa,
       author = {{Buchner}, J. and {Georgakakis}, A. and {Nandra}, K. and {Hsu}, L. and {Rangel}, C. and {Brightman}, M. and {Merloni}, A. and {Salvato}, M. and {Donley}, J. and {Kocevski}, D.},
        title = "{X-ray spectral modelling of the AGN obscuring region in the CDFS: Bayesian model selection and catalogue}",
      journal = {\aap},
         year = 2014,
        month = apr,
       volume = {564},
          eid = {A125},
        pages = {A125},
          doi = {10.1051/0004-6361/201322971},
archivePrefix = {arXiv},
       eprint = {1402.0004},
 primaryClass = {astro-ph.HE},
       adsurl = {https://ui.adsabs.harvard.edu/abs/2014A&A...564A.125B}
}

@ARTICLE{buchner2019-UltraNest,
       author = {{Buchner}, Johannes},
        title = "{UltraNest - a robust, general purpose Bayesian inference engine}",
      journal = {\jos},
         year = 2021,
        month = apr,
       volume = {6},
       number = {60},
          eid = {3001},
        pages = {3001},
          doi = {10.21105/joss.03001},
archivePrefix = {arXiv},
       eprint = {2101.09604},
 primaryClass = {stat.CO},
       adsurl = {https://ui.adsabs.harvard.edu/abs/2021JOSS....6.3001B}
}

@INPROCEEDINGS{arnaud1996XSPEC,
       author = {{Arnaud}, K.~A.},
        title = "{XSPEC: The First Ten Years}",
    booktitle = {Astronomical Data Analysis Software and Systems V},
         year = 1996,
       editor = {{Jacoby}, George H. and {Barnes}, Jeannette},
       series = {Astronomical Society of the Pacific Conference Series},
       volume = {101},
        month = jan,
        pages = {17},
       adsurl = {https://ui.adsabs.harvard.edu/abs/1996ASPC..101...17A}
}

@ARTICLE{kaastra2016-optimal-binning,
       author = {{Kaastra}, J.~S. and {Bleeker}, J.~A.~M.},
        title = "{Optimal binning of X-ray spectra and response matrix design}",
      journal = {\aap},
         year = 2016,
        month = mar,
       volume = {587},
          eid = {A151},
        pages = {A151},
          doi = {10.1051/0004-6361/201527395},
archivePrefix = {arXiv},
       eprint = {1601.05309},
 primaryClass = {astro-ph.IM},
       adsurl = {https://ui.adsabs.harvard.edu/abs/2016A&A...587A.151K}
}

@ARTICLE{spinelli2023,
       author = {{Spinelli}, Riccardo and {Gallo}, Elena and {Haardt}, Francesco and {Caldiroli}, Andrea and {Biassoni}, Federico and {Borsa}, Francesco and {Rauscher}, Emily},
        title = "{Planetary Parameters, XUV Environments, and Mass-loss Rates for Nearby Gaseous Planets with X-Ray-detected Host Stars}",
      journal = {\aj},
         year = 2023,
        month = may,
       volume = {165},
       number = {5},
          eid = {200},
        pages = {200},
          doi = {10.3847/1538-3881/acc336},
archivePrefix = {arXiv},
       eprint = {2208.01650},
 primaryClass = {astro-ph.EP},
       adsurl = {https://ui.adsabs.harvard.edu/abs/2023AJ....165..200S}
}

@ARTICLE{spinelli2019,
       author = {{Spinelli}, R. and {Borsa}, F. and {Ghirlanda}, G. and {Ghisellini}, G. and {Campana}, S. and {Haardt}, F. and {Poretti}, E.},
        title = "{The high-energy radiation environment of the habitable-zone super-Earth LHS 1140b}",
      journal = {\aap},
         year = 2019,
        month = jul,
       volume = {627},
          eid = {A144},
        pages = {A144},
          doi = {10.1051/0004-6361/201935636},
archivePrefix = {arXiv},
       eprint = {1906.08783},
 primaryClass = {astro-ph.EP},
       adsurl = {https://ui.adsabs.harvard.edu/abs/2019A&A...627A.144S}
}

@ARTICLE{smith2001-apec,
       author = {{Smith}, Randall K. and {Brickhouse}, Nancy S. and {Liedahl}, Duane A. and {Raymond}, John C.},
        title = "{Collisional Plasma Models with APEC/APED: Emission-Line Diagnostics of Hydrogen-like and Helium-like Ions}",
      journal = {\apjl},
         year = 2001,
        month = aug,
       volume = {556},
       number = {2},
        pages = {L91-L95},
          doi = {10.1086/322992},
archivePrefix = {arXiv},
       eprint = {astro-ph/0106478},
 primaryClass = {astro-ph},
       adsurl = {https://ui.adsabs.harvard.edu/abs/2001ApJ...556L..91S}
}

@ARTICLE{Ment_2019,
       author = {{Ment}, Kristo and {Dittmann}, Jason A. and {Astudillo-Defru}, Nicola and {Charbonneau}, David and {Irwin}, Jonathan and {Bonfils}, Xavier and {Murgas}, Felipe and {Almenara}, Jose-Manuel and {Forveille}, Thierry and {Agol}, Eric and {Ballard}, Sarah and {Berta-Thompson}, Zachory K. and {Bouchy}, Fran{\c{c}}ois and {Cloutier}, Ryan and {Delfosse}, Xavier and {Doyon}, Ren{\'e} and {Dressing}, Courtney D. and {Esquerdo}, Gilbert A. and {Haywood}, Rapha{\"e}lle D. and {Kipping}, David M. and {Latham}, David W. and {Lovis}, Christophe and {Newton}, Elisabeth R. and {Pepe}, Francesco and {Rodriguez}, Joseph E. and {Santos}, Nuno C. and {Tan}, Thiam-Guan and {Udry}, Stephane and {Winters}, Jennifer G. and {W{\"u}nsche}, Ana{\"e}l},
        title = "{A Second Terrestrial Planet Orbiting the Nearby M Dwarf LHS 1140}",
      journal = {\aj},
         year = 2019,
        month = jan,
       volume = {157},
       number = {1},
          eid = {32},
        pages = {32},
          doi = {10.3847/1538-3881/aaf1b1},
archivePrefix = {arXiv},
       eprint = {1808.00485},
 primaryClass = {astro-ph.EP},
       adsurl = {https://ui.adsabs.harvard.edu/abs/2019AJ....157...32M}
}

@ARTICLE{Vissapragada_2024,
       author = {{Vissapragada}, Shreyas and {McCreery}, Patrick and {Dos Santos}, Leonardo A. and {Espinoza}, N{\'e}stor and {McWilliam}, Andrew and {Matsunaga}, Noriyuki and {Redai}, J{\'e}a Adams and {Behr}, Patrick and {France}, Kevin and {Hamano}, Satoshi and {Hull}, Charlie and {Ikeda}, Yuji and {Katoh}, Haruki and {Kawakita}, Hideyo and {L{\'o}pez-Morales}, Mercedes and {Ortiz Ceballos}, Kevin N. and {Otsubo}, Shogo and {Sarugaku}, Yuki and {Takeuchi}, Tomomi},
        title = "{A High-resolution Non-detection of Escaping Helium in the Ultrahot Neptune LTT 9779b: Evidence for Weakened Evaporation}",
      journal = {\apjl},
         year = 2024,
        month = feb,
       volume = {962},
       number = {1},
          eid = {L19},
        pages = {L19},
          doi = {10.3847/2041-8213/ad23cf},
archivePrefix = {arXiv},
       eprint = {2401.16474},
 primaryClass = {astro-ph.EP},
       adsurl = {https://ui.adsabs.harvard.edu/abs/2024ApJ...962L..19V}
}

@ARTICLE{Oliva_2015,
       author = {{Oliva}, E. and {Origlia}, L. and {Scuderi}, S. and {Benatti}, S. and {Carleo}, I. and {Lapenna}, E. and {Mucciarelli}, A. and {Baffa}, C. and {Biliotti}, V. and {Carbonaro}, L. and {Falcini}, G. and {Giani}, E. and {Iuzzolino}, M. and {Massi}, F. and {Sanna}, N. and {Sozzi}, M. and {Tozzi}, A. and {Ghedina}, A. and {Ghinassi}, F. and {Lodi}, M. and {Harutyunyan}, A. and {Pedani}, M.},
        title = "{Lines and continuum sky emission in the near infrared: observational constraints from deep high spectral resolution spectra with GIANO-TNG}",
      journal = {\aap},
         year = 2015,
        month = sep,
       volume = {581},
          eid = {A47},
        pages = {A47},
          doi = {10.1051/0004-6361/201526291},
archivePrefix = {arXiv},
       eprint = {1506.09004},
 primaryClass = {astro-ph.IM},
       adsurl = {https://ui.adsabs.harvard.edu/abs/2015A&A...581A..47O}
}

@ARTICLE{McCann_2019,
       author = {{McCann}, John and {Murray-Clay}, Ruth A. and {Kratter}, Kaitlin and {Krumholz}, Mark R.},
        title = "{Morphology of Hydrodynamic Winds: A Study of Planetary Winds in Stellar Environments}",
      journal = {\apj},
         year = 2019,
        month = mar,
       volume = {873},
       number = {1},
          eid = {89},
        pages = {89},
          doi = {10.3847/1538-4357/ab05b8},
archivePrefix = {arXiv},
       eprint = {1811.09276},
 primaryClass = {astro-ph.EP},
       adsurl = {https://ui.adsabs.harvard.edu/abs/2019ApJ...873...89M}
}

@article{Robinson_2012,
	author = {Tyler D. Robinson and David C. Catling},
	doi = {10.1088/0004-637X/757/1/104},
	journal = {Astrophys. J.},
	month = {sep},
	number = {1},
	pages = {104},
	publisher = {The American Astronomical Society},
	title = {An analytic radiative--convective model for planetary atmospheres},
	volume = {757},
	year = {2012}
}

@ARTICLE{Murray-Clay_2009,
       author = {{Murray-Clay}, Ruth A. and {Chiang}, Eugene I. and {Murray}, Norman},
        title = "{Atmospheric Escape From Hot Jupiters}",
      journal = {\apj},
         year = 2009,
        month = mar,
       volume = {693},
       number = {1},
        pages = {23-42},
          doi = {10.1088/0004-637X/693/1/23},
archivePrefix = {arXiv},
       eprint = {0811.0006},
 primaryClass = {astro-ph},
       adsurl = {https://ui.adsabs.harvard.edu/abs/2009ApJ...693...23M}
}

@ARTICLE{de_Jager_1966,
       author = {{de Jager}, C. and {Namba}, O. and {Neven}, L.},
        title = "{The profile of the infrared He I lines over the undisturbed solar disk}",
      journal = {Bull. Astron. Inst. Neth.},
         year = 1966,
        month = jan,
       volume = {18},
        pages = {128},
       adsurl = {https://ui.adsabs.harvard.edu/abs/1966BAN....18..128D}
}

@incollection{Drake_1996,
  author    = {Drake, G. W. F.},
  title     = {High Precision Calculations for Helium},
  booktitle = {Atomic, Molecular, \& Optical Physics Handbook},
  publisher = {American Institute of Physics},
  address   = {Woodbury, New York},
  year      = {1996},
  pages     = {154--171}
}

@ARTICLE{Fuhrmeister_2019,
       author = {{Fuhrmeister}, B. and {Czesla}, S. and {Hildebrandt}, L. and {Nagel}, E. and {Schmitt}, J.~H.~M.~M. and {Hintz}, D. and {Johnson}, E.~N. and {Sanz-Forcada}, J. and {Sch{\"o}fer}, P. and {Jeffers}, S.~V. and {Caballero}, J.~A. and {Zechmeister}, M. and {Reiners}, A. and {Ribas}, I. and {Amado}, P.~J. and {Quirrenbach}, A. and {Bauer}, F.~F. and {B{\'e}jar}, V.~J.~S. and {Cort{\'e}s-Contreras}, M. and {D{\'\i}ez-Alonso}, E. and {Dreizler}, S. and {Galad{\'\i}-Enr{\'\i}quez}, D. and {Guenther}, E.~W. and {Kaminski}, A. and {K{\"u}rster}, M. and {Lafarga}, M. and {Montes}, D.},
        title = "{The CARMENES search for exoplanets around M dwarfs. The He I triplet at 10830 {\r{A}} across the M dwarf sequence}",
      journal = {\aap},
         year = 2019,
        month = dec,
       volume = {632},
          eid = {A24},
        pages = {A24},
          doi = {10.1051/0004-6361/201936193},
archivePrefix = {arXiv},
       eprint = {1911.00246},
 primaryClass = {astro-ph.SR},
       adsurl = {https://ui.adsabs.harvard.edu/abs/2019A&A...632A..24F}
}

@ARTICLE{Fuhrmeister_2023,
       author = {{Fuhrmeister}, B. and {Czesla}, S. and {Schmitt}, J.~H.~M.~M. and {Schneider}, P.~C. and {Caballero}, J.~A. and {Jeffers}, S.~V. and {Nagel}, E. and {Montes}, D. and {G{\'a}lvez Ortiz}, M.~C. and {Reiners}, A. and {Ribas}, I. and {Quirrenbach}, A. and {Amado}, P.~J. and {Henning}, Th. and {Lodieu}, N. and {Mart{\'\i}n-Fern{\'a}ndez}, P. and {Morales}, J.~C. and {Sch{\"o}fer}, P. and {Seifert}, W. and {Zechmeister}, M.},
        title = "{The CARMENES search for exoplanets around M dwarfs. Behaviour of the Paschen lines during flares and quiescence}",
      journal = {\aap},
         year = 2023,
        month = oct,
       volume = {678},
          eid = {A1},
        pages = {A1},
          doi = {10.1051/0004-6361/202347161},
archivePrefix = {arXiv},
       eprint = {2308.07685},
 primaryClass = {astro-ph.SR},
       adsurl = {https://ui.adsabs.harvard.edu/abs/2023A&A...678A...1F}
}

@ARTICLE{Wordsworth_2022,
       author = {{Wordsworth}, Robin and {Kreidberg}, Laura},
        title = "{Atmospheres of Rocky Exoplanets}",
      journal = {\araa},
         year = 2022,
        month = aug,
       volume = {60},
        pages = {159-201},
          doi = {10.1146/annurev-astro-052920-125632},
archivePrefix = {arXiv},
       eprint = {2112.04663},
 primaryClass = {astro-ph.EP},
       adsurl = {https://ui.adsabs.harvard.edu/abs/2022ARA&A..60..159W}
}

@ARTICLE{redfield2000,
       author = {{Redfield}, Seth and {Linsky}, Jeffrey L.},
        title = "{The Three-dimensional Structure of the Warm Local Interstellar Medium. II. The Colorado Model of the Local Interstellar Cloud}",
      journal = {\apj},
         year = 2000,
        month = may,
       volume = {534},
       number = {2},
        pages = {825-837},
          doi = {10.1086/308769},
       adsurl = {https://ui.adsabs.harvard.edu/abs/2000ApJ...534..825R}
}

@ARTICLE{cash1979,
       author = {{Cash}, W.},
        title = "{Parameter estimation in astronomy through application of the likelihood ratio.}",
      journal = {\apj},
         year = 1979,
        month = mar,
       volume = {228},
        pages = {939-947},
          doi = {10.1086/156922},
       adsurl = {https://ui.adsabs.harvard.edu/abs/1979ApJ...228..939C}
}

@ARTICLE{aslpund2009,
       author = {{Asplund}, Martin and {Grevesse}, Nicolas and {Sauval}, A. Jacques and {Scott}, Pat},
        title = "{The Chemical Composition of the Sun}",
      journal = {\araa},
         year = 2009,
        month = sep,
       volume = {47},
       number = {1},
        pages = {481-522},
          doi = {10.1146/annurev.astro.46.060407.145222},
archivePrefix = {arXiv},
       eprint = {0909.0948},
 primaryClass = {astro-ph.SR},
       adsurl = {https://ui.adsabs.harvard.edu/abs/2009ARA&A..47..481A}
}

@ARTICLE{Caldiroli_2022,
       author = {{Caldiroli}, Andrea and {Haardt}, Francesco and {Gallo}, Elena and {Spinelli}, Riccardo and {Malsky}, Isaac and {Rauscher}, Emily},
        title = "{Irradiation-driven escape of primordial planetary atmospheres. II. Evaporation efficiency of sub-Neptunes through hot Jupiters}",
      journal = {\aap},
         year = 2022,
        month = jul,
       volume = {663},
          eid = {A122},
        pages = {A122},
          doi = {10.1051/0004-6361/202142763},
archivePrefix = {arXiv},
       eprint = {2112.00744},
 primaryClass = {astro-ph.EP},
       adsurl = {https://ui.adsabs.harvard.edu/abs/2022A&A...663A.122C}
}

@ARTICLE{Diamond-Lowe_2020,
       author = {{Diamond-Lowe}, Hannah and {Berta-Thompson}, Zachory and {Charbonneau}, David and {Dittmann}, Jason and {Kempton}, Eliza M. -R.},
        title = "{Simultaneous Optical Transmission Spectroscopy of a Terrestrial, Habitable-zone Exoplanet with Two Ground-based Multiobject Spectrographs}",
      journal = {\aj},
         year = 2020,
        month = jul,
       volume = {160},
       number = {1},
          eid = {27},
        pages = {27},
          doi = {10.3847/1538-3881/ab935f},
archivePrefix = {arXiv},
       eprint = {1909.09104},
 primaryClass = {astro-ph.EP},
       adsurl = {https://ui.adsabs.harvard.edu/abs/2020AJ....160...27D}
}

@ARTICLE{judge2003,
       author = {{Judge}, Philip G. and {Solomon}, Stanley C. and {Ayres}, Thomas R.},
        title = "{An Estimate of the Sun's ROSAT-PSPC X-Ray Luminosities Using SNOE-SXP Measurements}",
      journal = {\apj},
         year = 2003,
        month = aug,
       volume = {593},
       number = {1},
        pages = {534-548},
          doi = {10.1086/376405},
       adsurl = {https://ui.adsabs.harvard.edu/abs/2003ApJ...593..534J}
}

@ARTICLE{Ginzburg_2018,
       author = {{Ginzburg}, Sivan and {Schlichting}, Hilke E. and {Sari}, Re'em},
        title = "{Core-powered mass-loss and the radius distribution of small exoplanets}",
      journal = {\mnras},
         year = 2018,
        month = may,
       volume = {476},
       number = {1},
        pages = {759-765},
          doi = {10.1093/mnras/sty290},
archivePrefix = {arXiv},
       eprint = {1708.01621},
 primaryClass = {astro-ph.EP},
       adsurl = {https://ui.adsabs.harvard.edu/abs/2018MNRAS.476..759G}
}

@ARTICLE{Wright_2018,
       author = {{Wright}, Nicholas J. and {Newton}, Elisabeth R. and {Williams}, Peter K.~G. and {Drake}, Jeremy J. and {Yadav}, Rakesh K.},
        title = "{The stellar rotation-activity relationship in fully convective M dwarfs}",
      journal = {\mnras},
         year = 2018,
        month = sep,
       volume = {479},
       number = {2},
        pages = {2351-2360},
          doi = {10.1093/mnras/sty1670},
archivePrefix = {arXiv},
       eprint = {1807.03304},
 primaryClass = {astro-ph.SR},
       adsurl = {https://ui.adsabs.harvard.edu/abs/2018MNRAS.479.2351W}
}

@ARTICLE{brown2023,
       author = {{Brown}, Alexander and {Schneider}, P. Christian and {France}, Kevin and {Froning}, Cynthia S. and {Youngblood}, Allison A. and {J. Wilson}, David and {Loyd}, R.~O. Parke and {Pineda}, J. Sebastian and {Duvvuri}, Girish M. and {Kowalski}, Adam F. and {Berta-Thompson}, Zachory K.},
        title = "{Coronal X-Ray Emission from Nearby, Low-mass, Exoplanet Host Stars Observed by the MUSCLES and Mega-MUSCLES HST Treasury Survey Projects}",
      journal = {\aj},
         year = 2023,
        month = may,
       volume = {165},
       number = {5},
          eid = {195},
        pages = {195},
          doi = {10.3847/1538-3881/acc38a},
archivePrefix = {arXiv},
       eprint = {2303.12929},
 primaryClass = {astro-ph.SR},
       adsurl = {https://ui.adsabs.harvard.edu/abs/2023AJ....165..195B}
}

@ARTICLE{cilley2024,
       author = {{Cilley}, Raven and {King}, George W. and {Corrales}, L{\'\i}a},
        title = "{Detecting Exoplanet Transits with the Next Generation of X-Ray Telescopes}",
      journal = {\aj},
         year = 2024,
        month = oct,
       volume = {168},
       number = {4},
          eid = {177},
        pages = {177},
          doi = {10.3847/1538-3881/ad6d60},
archivePrefix = {arXiv},
       eprint = {2408.06417},
 primaryClass = {astro-ph.EP},
       adsurl = {https://ui.adsabs.harvard.edu/abs/2024AJ....168..177C}
}

@ARTICLE{wilson2025-Mega-MUSCLES,
       author = {{Wilson}, David J. and {Froning}, Cynthia S. and {Duvvuri}, Girish M. and {Youngblood}, Allison and {France}, Kevin and {Brown}, Alexander and {Schneider}, P. Christian and {Berta-Thompson}, Zachory and {Buccino}, Andrea P. and {Linsky}, Jeffrey and {Loyd}, R.~O. Parke and {Miguel}, Yamila and {Newton}, Elisabeth and {Pineda}, J. Sebastian and {Redfield}, Seth and {Roberge}, Aki and {Rugheimer}, Sarah and {Vieytes}, Mariela C.},
        title = "{The Mega-MUSCLES Treasury Survey: X-Ray to Infrared Spectral Energy Distributions of a Representative Sample of M Dwarfs}",
      journal = {\apj},
         year = 2025,
        month = jan,
       volume = {978},
       number = {1},
          eid = {85},
        pages = {85},
          doi = {10.3847/1538-4357/ad9251},
archivePrefix = {arXiv},
       eprint = {2411.07394},
 primaryClass = {astro-ph.EP},
       adsurl = {https://ui.adsabs.harvard.edu/abs/2025ApJ...978...85W}
}

@ARTICLE{wilson2021-Mega-MUSCLES,
       author = {{Wilson}, David J. and {Froning}, Cynthia S. and {Duvvuri}, Girish M. and {France}, Kevin and {Youngblood}, Allison and {Schneider}, P. Christian and {Berta-Thompson}, Zachory and {Brown}, Alexander and {Buccino}, Andrea P. and {Hawley}, Suzanne and {Irwin}, Jonathan and {Kaltenegger}, Lisa and {Kowalski}, Adam and {Linsky}, Jeffrey and {Loyd}, R.~O. Parke and {Miguel}, Yamila and {Pineda}, J. Sebastian and {Redfield}, Seth and {Roberge}, Aki and {Rugheimer}, Sarah and {Tian}, Feng and {Vieytes}, Mariela},
        title = "{The Mega-MUSCLES Spectral Energy Distribution of TRAPPIST-1}",
      journal = {\apj},
         year = 2021,
        month = apr,
       volume = {911},
       number = {1},
          eid = {18},
        pages = {18},
          doi = {10.3847/1538-4357/abe771},
archivePrefix = {arXiv},
       eprint = {2102.11415},
 primaryClass = {astro-ph.SR},
       adsurl = {https://ui.adsabs.harvard.edu/abs/2021ApJ...911...18W}
}

@ARTICLE{Toledo_2019,
       author = {{Toledo-Padr{\'o}n}, B. and {Gonz{\'a}lez Hern{\'a}ndez}, J.~I. and {Rodr{\'\i}guez-L{\'o}pez}, C. and {Su{\'a}rez Mascare{\~n}o}, A. and {Rebolo}, R. and {Butler}, R.~P. and {Ribas}, I. and {Anglada-Escud{\'e}}, G. and {Johnson}, E.~N. and {Reiners}, A. and {Caballero}, J.~A. and {Quirrenbach}, A. and {Amado}, P.~J. and {B{\'e}jar}, V.~J.~S. and {Morales}, J.~C. and {Perger}, M. and {Jeffers}, S.~V. and {Vogt}, S. and {Teske}, J. and {Shectman}, S. and {Crane}, J. and {D{\'\i}az}, M. and {Arriagada}, P. and {Holden}, B. and {Burt}, J. and {Rodr{\'\i}guez}, E. and {Herrero}, E. and {Murgas}, F. and {Pall{\'e}}, E. and {Morales}, N. and {L{\'o}pez-Gonz{\'a}lez}, M.~J. and {D{\'\i}ez Alonso}, E. and {Tuomi}, M. and {Kiraga}, M. and {Engle}, S.~G. and {Guinan}, E.~F. and {Strachan}, J.~B.~P. and {Aceituno}, F.~J. and {Aceituno}, J. and {Casanova}, V.~M. and {Mart{\'\i}n-Ruiz}, S. and {Montes}, D. and {Ortiz}, J.~L. and {Sota}, A. and {Briol}, J. and {Barbieri}, L. and {Cervini}, I. and {Deldem}, M. and {Dubois}, F. and {Hambsch}, F. -J. and {Harris}, B. and {Kotnik}, C. and {Logie}, L. and {Lopez}, J. and {McNeely}, M. and {Ogmen}, Y. and {P{\'e}rez}, L. and {Rau}, S. and {Rodr{\'\i}guez}, D. and {Urquijo}, F.~S. and {Vanaverbeke}, S.},
        title = "{Stellar activity analysis of Barnard's Star: very slow rotation and evidence for long-term activity cycle}",
      journal = {\mnras},
         year = 2019,
        month = oct,
       volume = {488},
       number = {4},
        pages = {5145-5161},
          doi = {10.1093/mnras/stz1975},
archivePrefix = {arXiv},
       eprint = {1812.06712},
 primaryClass = {astro-ph.SR},
       adsurl = {https://ui.adsabs.harvard.edu/abs/2019MNRAS.488.5145T}
}

@ARTICLE{Bonfils_2018,
       author = {{Bonfils}, X. and {Almenara}, J. -M. and {Cloutier}, R. and {W{\"u}nsche}, A. and {Astudillo-Defru}, N. and {Berta-Thompson}, Z. and {Bouchy}, F. and {Charbonneau}, D. and {Delfosse}, X. and {D{\'\i}az}, R.~F. and {Dittmann}, J. and {Doyon}, R. and {Forveille}, T. and {Irwin}, J. and {Lovis}, C. and {Mayor}, M. and {Menou}, K. and {Murgas}, F. and {Newton}, E. and {Pepe}, F. and {Santos}, N.~C. and {Udry}, S.},
        title = "{Radial velocity follow-up of GJ1132 with HARPS. A precise mass for planet b and the discovery of a second planet}",
      journal = {\aap},
         year = 2018,
        month = oct,
       volume = {618},
          eid = {A142},
        pages = {A142},
          doi = {10.1051/0004-6361/201731884},
archivePrefix = {arXiv},
       eprint = {1806.03870},
 primaryClass = {astro-ph.EP},
       adsurl = {https://ui.adsabs.harvard.edu/abs/2018A&A...618A.142B}
}

@ARTICLE{Charbonneau_2002,
       author = {{Charbonneau}, David and {Brown}, Timothy M. and {Noyes}, Robert W. and {Gilliland}, Ronald L.},
        title = "{Detection of an Extrasolar Planet Atmosphere}",
      journal = {\apj},
         year = 2002,
        month = mar,
       volume = {568},
       number = {1},
        pages = {377-384},
          doi = {10.1086/338770},
archivePrefix = {arXiv},
       eprint = {astro-ph/0111544},
 primaryClass = {astro-ph},
       adsurl = {https://ui.adsabs.harvard.edu/abs/2002ApJ...568..377C}
}

@ARTICLE{Hu_2015,
       author = {{Hu}, Renyu and {Seager}, Sara and {Yung}, Yuk L.},
        title = "{Helium Atmospheres on Warm Neptune- and Sub-Neptune-sized Exoplanets and Applications to GJ 436b}",
      journal = {\apj},
         year = 2015,
        month = jul,
       volume = {807},
       number = {1},
          eid = {8},
        pages = {8},
          doi = {10.1088/0004-637X/807/1/8},
archivePrefix = {arXiv},
       eprint = {1505.02221},
 primaryClass = {astro-ph.EP},
       adsurl = {https://ui.adsabs.harvard.edu/abs/2015ApJ...807....8H}
}

@ARTICLE{Lammer_2025,
       author = {{Lammer}, Helmut and {Scherf}, Manuel and {Erkaev}, Nikolai V. and {Kubyshkina}, Daria and {Gorbunova}, Kseniia D. and {Fossati}, Luca and {Woitke}, Peter},
        title = "{Earth-mass planets with He atmospheres in the habitable zone of Sun-like stars}",
      journal = {\natast},
         year = 2025,
        month = jul,
       volume = {9},
        pages = {1022-1030},
          doi = {10.1038/s41550-025-02550-6},
       adsurl = {https://ui.adsabs.harvard.edu/abs/2025NatAs...9.1022L}
}

@ARTICLE{Lampon_2020,
       author = {{Lamp{\'o}n}, M. and {L{\'o}pez-Puertas}, M. and {Lara}, L.~M. and {S{\'a}nchez-L{\'o}pez}, A. and {Salz}, M. and {Czesla}, S. and {Sanz-Forcada}, J. and {Molaverdikhani}, K. and {Alonso-Floriano}, F.~J. and {Nortmann}, L. and {Caballero}, J.~A. and {Bauer}, F.~F. and {Pall{\'e}}, E. and {Montes}, D. and {Quirrenbach}, A. and {Nagel}, E. and {Ribas}, I. and {Reiners}, A. and {Amado}, P.~J.},
        title = "{Modelling the He I triplet absorption at 10 830 {\r{A}} in the atmosphere of HD 209458 b}",
      journal = {\aap},
         year = 2020,
        month = apr,
       volume = {636},
          eid = {A13},
        pages = {A13},
          doi = {10.1051/0004-6361/201937175},
archivePrefix = {arXiv},
       eprint = {2003.04872},
 primaryClass = {astro-ph.EP},
       adsurl = {https://ui.adsabs.harvard.edu/abs/2020A&A...636A..13L}
}

@ARTICLE{Lampon_2021,
       author = {{Lamp{\'o}n}, M. and {L{\'o}pez-Puertas}, M. and {Sanz-Forcada}, J. and {S{\'a}nchez-L{\'o}pez}, A. and {Molaverdikhani}, K. and {Czesla}, S. and {Quirrenbach}, A. and {Pall{\'e}}, E. and {Caballero}, J.~A. and {Henning}, T. and {Salz}, M. and {Nortmann}, L. and {Aceituno}, J. and {Amado}, P.~J. and {Bauer}, F.~F. and {Montes}, D. and {Nagel}, E. and {Reiners}, A. and {Ribas}, I.},
        title = "{Modelling the He I triplet absorption at 10 830 {\r{A}} in the atmospheres of HD 189733 b and GJ 3470 b}",
      journal = {\aap},
         year = 2021,
        month = mar,
       volume = {647},
          eid = {A129},
        pages = {A129},
          doi = {10.1051/0004-6361/202039417},
archivePrefix = {arXiv},
       eprint = {2101.09393},
 primaryClass = {astro-ph.EP},
       adsurl = {https://ui.adsabs.harvard.edu/abs/2021A&A...647A.129L}
}

@ARTICLE{Zhang_2023,
       author = {{Zhang}, Michael and {Knutson}, Heather A. and {Dai}, Fei and {Wang}, Lile and {Ricker}, George R. and {Schwarz}, Richard P. and {Mann}, Christopher and {Collins}, Karen},
        title = "{Detection of Atmospheric Escape from Four Young Mini-Neptunes}",
      journal = {\aj},
         year = 2023,
        month = feb,
       volume = {165},
       number = {2},
          eid = {62},
        pages = {62},
          doi = {10.3847/1538-3881/aca75b},
archivePrefix = {arXiv},
       eprint = {2207.13099},
 primaryClass = {astro-ph.EP},
       adsurl = {https://ui.adsabs.harvard.edu/abs/2023AJ....165...62Z}
}

@ARTICLE{Zhang_2025,
       author = {{Zhang}, Michael and {Bean}, Jacob L. and {Wilson}, David and {Duvvuri}, Girish and {Schneider}, Christian and {Knutson}, Heather A. and {Dai}, Fei and {Collins}, Karen A. and {Watkins}, Cristilyn N. and {Schwarz}, Richard P. and {Barkaoui}, Khalid and {Shporer}, Avi and {Horne}, Keith and {Sefako}, Ramotholo and {Murgas}, Felipe and {Palle}, Enric},
        title = "{Constraining Atmospheric Composition from the Outflow: Helium Observations Reveal the Fundamental Properties of Two Planets Straddling the Radius Gap}",
      journal = {\aj},
         year = 2025,
        month = apr,
       volume = {169},
       number = {4},
          eid = {204},
        pages = {204},
          doi = {10.3847/1538-3881/adb490},
archivePrefix = {arXiv},
       eprint = {2409.08318},
 primaryClass = {astro-ph.EP},
       adsurl = {https://ui.adsabs.harvard.edu/abs/2025AJ....169..204Z}
}

@ARTICLE{MacLeod_2022,
       author = {{MacLeod}, Morgan and {Oklop{\v{c}}i{\'c}}, Antonija},
        title = "{Stellar Wind Confinement of Evaporating Exoplanet Atmospheres and Its Signatures in 1083 nm Observations}",
      journal = {\apj},
         year = 2022,
        month = feb,
       volume = {926},
       number = {2},
          eid = {226},
        pages = {226},
          doi = {10.3847/1538-4357/ac46ce},
archivePrefix = {arXiv},
       eprint = {2107.07534},
 primaryClass = {astro-ph.EP},
       adsurl = {https://ui.adsabs.harvard.edu/abs/2022ApJ...926..226M}
}

@ARTICLE{Coulombe_2025,
       author = {{Coulombe}, Louis-Philippe and {Benneke}, Bj{\"o}rn and {Krissansen-Totton}, Joshua and {L'Heureux}, Alexandrine and {Piaulet-Ghorayeb}, Caroline and {Radica}, Michael and {Roy}, Pierre-Alexis and {Ahrer}, Eva-Maria and {Cadieux}, Charles and {Miguel}, Yamila and {Schlichting}, Hilke E. and {Delgado-Mena}, Elisa and {Monaghan}, Christopher and {Adamski}, Hanna and {Raul}, Eshan and {Cloutier}, Ryan and {Komacek}, Thaddeus D. and {Taylor}, Jake and {Gapp}, Cyril and {Allart}, Romain and {Bouchy}, Fran{\c{c}}ois and {Canto Martins}, Bruno L. and {Cook}, Neil J. and {Doyon}, Ren{\'e} and {Evans-Soma}, Thomas M. and {Larue}, Pierre and {Su{\'a}rez Mascare{\~n}o}, Alejandro and {Wardenier}, Joost P.},
        title = "{Possible Evidence for the Presence of Volatiles on the Warm Super-Earth TOI-270 b}",
      journal = {\aj},
         year = 2025,
        month = oct,
       volume = {170},
       number = {4},
          eid = {226},
        pages = {226},
          doi = {10.3847/1538-3881/adfc6a},
       adsurl = {https://ui.adsabs.harvard.edu/abs/2025AJ....170..226C}
}

@ARTICLE{Glidden_2025,
       author = {{Glidden}, Ana and {Ranjan}, Sukrit and {Seager}, Sara and {Espinoza}, N{\'e}stor and {MacDonald}, Ryan J. and {Allen}, Natalie H. and {Ca{\~n}as}, Caleb I. and {Grant}, David and {Gressier}, Am{\'e}lie and {Stevenson}, Kevin B. and {Batalha}, Natasha E. and {Lewis}, Nikole K. and {Long}, Douglas and {Wakeford}, Hannah R. and {Alderson}, Lili and {Challener}, Ryan C. and {Col{\'o}n}, Knicole and {Huang}, Jingcheng and {Lin}, Zifan and {Louie}, Dana R. and {Mullens}, Elijah and {Sotzen}, Kristin S. and {Valenti}, Jeff A. and {Valentine}, Daniel and {Clampin}, Mark and {Mountain}, C. Matt and {Perrin}, Marshall and {van der Marel}, Roeland P.},
        title = "{JWST-TST DREAMS: Secondary Atmosphere Constraints for the Habitable Zone Planet TRAPPIST-1 e}",
      journal = {\apjl},
         year = 2025,
        month = sep,
       volume = {990},
       number = {2},
          eid = {L53},
        pages = {L53},
          doi = {10.3847/2041-8213/adf62e},
archivePrefix = {arXiv},
       eprint = {2509.05407},
 primaryClass = {astro-ph.EP},
       adsurl = {https://ui.adsabs.harvard.edu/abs/2025ApJ...990L..53G}
}

@ARTICLE{Diamond-Lowe_2023,
       author = {{Diamond-Lowe}, Hannah and {Mendon{\c{c}}a}, Jo{\~a}o M. and {Charbonneau}, David and {Buchhave}, Lars A.},
        title = "{Ground-based Optical Transmission Spectroscopy of the Nearby Terrestrial Exoplanet LTT 1445Ab}",
      journal = {\aj},
         year = 2023,
        month = apr,
       volume = {165},
       number = {4},
          eid = {169},
        pages = {169},
          doi = {10.3847/1538-3881/acbf39},
archivePrefix = {arXiv},
       eprint = {2210.11809},
 primaryClass = {astro-ph.EP},
       adsurl = {https://ui.adsabs.harvard.edu/abs/2023AJ....165..169D}
}

@ARTICLE{Carnall_2017,
       author = {{Carnall}, A.~C.},
        title = "{SpectRes: A Fast Spectral Resampling Tool in Python}",
      journal = {arXiv e-prints},
         year = 2017,
        month = may,
          doi = {10.48550/arXiv.1705.05165},
archivePrefix = {arXiv},
       eprint = {1705.05165},
 primaryClass = {astro-ph.IM},
       adsurl = {https://ui.adsabs.harvard.edu/abs/2017arXiv170505165C}
}

@ARTICLE{Foreman-Mackey_2013,
       author = {{Foreman-Mackey}, Daniel and {Hogg}, David W. and {Lang}, Dustin and {Goodman}, Jonathan},
        title = "{emcee: The MCMC Hammer}",
      journal = {\pasp},
         year = 2013,
        month = mar,
       volume = {125},
       number = {925},
        pages = {306},
          doi = {10.1086/670067},
archivePrefix = {arXiv},
       eprint = {1202.3665},
 primaryClass = {astro-ph.IM},
       adsurl = {https://ui.adsabs.harvard.edu/abs/2013PASP..125..306F}
}

@ARTICLE{Hunten_1987,
       author = {{Hunten}, D.~M. and {Pepin}, R.~O. and {Walker}, J.~C.~G.},
        title = "{Mass fractionation in hydrodynamic escape}",
      journal = {\icarus},
         year = 1987,
        month = mar,
       volume = {69},
       number = {3},
        pages = {532-549},
          doi = {10.1016/0019-1035(87)90022-4},
       adsurl = {https://ui.adsabs.harvard.edu/abs/1987Icar...69..532H}
}

@ARTICLE{Fuhrmeister_2020,
       author = {{Fuhrmeister}, B. and {Czesla}, S. and {Hildebrandt}, L. and {Nagel}, E. and {Schmitt}, J.~H.~M.~M. and {Jeffers}, S.~V. and {Caballero}, J.~A. and {Hintz}, D. and {Johnson}, E.~N. and {Sch{\"o}fer}, P. and {Zechmeister}, M. and {Reiners}, A. and {Ribas}, I. and {Amado}, P.~J. and {Quirrenbach}, A. and {Nortmann}, L. and {Bauer}, F.~F. and {B{\'e}jar}, V.~J.~S. and {Cort{\'e}s-Contreras}, M. and {Dreizler}, S. and {Galad{\'\i}-Enr{\'\i}quez}, D. and {Hatzes}, A.~P. and {Kaminski}, A. and {K{\"u}rster}, M. and {Lafarga}, M. and {Montes}, D.},
        title = "{The CARMENES search for exoplanets around M dwarfs. Variability of the He I line at 10 830 {\r{A}}}",
      journal = {\aap},
         year = 2020,
        month = aug,
       volume = {640},
          eid = {A52},
        pages = {A52},
          doi = {10.1051/0004-6361/202038279},
archivePrefix = {arXiv},
       eprint = {2006.09372},
 primaryClass = {astro-ph.SR},
       adsurl = {https://ui.adsabs.harvard.edu/abs/2020A&A...640A..52F}
}

@ARTICLE{Savanov_2012,
       author = {{Savanov}, I.~S.},
        title = "{Activity cycles of M dwarfs}",
      journal = {Astron. Rep.},
         year = 2012,
        month = sep,
       volume = {56},
       number = {9},
        pages = {716-721},
          doi = {10.1134/S1063772912090077},
       adsurl = {https://ui.adsabs.harvard.edu/abs/2012ARep...56..716S}
}

@ARTICLE{Suarez_2016,
       author = {{Su{\'a}rez Mascare{\~n}o}, A. and {Rebolo}, R. and {Gonz{\'a}lez Hern{\'a}ndez}, J.~I.},
        title = "{Magnetic cycles and rotation periods of late-type stars from photometric time series}",
      journal = {\aap},
         year = 2016,
        month = oct,
       volume = {595},
          eid = {A12},
        pages = {A12},
          doi = {10.1051/0004-6361/201628586},
archivePrefix = {arXiv},
       eprint = {1607.03049},
 primaryClass = {astro-ph.SR},
       adsurl = {https://ui.adsabs.harvard.edu/abs/2016A&A...595A..12S}
}
\bibliographystyle{sciencemag}

%
%
%
%
%
%

\section*{Acknowledgments}
We thank Evgenya Shkolnik for helpful comments on stellar helium lines. This paper is based on WINERED data gathered with the 6.5 m Clay/Magellan II Telescope located at Las Campanas Observatory, Chile. We thank the staff at Las Campanas Observatory for their efforts and particularly thank Carla Fuentes and Hernan Nuñez for assistance with telescope operations on the nights of our observations. We thank the WINERED team: Noriyuki Matsunaga, Shogo Otsubo, Yuki Saragaku, and Tomomi Takeuchi, for their assistance and support of instrument operations.
\paragraph*{Funding:}
A.~McW. and J.~T. acknowledge a Carnegie Venture Grant, which funded the installation and fabrication of ancillary equipment to enable WINERED at Las Campanas Observatory.
T.~C. was supported by NASA through the NASA Hubble Fellowship grant HST-HF2-51527.001-A awarded by the Space Telescope Science Institute, which is operated by the Association of Universities for Research in Astronomy, Inc., for NASA, under contract NAS5-26555.
A.~H. was supported by the National Science Foundation Graduate Research Fellowship under Grant No. 2141064 and the MIT Dean of Science Fellowship.
M.~Z. and J.~A.~D. were supported by the Heising-Simons Foundation's 51 Pegasi b fellowship (FP107579).
S.~V. acknowledges support from the Mt. Cuba Astronomical Foundation.
R.~W. acknowledges support from Leverhulme Center for Life in the Universe grant G119167, LBAG/312.
W.~M. acknowledges support from the AEThER program, funded in part by the Alfred P.\ Sloan Foundation under grant \#G202114194, and the Carnegie Postdoctoral Fellowship.
N.~L.~W. was supported by NASA through a grant (for program \#2512) from the Space Telescope Science Institute, which is operated by the Association of Universities for Research in Astronomy, Inc., under NASA contract NAS 5-03127.
M.~L.-M. was supported by NASA contracts NAS 5-26555 and NAS 5-03127 to the Associated Universities for Research in Astronomy for the operation of the Hubble and James Webb Space Telescope Science Operations Centers at STScI.

\paragraph*{Author contributions:}
C.~C., S.~V., A.~G.~M., and A.~H. planned and performed the WINERED observations. C.~C. led the data analysis and interpretation. S.~V. wrote data analysis code. J.~A.~D. proposed and collected the XMM-Newton data, which T.~C. analyzed. T.~C. also wrote the X-ray observations and analysis section. R.~W. and D.~C. supported the data reduction and interpretation. A.~G.~M. and M.~L.-M. contributed to the data analysis and interpretation, particularly the Gaussian process fitting. L.~A.~D.~S. supported the atmospheric retrieval. M.~Z. performed the independent data reduction. W.~M. and Z.~L. contributed theoretical interpretation. N.~L.~W. and J.~T. contributed to observational protocols followed in this study. A.~McW. investigated stellar contamination. C.~C. led manuscript writing, with contributions from J.~T., A.~McW., and T.~C.
\paragraph*{Competing interests:}
There are no competing interests to declare.
\paragraph*{Data, code, and materials availability:}




The WINERED observations and all code necessary to reproduce our findings and main figures are archived at Zenodo~\cite{zenodo}.
No physical materials were generated in this work.


\subsection*{Supplementary materials}
Materials and Methods\\
Figures S1-S13\\
Table S1\\
References \textit{(50-\arabic{enumiv})} 


\newpage


\renewcommand{\thefigure}{S\arabic{figure}}
\renewcommand{\thetable}{S\arabic{table}}
\renewcommand{\theequation}{S\arabic{equation}}
\renewcommand{\thepage}{S\arabic{page}}
\setcounter{figure}{0}
\setcounter{table}{0}
\setcounter{equation}{0}
\setcounter{page}{1} 


\begin{center}
\section*{Supplementary Materials for\\ \scititle}

\author{
Collin Cherubim$^\ast$,
Shreyas Vissapragada,
Tim Cunningham,
Annabella G. Meech,
David Charbonneau,
Robin Wordsworth,
Aaron Householder,
Johanna Teske,
Leonardo A. Dos Santos,
Nicole L. Wallack,
William Misener,
Zifan Lin,
Andrew McWilliam,
Michael Zhang,
Jason A. Dittmann,
Mercedes L\'opez-Morales 
\and
\small$^\ast$Corresponding author. Email: ccherubim@g.harvard.edu
}
\end{center}

\subsubsection*{This PDF file includes:}
Materials and Methods\\
Supplementary Text\\
Figures S1 to S13\\
Table S1


\newpage


\subsection*{Materials and Methods}






\subsubsection*{Observations and data reduction} \label{sec:reduction}
We observed one transit each for LHS~1140b and LHS~1140c on Universal Time (UT) 2024 September 23 and an additional transit of LHS~1140b on UT 2025 September 29 using the WINERED spectrograph mounted on the 6.5 m Clay/Magellan II Telescope at Las Campanas Observatory in Chile~\cite{Ikeda_2016, Ikeda_2022}. Data were obtained in HIRES-Y mode~\cite{Otsubo_2016} with the 100 micrometer slit with a resolving power of 68,000. We took 300 s exposures in an ABBA nod pattern to correct for telluric OH lines. Some residuals remained after OH subtraction at 10,834 \AA\, which we later masked by excluding these data by eye. In 2024, we took 70 exposures between UT 03:20:51 and UT 09:49:13 with a starting airmass of 1.161 and a final airmass of 2.146. In 2025, we took 51 exposures between UT 05:00:01 and 09:46:46 with a starting airmass of 1.031 and final airmass of 2.810.

We reduced the spectra using WARP~\cite{Hamano_2024}. WARP performs a flat-field correction and subtracts sky emission and scattered light, then performs aperture extraction and wavelength calibration using a ThAr lamp spectrum. For our analysis, we used the fluxes and wavelengths for echelle order 163 which spans wavelengths 10,793 to 10,859\,\AA\ and contains the metastable helium triplet. In 2024, the median signal-to-noise ratio (SNR) was 105; we discarded one frame with SNR $<$ 20 which was collected during post-egress of LHS~1140b. In 2025, the median SNR was 131. In 2024, we noticed persistence in the pixel counts between positions A and B. Internal WINERED tests have shown that persistence drops to negligible levels after 30 minutes~\cite{Vissapragada_2024}. We therefore discarded the first 30 minutes, corresponding to 6 exposures, leaving 62 spectra for our analysis.

We continuum-normalized the spectra by fitting a fourth-degree Chebyshev polynomial to the regions with the fewest visible lines and dividing each spectrum by the fitted continuum. We observed a slight continuum offset between spectra collected at the A position vs. the B position at the edges of the order. This was corrected by multiplying all B spectra by a continuum offset spectrum produced by dividing the sum of all A spectra by the sum of all B spectra. We then refined the wavelength solution produced by WARP by comparing our spectra to a line-by-line telluric transmission spectrum calculated with the High Resolution Transmission (HITRAN) \texttt{hapi} software~\cite{Kochanov_2016, Gordon_2022}. After cross-correlating the telluric template with the mean spectrum from each night, we shifted the wavelength grids to ensure all data were aligned in the telluric rest frame.

After normalizing and aligning the spectra, we masked the telluric features, following previous methods~\cite{Vissapragada_2024}. We used a synthetic telluric transmission spectrum generated with \texttt{hapi} to identify two water absorption lines redward of the helium triplet based on a mask threshold for features exceeding 2.5\% absorption of the stellar flux. We also identified and masked a residual from OH sky subtraction~\cite{Oliva_2015}.

The data were then shifted into the stellar rest frame to create a time series of excess absorption. First, we used the \texttt{astropy} software~\cite{astropy_2013, astropy_2018, astropy_2022} to shift the spectra first into the Solar System barycenter frame then into the stellar rest frame using the systemic radial velocity of -13.23~$\pm$~0.60 km s$^{-1}$~\cite{Dittmann_2017}. In the stellar rest frame, we constructed a combined out-of-transit stellar template spectrum for each night using ephemerides of LHS~1140b's orbit \cite{Cadieux_2024a} which had a precision of 178 s at the time of our observations. We then normalized each spectroscopic time series by the stellar template spectrum, subtracted 1, then multiplied by 100\% to calculate excess absorption percentage relative to the stellar template spectrum shown in fig. \ref{fig1}. 

The planetary transmission spectrum was determined by shifting the time series spectra into the planetary rest frame, assuming a circular orbit, using the previously measured orbital inclination of 89.96 {$\pm$ 0.04} degrees and stellar mass of 0.1844 {$\pm$ 0.0045} Solar masses ($M_\mathrm{Sun}$)~\cite{Cadieux_2024a}, and taking the mean of all in-transit spectra (fig. \ref{fig2}). For all wavelength shifts, we used the \texttt{spectres} software~\cite{Carnall_2017} to resample fluxes and their uncertainties onto the revised wavelength grids.

\subsubsection*{Model fitting}\label{sec:model_fit}

To characterize the helium absorption feature at 10,833\,\AA, we adopted a model comparison framework to minimize model complexity. We assumed a line profile consisting of three delta functions at the laboratory vacuum wavelengths of the metastable helium triplet 
[10,832.057, 10,833.217, and 10,833.306 \AA; \cite{Drake_1996}] 
convolved with two Gaussian profiles centered on the blue-most line and on the mean wavelengths of the two red-most lines, which are not resolved. To incorporate instrumental broadening, we assumed a Gaussian line spread function with FWHM of \( 10,833\,\text{\AA} / R \), where \( R = 68{,}000 \) is the resolving power of the WINERED spectrograph in HIRES-Y mode. We validated this kernel by comparing it to nearby telluric features, finding that it reproduced the observed widths. The observed helium absorption feature has a larger FWHM than the instrument broadening profile, therefore thermal broadening dominates over instrumental and natural (Lorentzian) broadening. This is consistent with our expectations for pressures of microbar to nanobar in the upper atmosphere probed by the metastable helium triplet.

We compared models convolved using Gaussian or Voigt profiles using MCMC sampling with the \texttt{emcee} software \cite{Foreman-Mackey_2013}. Both profiles included parameters for a common Doppler shift of the helium lines, a common width (Gaussian \( \sigma \) and Lorentzian \( \gamma \)), and amplitudes for the blue component and blended red components of the triplet. The Voigt profile would provide a better fit to the data if there was substantial pressure broadening, which we did not expect in the low-pressure regime probed by metastable helium. We compared the models via the Bayesian information criterion $\mathrm{BIC} \equiv k \mathrm{ln}(n) - 2\mathrm{ln}(\hat{L})$ where $k$ is the number of model parameters, $n$ is the number of data points, and $\hat{L}$ is the maximum likelihood. The Gaussian model was favored, with \( \Delta \mathrm{BIC} = +8.1 \), indicating strong evidence against the presence of pressure broadening.

To assess whether the data contained absorption in the blue component of the triplet, we compared a single-peak model containing only the two blended red components of the triplet to the previous two-peak model, which included the blue component. The two-peak model was favored (\( \Delta \mathrm{BIC} = -7.4 \)), indicating evidence for the blue component, but with low significance $\sim 2\sigma$ ($\delta = 0.24^{+0.14}_{-0.12}\%$). Based on these tests, we determined that a Gaussian model that includes the blue peak is preferred by the data. We also tested a three-peak model, with one Gaussian convolved with each delta function centered on the three line positions in the helium triplet. The MCMC results were negligibly different from those obtained with the two-peak model. Although our data do not resolve the two blended red peaks, the three-peak model is more physically motivated. Hence, all subsequent analysis used the three-peak model.

After subtracting the best-fitting model from the MCMC analysis, the residuals exhibited correlated noise. To account for this, we incorporated a GP in the likelihood function to model correlated noise. 
We could not identify the source of the correlated noise but it is probably dominated by instrument systematics arising from spectral over-sampling at $\sim$5 pixels per resolution element, particularly during our re-sampling of the wavelength scale when shifting to the stellar and planetary rest frames.
For the GP, we used the Matérn-3/2 kernel in the \texttt{celerite} software~\cite{celerite}, which includes amplitude and length-scale hyperparameters. These were optimized on continuum regions of the same echelle order (10,793 to 10,859\,\AA) excluding the 10,833\,\AA\ absorption window and visible outliers, following previous methods \cite{McCreery_2025}. The optimized hyperparameters were then fixed in the MCMC sampling to avoid degeneracies with the absorption model. The MCMC posteriors are shown in fig. \ref{fig:gauss_corner}.


\subsubsection*{Independent re-reduction}
We re-reduced both LHS~1140b datasets from 2024 and 2025 using a fully independent data reduction pipeline~\cite{zenodo}. This analysis also shows helium absorption in 2024, but not in 2025 (fig. \ref{fig:MZ_reduction}). The re-reduction is based on a previously published pipeline~\cite{Zhang_2023} with several changes. A summary is provided here, with more detail provided in our data repository~\cite{zenodo}.

Calibration exposures for each night were obtained from the instrument team~\cite{zenodo} constructed from flat field images stacked into a master flat. We masked all regions except the order containing the helium triplet (m = 163), as well as pixels below half or above twice the typical pixel values in the neighborhood.
From each A/B nod pair we constructed two gain-scaled, flat-fielded difference images, $I_A=g(A-B)/F$ and $I_B=g(B-A)/F$, where $g=2.27~{\rm e^-\,DN^{-1}}$ (digital number) and $F$ is the master flat. Pixels marked as bad were repaired row-by-row by interpolating across neighboring good pixels. 

Because WINERED's trace sits 5 degrees from horizontal and its iso-wavelength curves 48 degrees from vertical, the order containing the helium line was rectified in two steps: first the trace was flattened column by column, then each row was shifted so the iso-wavelength contours became vertical, using a column-dependent shift calibrated from manually measured emission-line slopes. The science, variance, and bad-pixel images were all rectified. Residual OH airglow was removed by subtracting each column's median. The spectrum was then extracted using an optimal extraction algorithm~\cite{horne_1986} over a 9-pixel window, using a fifth-order profile fit and 5$\sigma$ clipping. A wavelength solution is found per spectrum by fitting a \texttt{PHOENIX}~\cite{Husser_2013} stellar spectrum multiplied by an \texttt{ATRAN}~\cite{Lord_1992} telluric model (shifted into the stellar rest frame) to the data via differential evolution, parameterizing wavelength with a cubic Chebyshev polynomial and the continuum with a quadratic one.

Telluric absorption lines were removed with \texttt{molecfit}~\cite{Smette_2015}, fitting the water column density and line-spread-function width over narrow telluric-dominated windows, though the helium line fortunately never overlaps strong tellurics here. All spectra were then interpolated onto a common log-uniform grid (10,810 to 10,850 \AA), with interpolation covariances neglected. WINERED's time-variable fringing was suppressed with an infinite impulse response (IIR) notch filter tuned to the dominant ~0.091 pixel$^{-1}$ frequency identified from a Lomb-Scargle periodogram. Systematic A/B nod differences were divided out multiplicatively using median spectra at each nod position. Finally, the spectra were divided by their fitted continuum, converted into a residuals grid via negative-log and row/column median subtraction (so planetary absorption appears positive), and continuum-normalized by masking strong stellar lines and subtracting a cubic polynomial fit from each exposure.

\subsubsection*{Retrieving atmospheric properties} \label{sec:pwinds}
We used the \texttt{p-winds} code~\cite{Dos_Santos_2022} to estimate physical properties of the escaping upper atmosphere of LHS~1140b. The \texttt{p-winds} code models the metastable helium triplet in a planetary atmospheric outflow as a Parker wind~\cite{Parker_1958, Oklopcic_2018}. The code forward models an atmospheric transmission spectrum for the helium triplet for given values of the atmospheric escape rate, outflow temperature, H:He atomic number ratio, line-of-sight wind speed, top-of-atmosphere XUV spectrum, planetary mass and radius, and the instrumental line spread function which we modeled as a Gaussian with FWHM = 4.4 km $\mathrm{s}^{-1}$ (equivalent to R = 68,000). 

The \texttt{p-winds} model assumes a spherical, homogeneous outflow. Although such a one-dimensional model cannot fully capture the three-dimensional outflow, previous work has shown that it can retrieve escape parameters~\cite{Lampon_2020, Lampon_2021, Zhang_2023, Zhang_2025}. One- and three-dimensional models are typically consistent in cases of mild to moderate stellar wind interactions~\cite{MacLeod_2022}, a condition expected for temperate planets around inactive stars like LHS~1140b. When fitting our observed data with \texttt{p-winds}, we included the in-transit exposures corresponding to the white-light transit of planet b only (fig. \ref{fig3}b). The outflow probed in these exposures is assumed to be dominated by one-dimensional effects originating from the slow-moving thermosphere. We constructed two different stellar SEDs to compare in our analysis as described in the next section. 

We used the \texttt{p-winds} models to constrain the mass-loss rate, outflow temperature, and H:He ratio. We performed a grid search over a wide range of values for these three free parameters and evaluated the goodness-of-fit of each model to our data with a chi-squared test (fig. \ref{fig5}). For both SEDs, we found that only models with low H:He ratios fit the data. This is because, given the low XUV flux at LHS~1140b's orbit, if H:He $\gtrsim 1 \%$, XUV radiation that ionizes ground-state helium and populates metastable helium through recombination is attenuated by hydrogen absorption, preventing the production of sufficient metastable helium to explain the observed absorption depth at 10,833 \AA. We confirmed that the model was numerically stable for low H:He ratios by evaluating calculated reaction rates and number densities for a range of small H:He values (H:He $\geq 1 \times 10^{-8}$ for the GJ~1132 SED and H:He $\geq 1 \times 10^{-3}$ for the GJ~699 SED). We tested the assumption in the \texttt{p-winds} code that the electrons from helium ionization do not contribute to the atmospheric mean molecular weight by fixing the mean molecular weight at 4 amu, the theoretical maximum for a pure, neutral helium wind. The transmission spectra were unchanged by this assumption.

After verifying that the model was stable for low H:He ratios, we performed an MCMC analysis using the \texttt{emcee} package~\cite{Foreman-Mackey_2013} to estimate atmospheric escape rate, outflow temperature, H:He ratio, and line-of-sight bulk wind velocity. Fig. \ref{fig5} shows the results and the full MCMC posterior probability distributions are shown in figs. \ref{fig:pwinds_corner} and \ref{fig:pwinds_corner_699}. We performed separate runs for the two SEDs with the stellar flux increased by a factor of ten to test sensitivity to the XUV flux. The resulting posterior probability distributions were consistent with a helium-dominated outflow (fig. \ref{fig5}).

\subsubsection*{X-ray observations and analysis}
\label{sec:xray}
We examined archival observations of LHS\,1140 with XMM-Newton~\cite{jansen2001} taken on 2018 December 21 (Obs ID: 0822600101). The observation was a single visit with a duration of 64\,ks. LHS 1140 is contaminated by a brighter nearby source (Gaia DR3 2371032989200665984). The blended sources are detected in all three cameras of the European Photon Imaging Camera (EPIC), including the pn camera \cite{struder2001}, and both MOS (Metal Oxide Semi-conductor) cameras MOS\,1 and MOS\,2 \cite{turner2001}. We reduced the EPIC data using the Science Analysis Software [\texttt{SAS}; v21.0.0 \cite{gabriel2004-sas}]. We filtered the event files using the routine \texttt{evselect} with default pattern selection for the pn and MOS data. We defined good time intervals (GTI) using the routine \texttt{tabgtigen}, with threshold rates of 0.5, 0.15, and 0.2 counts\,s$^{-1}$ for pn, MOS\,1 and MOS\,2, respectively, determined from visual inspection of the total time-dependent count rate. This procedure resulted in a total effective exposure time of 31.3, 47.5, and 48.9\,ks for pn, MOS\,1 and MOS\,2, respectively. We extracted spectra for the source, the contaminant and a nearby background region using \texttt{evselect}. We adopt two approaches to constrain the X-ray flux of LHS\,1140. Firstly, we determined an upper limit on the flux by adopting an elliptical source aperture containing both LHS\,1140, and the contaminant. The ellipse was chosen to encompass two 20\,arcsec circular apertures, one centered on the position of LHS\,1140 \cite{brown2023-gaiaDR3}, the other centered on the Gaia position of the contaminant. This procedure provides an upper limit on the X-ray flux, but includes a contribution from the contaminant. In the second approach, we performed point spread function (PSF) photometry on the pn, MOS\,1 and MOS\,2 events files, to estimate the source aperture sizes for LHS\,1140 and the contaminant that minimize the contribution in each aperture from the blended source. We found optimal source apertures have radii 6.5 and 10\,arcsec, for LHS\,1140 and the contaminant respectively. We estimate that the contaminant still contributes $\sim$10\% of the flux in the 6.5-arcsec source aperture of LHS\,1140. 

Using the extracted source spectra, we performed a spectral analysis using the Bayesian X-ray Analysis (\texttt{BXA}) package~\cite{buchner2016-bxa} which integrates the nested sampling algorithm \texttt{ULTRANEST}~\cite{buchner2019-UltraNest} with an X-ray spectral fitting platform. For the latter, we utilize the \texttt{PyXspec} implementation of the X-ray spectral fitting package (\texttt{XSPEC}) \cite{arnaud1996XSPEC}. We simultaneously fitted all EPIC data using one spectrum for the pn data and one spectrum containing both MOS datasets, which were merged using the \texttt{SAS} routine \texttt{epicspeccombine}. Prior to fitting, the data were binned using a previously described algorithm \cite{kaastra2016-optimal-binning}. Due to the low number of source counts ($\sim$25 in pn and MOS), the spectra were fitted using \textit{C}-statistics~\cite{cash1979}. For the abundance profile of the emitting region, we assumed solar abundances~\cite{aslpund2009}.
Our spectral analysis used the Astrophysical Plasma Emission Code [\texttt{APEC}~\cite{smith2001-apec}], suitable for isothermal, optically-thin plasmas. We combine two \texttt{APEC} components additively to produce a 2-temperature, optically thin plasma model. 
At the distance to LHS\,1140, the interstellar neutral hydrogen column density ($N_{\rm H}$) is expected~\cite{redfield2000} to be $N_{\rm H}$\,=\,$4.6 \times 10^{18}\,\mathrm{cm}^{-2}$, which has a negligible effect in the XMM EPIC-pn and EPIC-MOS passbands. 
Nonetheless, 
we include an absorption component to account for any absorption from the Galactic neutral hydrogen column density, which we allow to vary from $N_{\rm H}$\,=\,0 to \,$4.6 \times 10^{18}\,\rm{cm}^{-2}$. 
We fitted five free parameters -- $N_{\rm H}$, two \texttt{APEC} plasma temperatures (times the Boltmann constant, $k$) $kT$, and two normalizations $N$. For the latter two, we fit the log normalization.
We adopt uniform priors on all parameters.
Using the spectrum of LHS\,1140 from the 6.5\,arcsec aperture yields plasma temperatures $kT$\,=\,$0.22^{+0.019}_{-0.017}$ and $3.7^{+1.6}_{-1.8}$\,keV, with associated normalizations $N\,=\,1.93_{-0.28}^{+0.27}\times10^{-6}$ and $5.78_{-4.1}^{+17.0}\times10^{-8}$, respectively, and Galactic neutral hydrogen column density of $N_{\rm H}=7.4_{-4.9}^{+5.1}\times10^{-8}\,\rm{cm}^{-2}$.
A previous analysis of the same data also measured X-ray flux and temperature \cite{spinelli2023}. The two temperatures we infer from the EPIC data are consistent with the previous analysis. We recover an X-ray flux ($F_{\rm X}$) in the 0.25 to 10.0\,keV band of $F_{\rm X}$\,=\,$3.15^{+0.50}_{-0.42} \times 10^{-15}$\,erg\,s$^{-1}$\,cm$^{-2}$. For comparison, the previous work found $F_{\rm X}$\,=\,$5.0^{+0.7}_{-0.8} \times 10^{-15}$\,erg\,s$^{-1}$\,cm$^{-2}$ using a 10 arcsec aperture and 0.3 to 10 keV band~\cite{spinelli2023}. Our derived flux is slightly lower, but the difference is not statistically significant  ($<$ 2$\sigma$). If we had used a source aperture radius of 10\,arcsec, the contamination would have been $\sim25$\%, compared to the $\sim$10\% estimated for the 6.5-arcsec aperture. 

An X-ray flux upper limit for LHS\,1140 has previously been reported using independent X-ray observations \cite{spinelli2019}. They set an upper limit of $F_{\rm X}\lesssim 1.29 \times 10^{-14}$\,erg\,s$^{-1}$\,cm$^{-2}$, in the 0.3 to 10.0\,keV band, fully consistent with our results. 
%
We present the measured X-ray flux and luminosity of LHS~1140 in Table\,\ref{table2}. The best-fitting spectral models and parameter posterior probability distributions for LHS~1140 and the contaminant are shown in figs.\,\ref{fig:bxa1} and \ref{fig:bxa2}.

We used the best-fitting X-ray flux in the 0.25 to 2.0\,keV band (Table \ref{table2}) and the distance to LHS~1140 of $D$\,=\,14.97$\,\pm\,$0.01\,pc \cite{brown2023-gaiaDR3} to derive the X-ray luminosity  
$L_{\rm X}$\,=\,$7.8^{+1.1}_{-1.3}\,\times 10^{25}\, \mathrm{erg\,s^{-1}}$. 
Using the bolometric luminosity $L_{\rm bol}=0.0038\pm0.0003$~times the Solar luminosity~\cite{Cadieux_2024a}, we estimate 
$L_{\rm X}/L_{\rm bol}$\,=\,$6.49^{+1.34}_{-1.25}\times10^{-6}$, consistent with the expectation for an inactive mid-M dwarf~\cite{Wright_2018}. Using the semi-major axes of LHS~1140b and LHS~1140c, we estimate that the planets receive X-ray fluxes in the 0.25 to 2.0 keV band of $2.94^{+0.61}_{-0.57}$ and $36.1^{+7.5}_{-7.0}$\,$\mathrm{erg\,s^{-1}\,cm^{-2}}$, respectively. The X-ray luminosity of the Sun in the same bandpass varies between $6.3\times10^{26}$ to $2.5\times10^{27}\,\mathrm{erg\,s^{-1}}$ \cite{judge2003} over its activity cycle. Thus LHS~1140b and LHS~1140c are receiving $3.3^{+0.7}_{-0.6}$ to $13.1^{+2.7}_{-2.5}$ and $40.4^{+8.3}_{-7.8}$ to $161^{+33}_{-31}$ times the solar X-ray flux at Earth, respectively. 

We estimate the XUV SED of LHS~1140 by comparison to two M\,dwarf stars -- GJ~699 and GJ~1132 -- that are of similar spectral type (M4 and M3.5, respectively) and rotation period (145 and 125 days, respectively) to LHS~1140~\cite{Bonfils_2018, Toledo_2019}. Both stars have measured X-ray fluxes and panchromatic SEDs from X-ray to infrared wavelengths from the Mega-Measurements of the Ultraviolet Spectral Characteristics of Low-mass Exoplanetary Systems (Mega-MUSCLES) Treasury Survey \cite{wilson2025-Mega-MUSCLES}, including a reconstruction of the unobservable EUV region~\cite{wilson2021-Mega-MUSCLES}.
 GJ~699 has an X-ray flux in the 0.3 to 10.0\,keV band of $F_{\rm X,699} = (4.83 \pm 0.52) \times 10^{-14}\,\mathrm{erg\,s^{-1}\,cm^{-2}}$ \cite{brown2023}. 
The X-ray flux of GJ~1132 in the 0.2 to 2.4\,keV band is $F_{\rm X,1132} = (5.4 \pm 1.4) \times 10^{-15}\,\mathrm{erg\,s^{-1}\,cm^{-2}}$ \cite{cilley2024}. We integrate our best-fitting X-ray spectral model for LHS~1140 across the same bands, finding integrated fluxes of 
 $F_{\rm X} (0.3\,\rm{to}\,10.0\,\mathrm{keV}) = (2.7\pm0.5) \times 10^{-15}\,\mathrm{erg\,s^{-1}\,cm^{-2}}$ and  
 $F_{\rm X} (0.2\,\rm{to}\,2.4\,\mathrm{keV}) = (3.2\pm0.6) \times 10^{-15}\,\mathrm{erg\,s^{-1}\,cm^{-2}}$. We normalize the panchromatic SEDs of GJ~699 and GJ~1132 so their integrated X-ray flux equals that measured for LHS~1140. The multiplicative normalization factor ($A$) for GJ~699 ($A_{699}$) and GJ~1132 ($A_{1132}$) are $A_{699}=0.056$ and $A_{1132}=0.59$, respectively. The predicted stellar flux for LHS~1140 is $A_{699}F_{699}(\lambda)$ and $A_{1132}F_{1132}(\lambda)$. This procedure provides an estimate of the panchromatic SED of LHS~1140, constrained by the measured X-ray flux.


\subsection*{Supplementary Text}




\subsubsection*{Exclusion of spurious signals}
\label{sec:spurious}

We investigated whether the helium signal observed in 2024 could have resulted from the stellar photosphere rather than atmospheric material from the planet. {While the stellar metastable helium line typically does not vary for inactive mid-M dwarfs \cite{Fuhrmeister_2020},} stellar flares can have a range of time-varying effects on the helium line, from enhanced absorption to decreased absorption relative to baseline~\cite{Andretta_2008, Fuhrmeister_2020}, and emission in some cases~\cite{Judge_2015}. It is therefore conceivable that a planetary transit during, or temporally near, a stellar flare could cause excess metastable helium absorption. Model predictions and observations of stellar flares
demonstrate that flares have time-varying morphologies in spectra marked by sharp flux increases followed by exponential decays~\cite{Fuhrmeister_2019, Vissapragada_2021, Wang_2021, Fuhrmeister_2023}, which are not consistent with the smooth boundaries seen for LHS~1140b during pre-ingress and post-egress. LHS~1140 flares infrequently at a rate of one major flare 
every 5,770 days~\cite{Medina_2022}. 
Hence it is unlikely that LHS~1140 produced a flare during the six and a half hours of observations.

If the stellar helium absorption line varied across the stellar surface, and if the planet transited regions of the star where this absorption was weaker, this could produce an apparent increase in helium absorption during transit [the transit light source, TLS, effect~\cite{Rackham_2018}]. For the TLS effect to produce our observed excess absorption depth of 1.24\%, the transit chord would need to be approximately $3.3\times$ brighter at 10,833 \AA\ relative to the average flux at this wavelength across the stellar disk, given the white-light transit depth of 0.55\%~\cite{Cadieux_2024b}. This would require intense, localized helium emission along the transit chord or a perfectly helium-absorbing, unocculted area covering 70\% of the stellar surface. We consider these scenarios implausible given that metastable helium is observed in absorption for mid-M stars in quiescence~\cite{Fuhrmeister_2019}. The TLS effect is also unlikely because we observe helium absorption prior to the transit, when the planet is not blocking any part of the stellar disk.

Because we observed LHS~1140b and LHS~1140c transit in the same night, if the helium feature observed during planet b’s transit were due to the TLS effect, we would expect to see a similar effect during the transit of planet c scaled by the relative projected areas of the two planets \cite{Coulombe_2025}. Their transit chords have similar impact parameters, $0.23_{-0.07}^{+0.05}$ and $0.09_{-0.06}^{+0.09}$~\cite{Cadieux_2024a}. If the entire signal from planet b's transit were due to the TLS effect, we would expect to see a TLS signal strength for planet c of $\delta_c = (R_c^2/R_b^2) \delta_b = 0.67 \%$, where $\delta$ and $R$ are excess absorption depths and planetary radii and the subscripts b and c indicate LHS~1140b and LHS~1140c respectively. We did not observe such a signal (fig.~\ref{fig1}). We constructed a transmission spectrum from the 2024 data for planet c following the same steps described for planet b, injected the planet c absorption signal, $\delta_c$, subtracted residual absorption due to the pre-ingress of planet b, and attempted to recover the signal by fitting the same Gaussian model as for planet b. The helium signal was recovered with 20$\sigma$ confidence (fig. \ref{fig:TLS}).

We also investigated whether the helium feature could arise from random variability in the stellar helium line during the transit of planet b. We fitted the three-Gaussian model (described above) to the transmission spectrum composed of the in-transit spectra (fig. \ref{fig3}B) and to the transmission spectrum composed of the pre-ingress spectra (fig. \ref{fig3}C), without using a GP. We then fitted the same three-Gaussian model to the out-of-transit stellar template spectrum, which has a stellar helium absorption line (fig. \ref{fig1}A). We used the same MCMC process as before to determine the posterior probability distributions of the peak amplitudes, Gaussian standard deviations (corresponding to a FWHM), and Doppler shifts. The FWHM of the stellar helium line (FWHM$_\mathrm{stellar}$) is 0.22176$^{+0.0027}_{-0.00094}$~\AA\, the FWHM of the in-transit feature (FWHM$_\mathrm{in-transit}$) is 0.323$^{+0.012}_{-0.011}$~\AA\, and the FWHM of the pre-ingress feature (FWHM$_\mathrm{pre-ingress}$) is 0.392$^{+0.018}_{-0.017}$~\AA\ (fig. \ref{fig:He_line_comparison}). The median FWHM values for the planetary and stellar signals are $\sim 9 \sigma$ apart. We therefore consider it unlikely that the change in FWHM that occurs during transit is due to random variation of a stable out-of-transit baseline, so exclude stellar variability as an explanation for the observed absorption signal. This is consistent with our expectations, as the stellar metastable helium line is typically stable for inactive mid-M dwarfs~\cite{Fuhrmeister_2020} like LHS~1140.

We also investigated whether the helium signal could arise from telluric contamination. We examined all sky spectra--exposures taken from slit position A when the target was at position B, and vice versa. We identified nearby OH emission lines~\cite{Oliva_2015, Czesla_2022} in the sky spectra just blueward of the helium triplet at 10,832.6 and 10,832.9 \AA (vacuum wavelength) in the stellar rest frame. The peaks of the helium triplet appear between the OH lines in the sky spectra, and the primary helium line peaks do not overlap with any telluric line peaks (fig. \ref{fig:tellurics}). 

All sky spectra were examined to search for time-correlated variability of telluric features near the helium triplet.
We found helium airglow in the last 12 exposures of the sky spectra, just before twilight. The OH lines correlate with each other in time while the helium emission line only appears in the last 12 exposures. The helium triplet does not overlap substantially with the telluric helium feature, due to the radial velocity shift of the LHS~1140 system. Of the 12 exposures containing the helium emission line, 5 are in the post-egress exposure group, and 7 are in the out-of-transit stellar template spectrum exposures used to compare against the in-transit spectra when creating the planetary transmission spectrum. To test whether the telluric helium feature impacts the planetary signal, we removed these 7 exposures from the out-of-transit template, and constructed a transmission spectrum for the in-transit spectra, thereby excluding all 12 exposures that show helium airglow. The planetary helium feature persisted in the 2D spectra and transmission spectrum with an unchanged absorption depth (1.24\%). The 2D sky spectra show that no telluric features, including the helium airglow, are temporally correlated with the transit of either planet (fig. \ref{fig:tellurics}).

We investigated whether the helium absorption line follows the radial velocity (RV) of LHS~1140b in the stellar rest frame, for exposures corresponding to the expected white light transit (fig. \ref{fig:planet_RV}). We fitted a straight line to the RV of the helium line center for the in-transit spectra and binned spectra by a factor of four to boost the SNR, which also provides a better fit (reduced chi squared of 0.78 compared to 11.76 without binning). We compared the linear fit to a flat line using a chi-squared test and found that they differ by $\chi^2$ = 6.3, p=0.012, equivalent to $2.5 \sigma$ significance. 
We then fitted the expected RV to the data and calculated chi-squared. We compared this chi-squared with the chi-squared from the best fitting linear model to see if one is favored.
The difference in chi-squared is 4.2 (p=0.046, roughly $2 \sigma$ significance). If we consider the offset a nuisance parameter, then compare only the slopes of the best-fitting linear model and expected planetary RV, they differ by $\Delta$BIC = 2.2, which is weak evidence that the linear model matches the data better than the expected planetary RV does. These tests do not provide statistically significant evidence that support conclusions on whether the helium signal follows the expected RV of the planet.

We also consider our prior expectations. In the 2024 dataset, the helium absorption signal gradually appears in the nine exposures leading up to the transit, remains stable for the duration of the transit, and gradually fades in the eight exposures post-transit. This behavior has been predicted by hydrodynamic models of metastable helium absorption~\cite{MacLeod_2022} and seen in previous observations of atmospheric escape~\cite{Czesla_2022, Gully-Santiago_2024, Krishnamurthy_2026}. The escape rate inferred from the observations matches the theoretical energy-limited escape rate within 1 or 3 $\sigma$, for the GJ~699 vs. GJ~1132-normalized SEDs respectively.

\subsubsection*{Non-detection for planet c}
\label{sec:SI_planetc}
\begin{figure}
\centering
\includegraphics[width=1.0\textwidth]{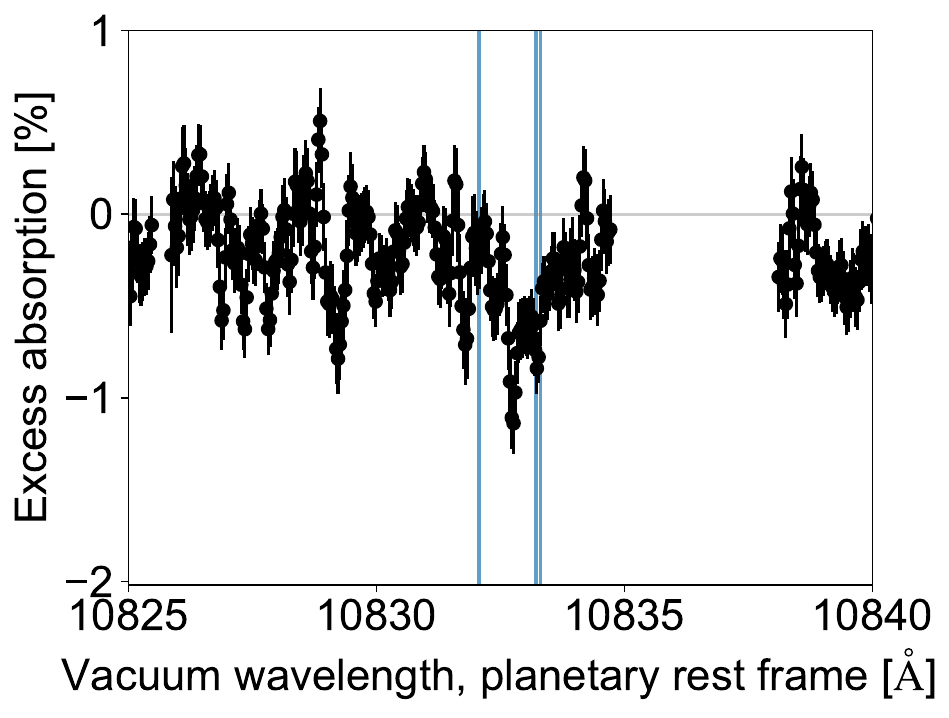}
\caption{\textbf{Same as fig. \ref{fig2} but for LHS~1140c in 2024.}}\label{fig:LHS1140c_2024}
\end{figure}

We constructed a transmission spectrum for LHS~1140c using the same method described for LHS~1140b and did not detect metastable helium absorption (fig.~\ref{fig:LHS1140c_2024}). The MCMC posterior probability distribution indicates an in-transit excess absorption depth of $0.59^{+0.37}_{-0.27}$\% for the two blended red peaks and a depth of $0.34^{+0.16}_{-0.13}\%$ for the single blue peak. We attribute the non-zero excess helium absorption during the transit of planet c to a pre-ingress tail of absorbing material originating from planet b, overlapping with the last three exposures of planet c's white-light transit (fig. \ref{fig1}B).

\subsubsection*{Variable helium escape from LHS~1140b}
\label{sec:SI_variable_escape}

\begin{figure}
\centering
\includegraphics[width=1.0\textwidth]{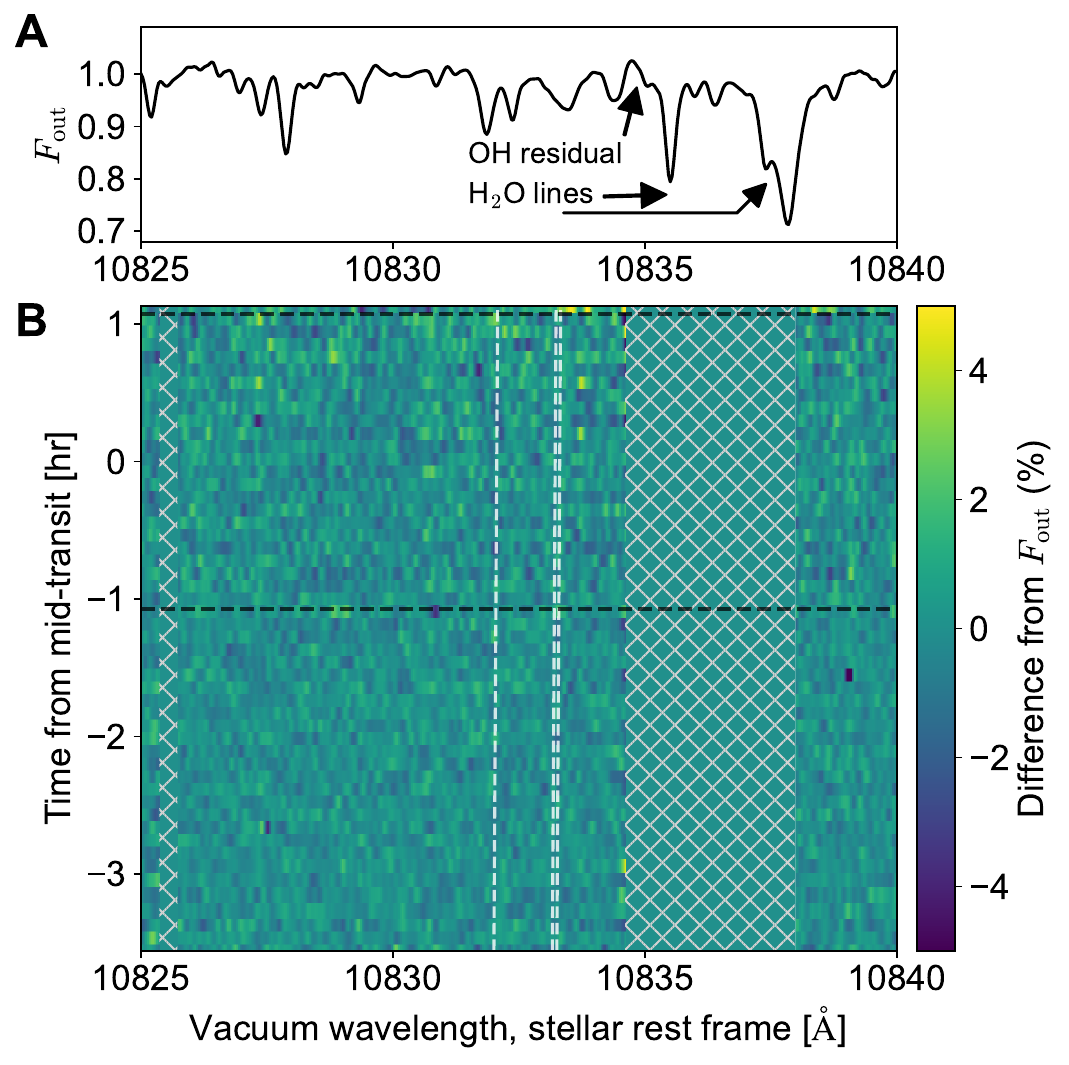}
\caption{\textbf{Same as fig. 1, but for the 2025 observations of LHS~1140b.}}\label{fig:timeseries_2025}
\end{figure}

\begin{figure}
\centering
\includegraphics[width=1.0\textwidth]{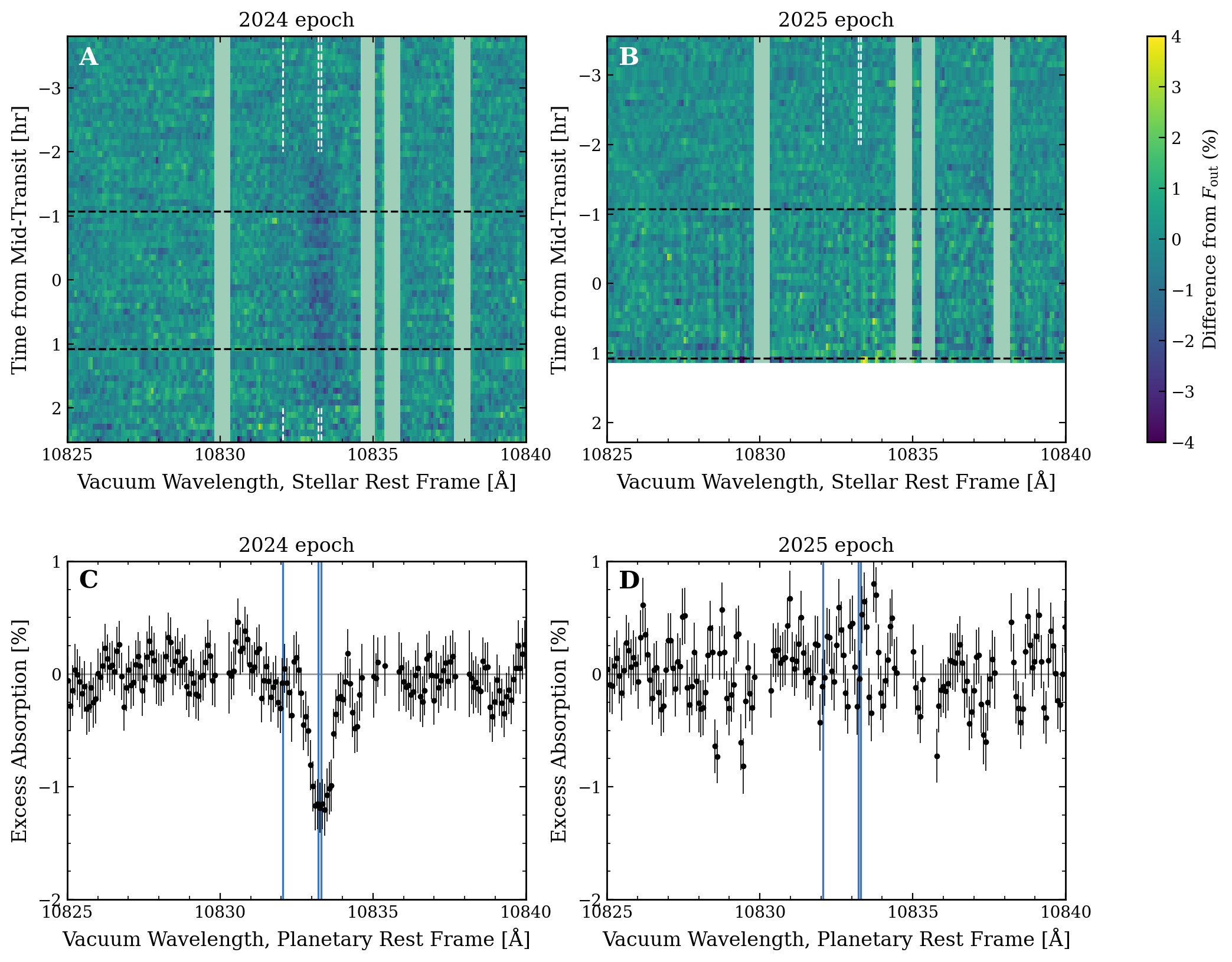}
\caption{\textbf{Same as figs. \ref{fig1} and \ref{fig2} but for the re-reduction for LHS~1140b.}}\label{fig:MZ_reduction}
\end{figure}

\begin{figure}
\centering
\includegraphics[width=1.0\textwidth]{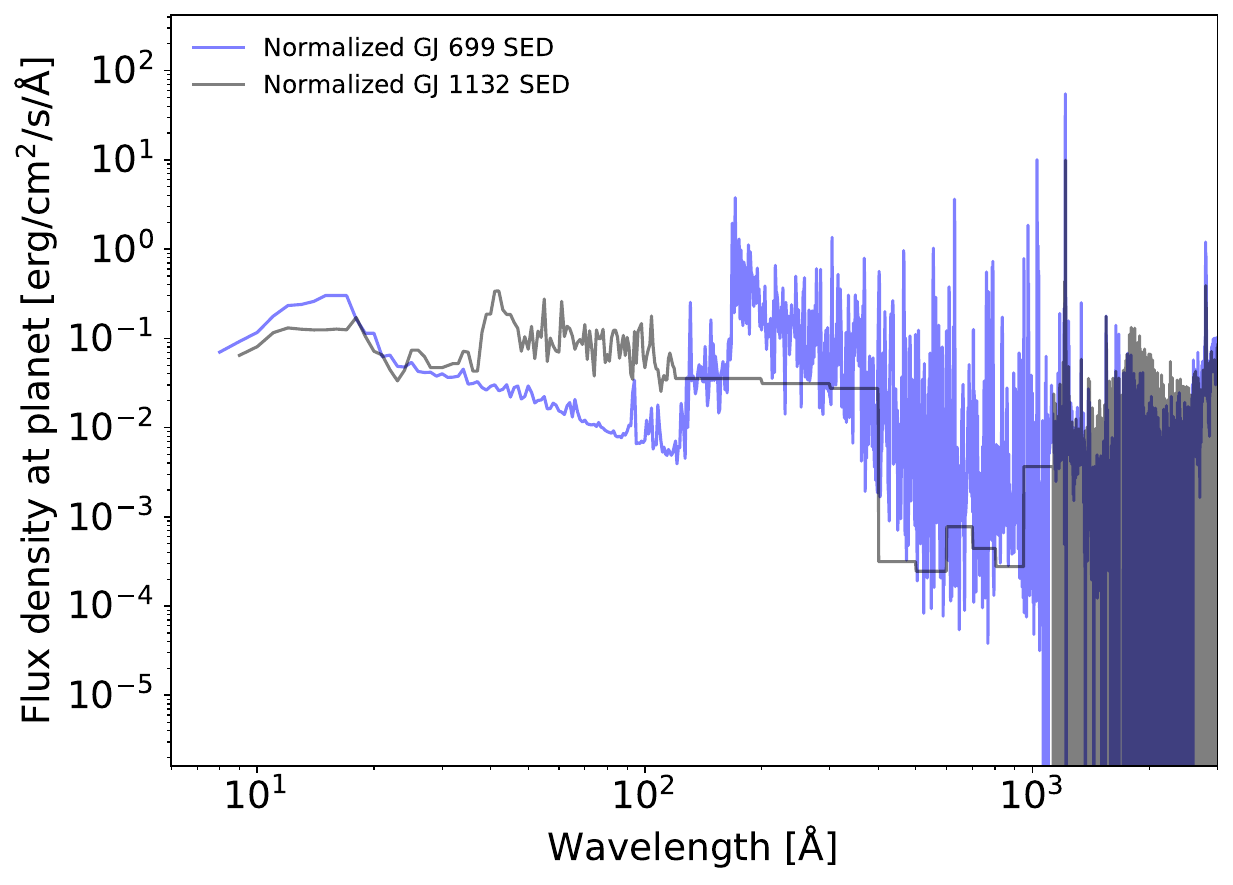}
\caption{\textbf{Spectral energy distributions used in our \texttt{p-winds} analysis.} The purple line shows the GJ~699 SED and the grey line shows the GJ~1132 SED.}\label{fig:SEDs}
\end{figure}

\begin{figure}
\centering
\includegraphics[width=1.0\textwidth]{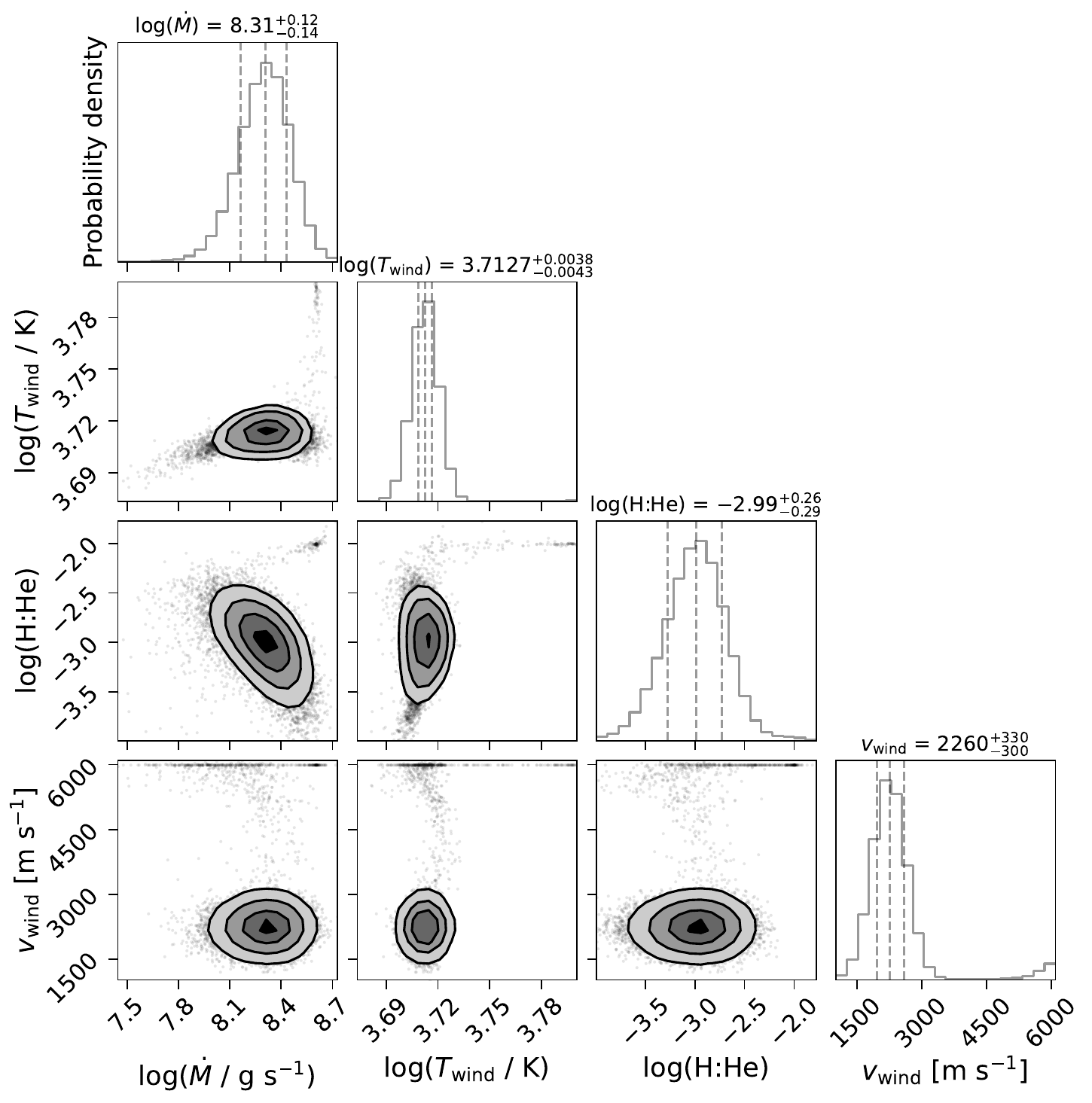}
\caption{\textbf{\texttt{p-winds} model fit using the GJ~1132-normalized SED.} Posterior probability distributions from the MCMC analysis of the \texttt{p-winds} model parameters fitted to the LHS~1140b transmission spectrum in 2024 are shown. Marginalized distributions for each parameter are plotted as solid lines in the diagonal histograms. Vertical grey dashed lines show the median and 1$\sigma$ values. Correlated distributions are shown in the off-diagonal plots where each point represents one sample in the MCMC chain, corresponding to one model. The grey contours show the 1, 2, and 3$\sigma$ boundaries from dark to light grey. $\dot{M}$ is the atmospheric escape rate.}\label{fig:pwinds_corner}
\end{figure}

\begin{figure}
\centering
\includegraphics[width=1.0\textwidth]{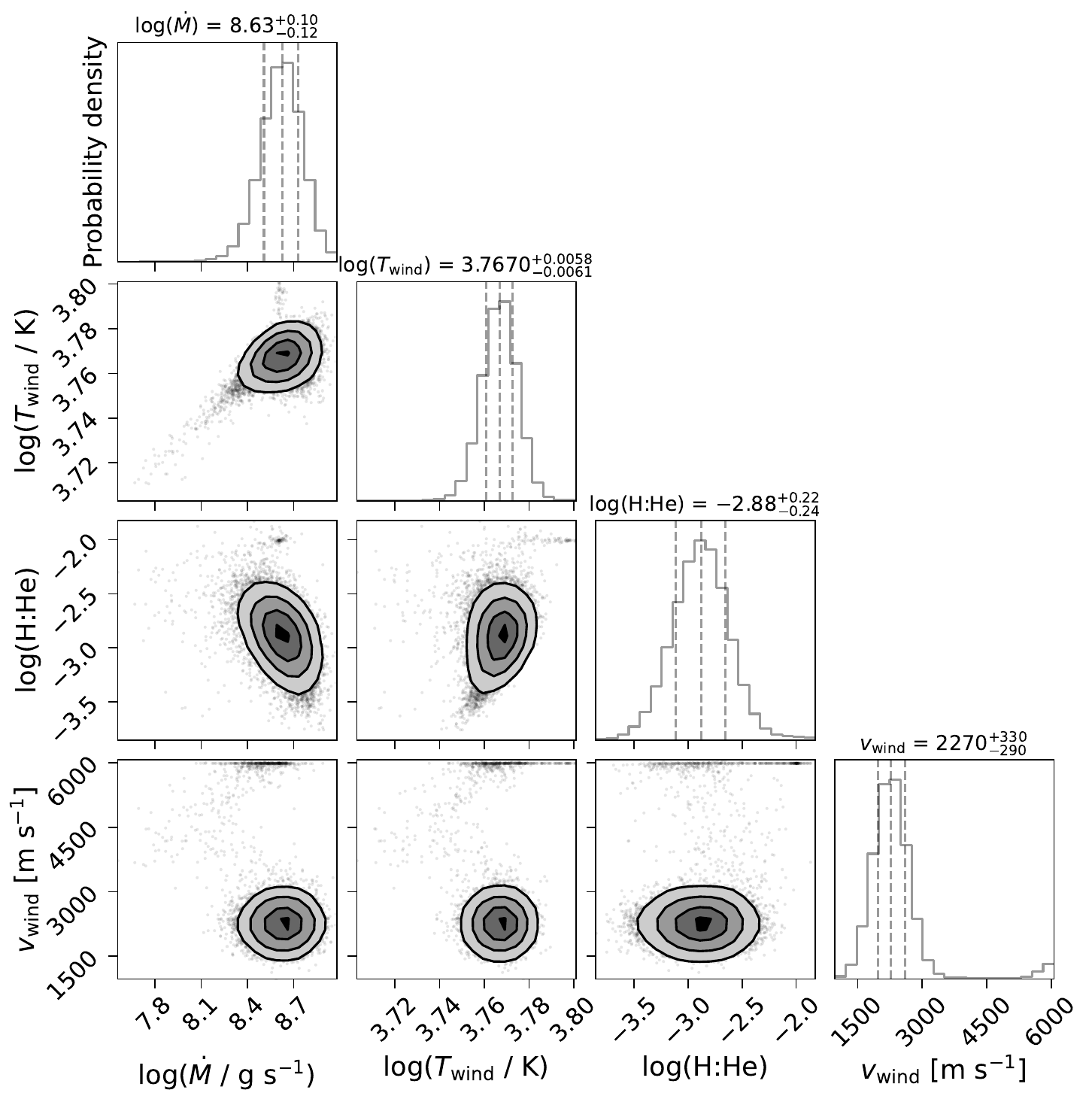}
\caption{\textbf{\texttt{p-winds} model fit using the GJ 699-normalized spectrum.} Same as fig. \ref{fig:pwinds_corner} but using the SED constructed from the GJ~699 spectrum.}\label{fig:pwinds_corner_699}
\end{figure}

\begin{figure}
\centering
\includegraphics[width=1.0\textwidth]{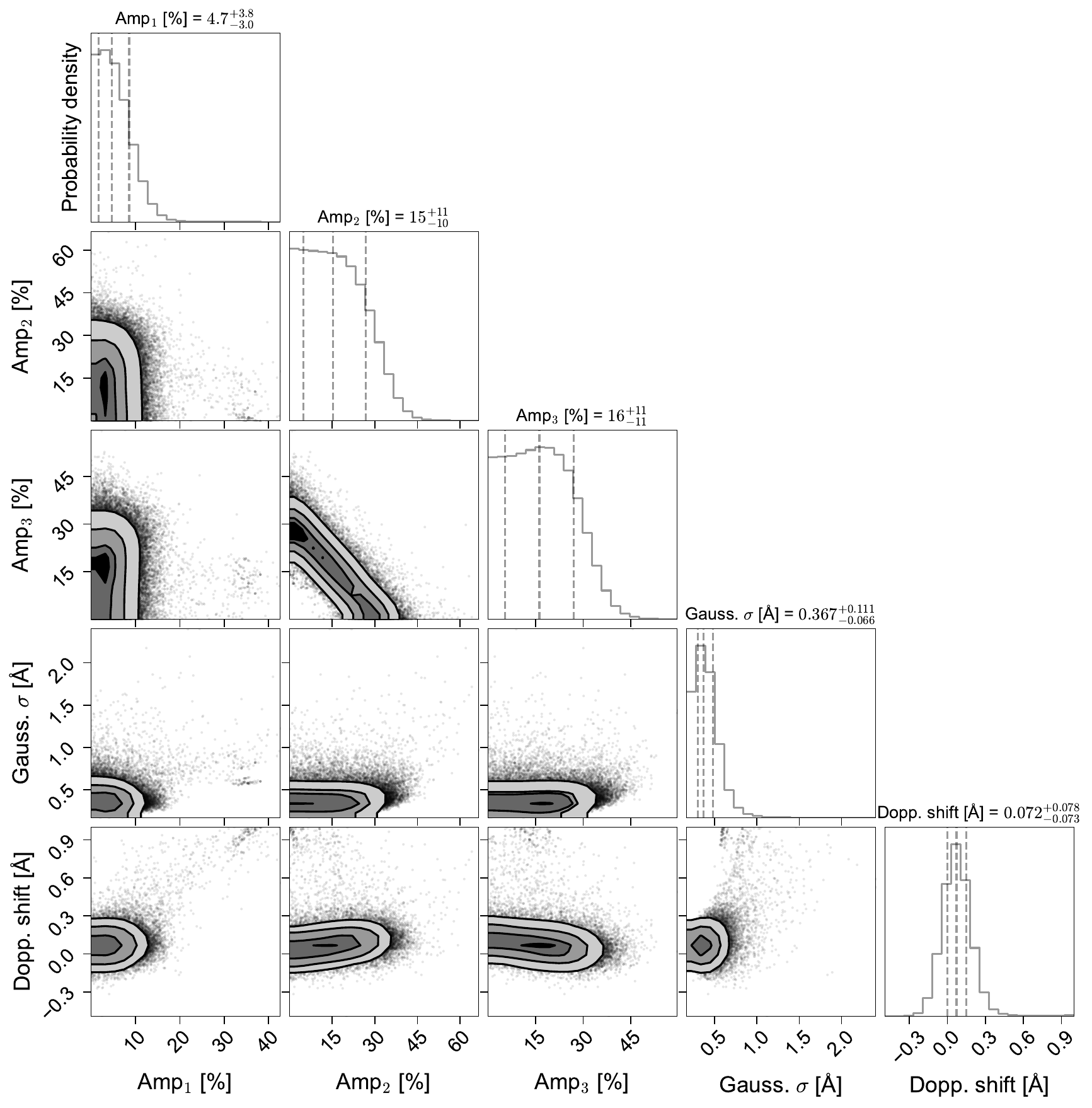}
\caption{\textbf{LHS~1140b transmission spectrum model in 2024.} Same as fig. \ref{fig:pwinds_corner} but for posterior probability distributions from the MCMC analysis of the Gaussian model parameters fitted to the 2024 LHS~1140b transmission spectrum. Amp$_1$, Amp$_2$, and Amp$_3$ are the peak amplitudes for the three helium triplet lines. }\label{fig:gauss_corner}
\end{figure}

\begin{figure}
\centering
\includegraphics[width=1.0\textwidth]{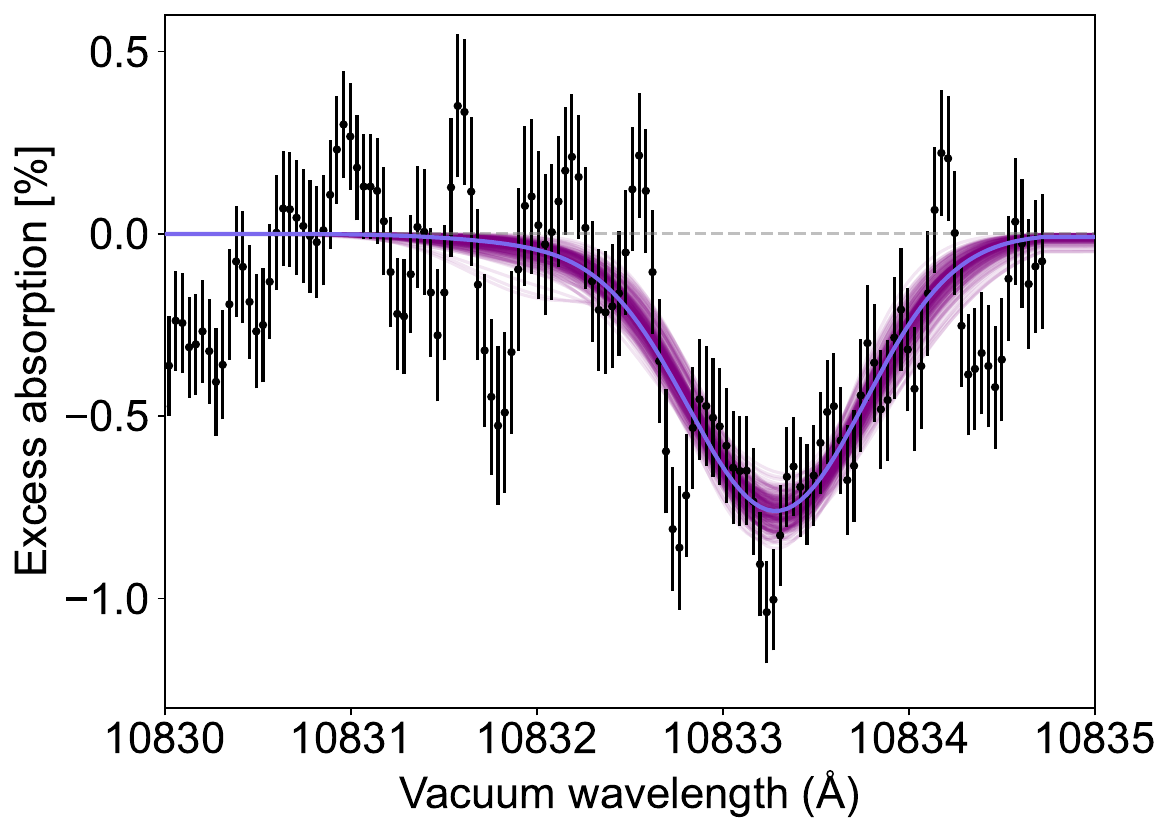}
\caption{\textbf{LHS~1140c TLS injection-recovery test.} Same as fig. \ref{fig4}B-D but for the TLS signal injected in the LHS~1140c transmission spectrum in 2024.
}\label{fig:TLS}
\end{figure}

\begin{figure}
\centering
\includegraphics[width=1.0\textwidth]{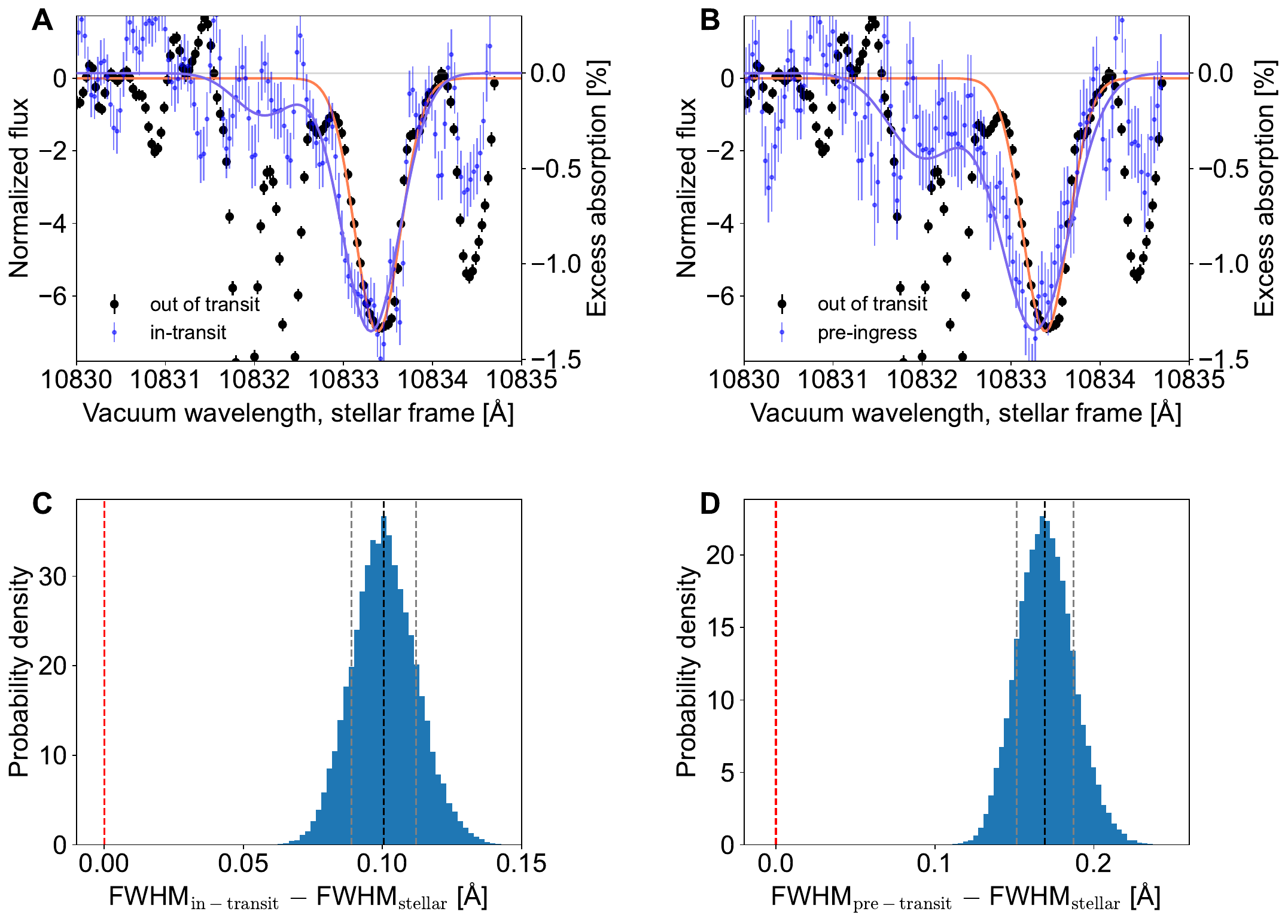}
\caption{\textbf{Comparison of the stellar and planetary metastable helium lines.} (A) The black points are the mean out-of-transit stellar template spectrum, shown in continuum-normalized flux units [left axis \cite{methods}]. The orange line is the best fitting three-Gaussian model to those points. The purple points are the in-transit transmission spectrum with 1~$\sigma$ error bars, plotted as excess absorption (right axis). The purple line is the best fitting three-Gaussian model to those points. The horizontal gray line shows zero absorption. (B) Same as panel A but for the pre-ingress spectrum. (C) Posterior probability distribution of the difference in FWHM for the planetary (FWHM$_\mathrm{in-transit}$) and stellar (FWHM$_\mathrm{stellar}$) helium lines shown in panel A, obtained from the MCMC analysis. The median and standard deviation are shown as vertical dashed black and gray lines, respectively. The dashed red vertical line shows zero difference in FWHM. (D) Same as panel C, but for the pre-ingress spectrum. The median offset from zero is 8.7 $\sigma$ and 9.5 $\sigma$ for the in-transit and pre-ingress spectra, respectively, indicating that the planetary helium line is broader than the stellar helium line.
}\label{fig:He_line_comparison}
\end{figure}

\begin{figure}
\centering
\includegraphics[width=1.0\textwidth]{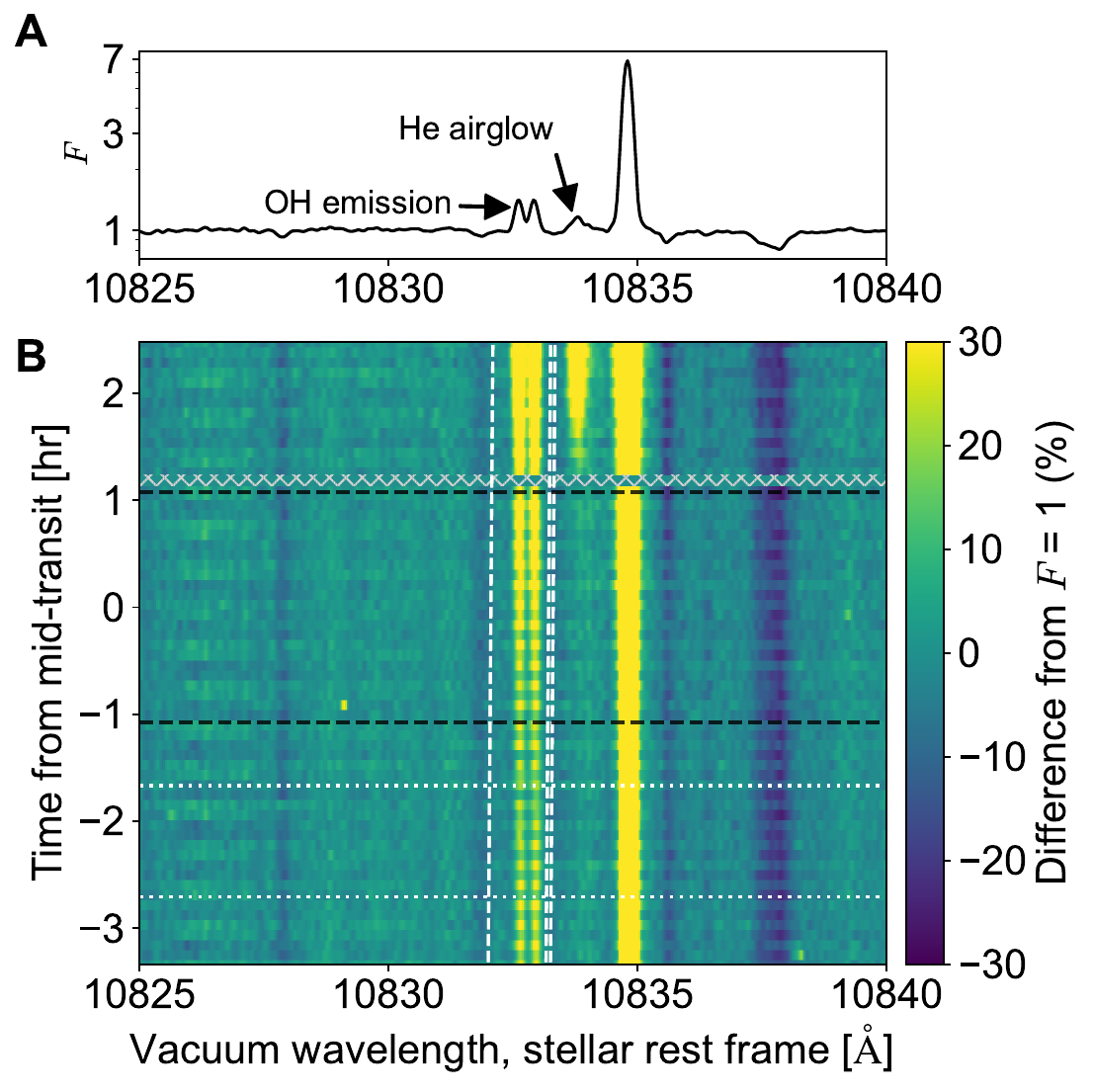}
\caption{\textbf{Sky spectra in the 2024 observation.} (A) Mean of all sky spectra. $F$ is the continuum-normalized flux~\cite{methods}. Arrows indicate OH and He emission features from Earth's atmosphere. (B) Colors indicate the percentage difference from $F = 1$. Other symbols are the same as in fig. \ref{fig1}B. The helium absorption peaks do not overlap with telluric features. Helium airglow is observed in the last 12 exposures.
}\label{fig:tellurics}
\end{figure}

\begin{figure}
\centering
\includegraphics[width=1.0\textwidth]{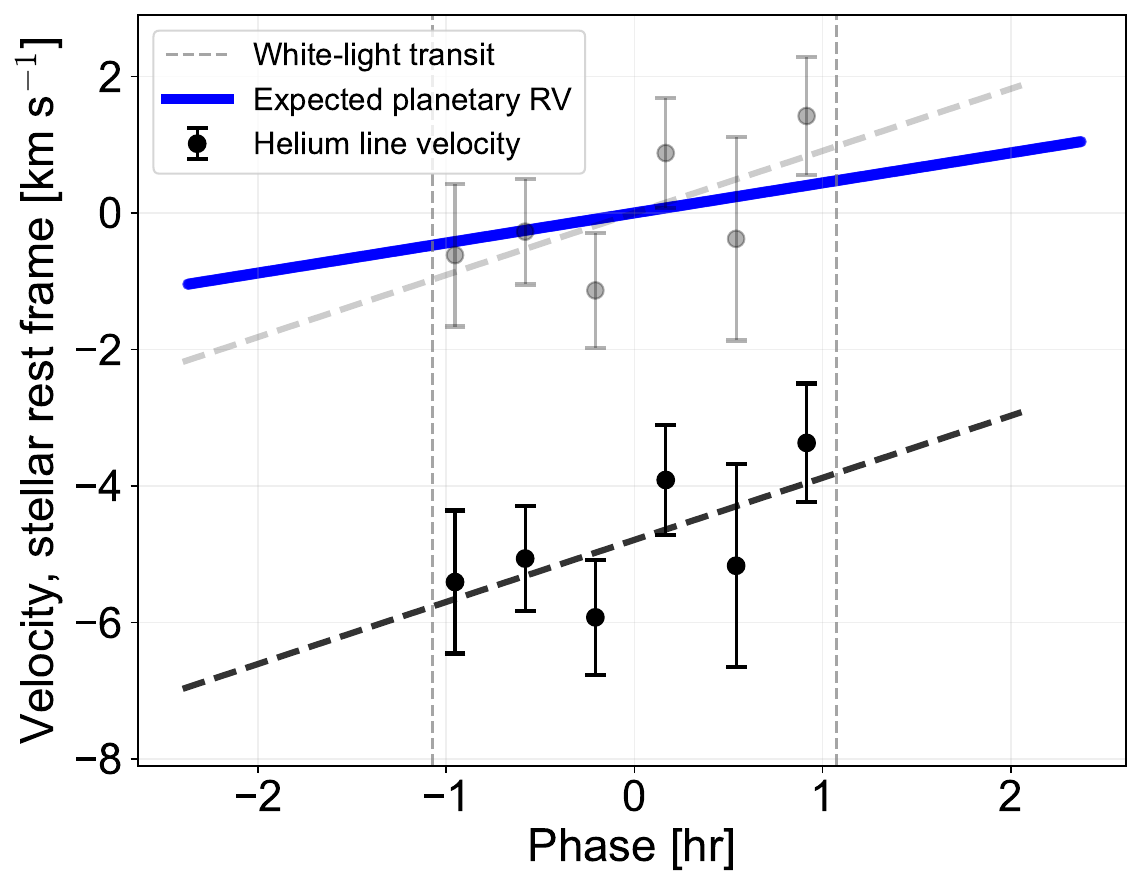}
\caption{\textbf{Time-dependent radial velocity shift of the in-transit helium absorption feature observed in 2024.} The blue line shows the expected RV of the planet due to its orbital motion, with zero phase corresponding to mid-transit. The black data points show the measured velocity from the in-transit data, binned in groups of four, with $1 \sigma$ error bars. Vertical dashed gray lines show the in-transit phase. The dashed black line shows the best fitting least squares regression line to the data.
}\label{fig:planet_RV}
\end{figure}

\begin{figure}
\centering
\includegraphics[width=\textwidth]{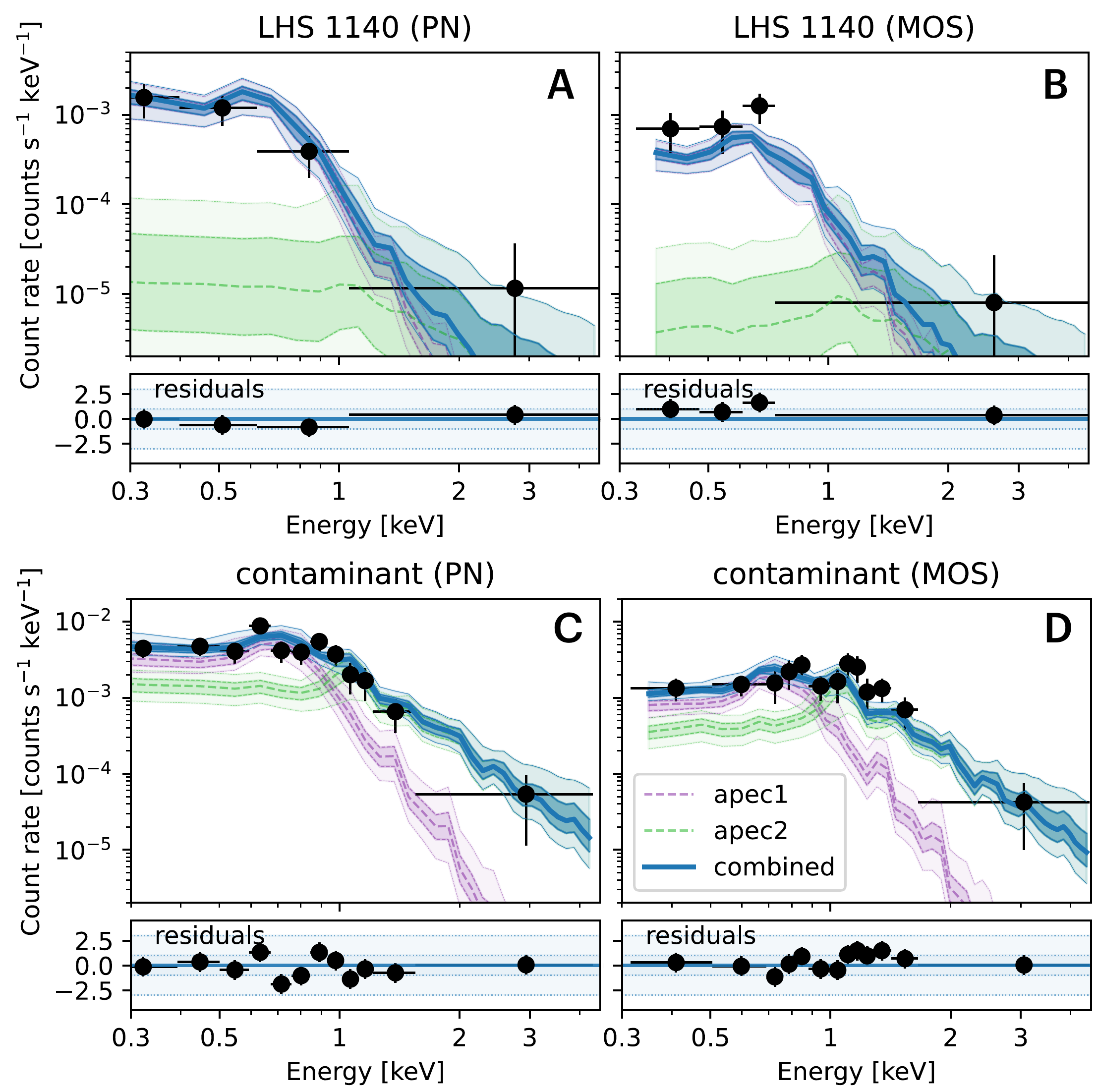}
\caption{\textbf{X-ray spectral analysis.} (A) X-ray spectrum of LHS\,1140 from the EPIC-pn instrument of XMM-Newton. The black points show the data, binned such that each bin has a significance of 1$\sigma$. The best-fitting two-temperature model (labelled ``combined'') is shown in blue, and the individual temperature components (labeled ``apec1'' and ``apec2'') are shown in magenta and green. The shaded regions indicate the 1 and 2$\sigma$ uncertainty on the best-fitting model and model components. The lower panel shows the residual between the best-fitting model and the data.
(B) Same as panel A, but for the EPIC-MOS data. (C-D) Same as panels A and B, but for the contaminating source.}\label{fig:bxa1}
\end{figure}

\begin{figure}
\centering
\includegraphics[width=\textwidth]{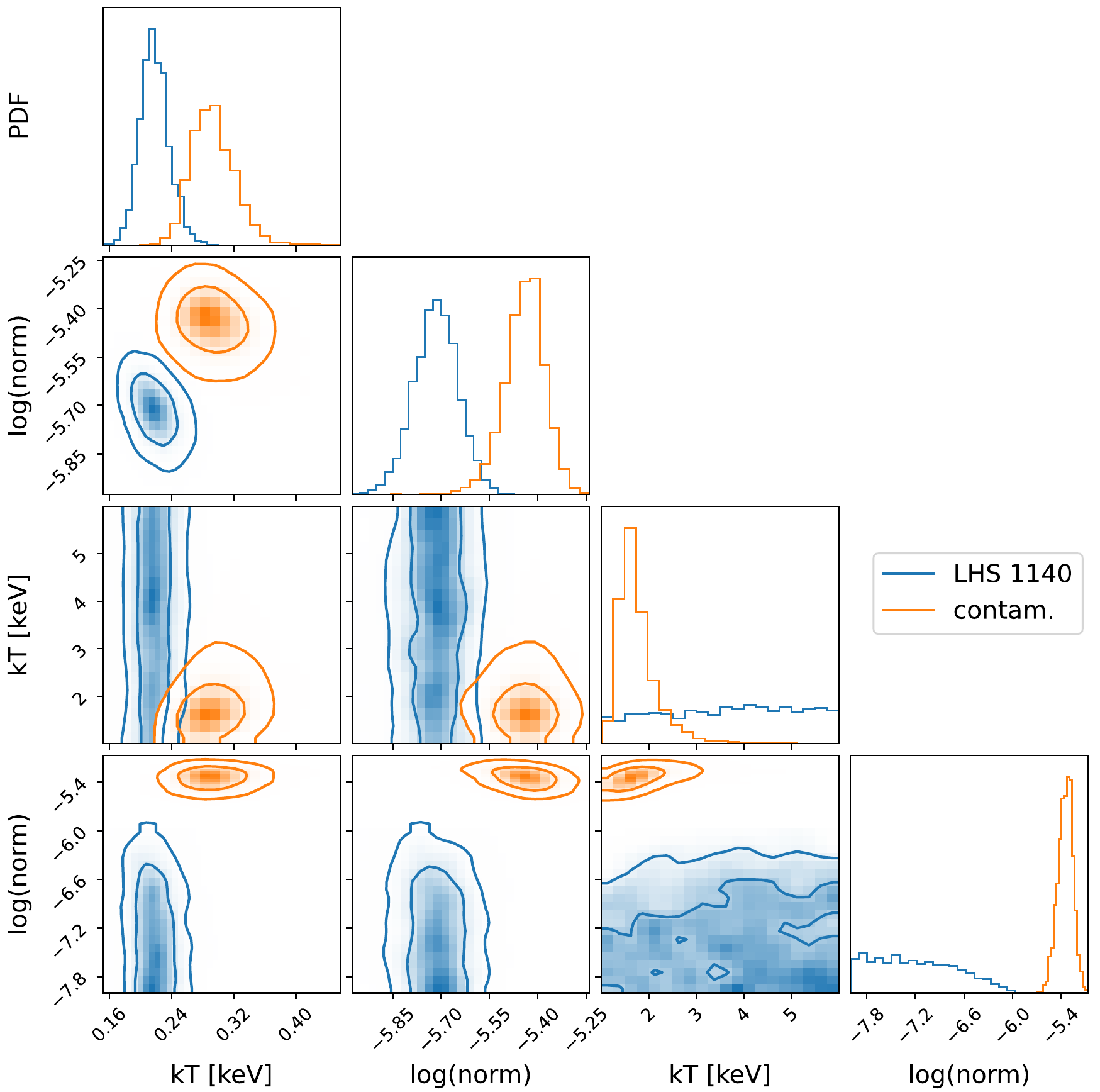}
\caption{\textbf{X-ray spectral analysis probability distributions.} Similar to Fig\,\ref{fig:gauss_corner}, but showing the posterior probability distributions of the 5 fitted parameters in the X-ray spectral modeling; absorption, 2 temperatures and 2 normalizations, one for each of the \texttt{APEC} components. The results of the X-ray spectral fitting for LHS\,1140 are shown in blue, and those for the contaminant are shown in orange.}\label{fig:bxa2}
\end{figure}

We reduced the 2025 data in the same way as the 2024 data but did not detect any helium absorption (fig. \ref{fig:timeseries_2025}). 
To determine the detection limit of the 2025 observation, we subtracted the best-fitting helium triplet Gaussian model and GP from our 2024 transmission spectrum (absorption depth 1.24\%) and injected the Gaussian model into the 2024 data with varying peak amplitudes. We did not use the 2025 dataset because if there was helium absorption in the 2025 dataset below the detection limit, any injected signal would be added to it. We attempted to recover each signal with the same MCMC process and found that absorption depths $\leq$0.6\% would not have been detectable (fig. \ref{fig4}). We consider this a conservative upper limit on the 2025 transmission spectrum, which has a lower signal-to-noise ratio (root mean square 0.34\% in 2024 vs. 0.42\% in 2025). 
X-ray flux variability of $\lesssim $5$\times$ has been observed on timescales ranging from several days to $\sim$6 years in other systems, including M dwarfs \cite{Etangs_2012, Levine_2024, Wargelin_2024, Pillitteri_2026}. Other potential sources of helium absorption variation include stellar wind variability and shear instabilities, which could vary the line depth at constant XUV flux \cite{Wang_2021}.

Variable escape of neutral hydrogen and helium from planetary atmospheres has been observed in other systems and is predicted from escape models. In the case of HD\,189733b, excess Lyman-alpha absorption was observed during transit at a depth of 14\% in September 2011, following a clear non-detection in April 2010 \cite{Etangs_2012}. A subsequent survey of HD\,189733b found variation in helium absorption depth of $\sim$32\% and a predicted 60\% variation with an XUV flux variation of only 33\%, which is typical for the star \cite{Zhang_2022}. X-ray flux variations of $\sim$50\%, and corresponding helium absorption variations, have also been observed for WASP-69b between 2017 and 2023 \cite{Levine_2024}. Helium absorption variability of $\sim$50\% has also been observed for the Neptune-mass planet HAT-P-11b, over five days \cite{Allart_2018}. Comprehensive hydrodynamic escape and radiative transfer models predict helium absorption depth variability of $\geq5 \times$ between minima and maxima in a solar-like activity cycle for HD 209458b \cite{Zhang_2022, Taylor_2025}. 

M dwarfs typically exhibit activity cycles ranging from 2 to 20 years \cite{Suarez_2016, Alonso_2019, Mignon_2023}. There is no correlation between stellar rotation period and activity cycle period for M dwarfs, likely due to random, irregular magnetic surface activity. \cite{Savanov_2012, Suarez_2016, Alonso_2019}. We therefore cannot predict the period and magnitude of LHS~1140's activity cycle. However, activity cycle minima and maxima can be as close as one year apart, so activity-driven XUV variability could potentially explain the variable helium absorption.




\begin{table}[htbp]
\def\arraystretch{1.2}
\centering
\caption{\textbf{X-ray parameters of LHS\,1140 derived from the X-ray spectral modeling}: X-ray flux, $F_{\rm X}$, X-ray luminosity, $L_{\rm X}$, and the ratio of the X-ray luminosity to the bolometric luminosity, $L_{\rm X}/L_{\rm bol}$, for LHS\,1140 ($1\sigma$ uncertainties; 90\% confidence limit in parentheses).}
\begin{tabular}{lll}
\hline
 & band [keV] & \\
\hline
$F_{\rm{X}}$ 
    & 0.25 to 2.0
    & $3.07^{+0.45}_{-0.41}$ (2.40, 3.83) $\times\, 10^{-15}\,\mathrm{erg\,cm^{-2}\,s^{-1}}$\\
$F_{\rm{X}}$ 
    & 0.25 to 10.0
    & $3.15^{+0.50}_{-0.42}$ (2.44, 3.99) $\times\, 10^{-15}\,\mathrm{erg\,cm^{-2}\,s^{-1}}$\\
$L_{\rm{X}}$ 
& 0.25 to 2.0
    & $7.77^{+1.14}_{-1.03}$ (6.08, 9.71) $\times\,
    10^{25}\,\mathrm{erg\,s^{-1}}$\\
$L_{\rm{X}}$
    & 0.25 to 10.0
    & $7.98^{+1.26}_{-1.06}$ (6.18, 10.09) $\times\,
    10^{25}\,\mathrm{erg\,s^{-1}}$\\
$L_{\rm{X}}/L_{\rm bol}$ 
    & 0.25 to 2.0
    & $6.49^{+1.34}_{-1.25}$ (4.9, 8.2) $\times\, 10^{-6}$\\
$L_{\rm{X}}/L_{\rm bol}$     
    & 0.25 to 10.0
    & $6.50^{+1.34}_{-1.26}$ (4.9, 8.2) $\times\, 10^{-6}$\\
\hline
\label{table2}
\end{tabular}
\end{table}








\end{document}